\documentclass[rmp,superscriptaddress,twocolumn,preprint]{revtex4-1}

\usepackage{amsmath, amsthm, amscd, amssymb}
\usepackage{graphicx, braket}
\usepackage{graphics}
\usepackage{bm}
\usepackage{bbm}\usepackage{natbib}
\usepackage{setspace}
\fussy

\begin{document}

\setlength{\belowdisplayskip}{4pt} \setlength{\belowdisplayshortskip}{0pt}
\setlength{\abovedisplayskip}{4pt} \setlength{\abovedisplayshortskip}{0pt}

\title{Models of Wave-function Collapse, Underlying Theories, and Experimental Tests}

\author{\large \bf Angelo Bassi}
\email{bassi@ts.infn.it}
\affiliation{Department of Physics \ University of Trieste\ Strada Costiera 11 \ 34151 Trieste Italy} \affiliation{Istituto Nazionale di Fisica Nucleare \ Trieste Section \ Via Valerio 2\ 34127 Trieste Italy}

\author{\large \bf Kinjalk Lochan}
\email{kinjalk@tifr.res.in} \affiliation{Tata Institute of Fundamental Research\  Homi Bhabha Road\ Mumbai 400005\ India}

\author{\large \bf Seema Satin}
\email{satin@imsc.res.in}
\affiliation{Inst. of Math. Sciences\ IV Cross Road\ CIT Campus\ Taramani\ Chennai 600 113
\ India}

\author{\large \bf Tejinder P. Singh}
\email{tpsingh@tifr.res.in}
\affiliation{Tata Institute of Fundamental Research\ Homi Bhabha Road\  Mumbai 400005\ India}

\author{\large \bf Hendrik Ulbricht}
\email{h.ulbricht@soton.ac.uk}
\affiliation{School of Physics and Astronomy\ University of Southampton\ SO17 1BJ\ UK}


\maketitle


\smallskip

\centerline{\bf ABSTRACT}
\noindent Quantum mechanics is an extremely successful theory that agrees with every experiment. However, the principle of linear superposition, a central tenet of the theory, apparently contradicts a commonplace observation: macroscopic objects are never found in a linear superposition of position states.  Moreover, the theory does not explain why during a quantum measurement, deterministic evolution is replaced by probabilistic evolution, whose random outcomes obey the Born probability rule. In this article we review an experimentally falsifiable phenomenological proposal, known as Continuous Spontaneous Collapse: a stochastic non-linear modification of the Schr\"{o}dinger equation, which resolves these problems, while giving the same experimental results as quantum theory in the microscopic regime.  Two underlying theories for this phenomenology are reviewed: Trace Dynamics, and gravity induced collapse.  As one approaches the macroscopic scale, predictions of this proposal begin to differ appreciably from those of quantum theory, and are being confronted by ongoing laboratory experiments that include molecular interferometry and optomechanics. These experiments, which test the validity of linear superposition for large systems, are reviewed here, and their technical challenges, current results, and future prospects  summarized.  We  conclude that it is likely that over the next two decades or so, these experiments can verify or rule out the proposed stochastic modification of quantum theory.


\tableofcontents

\section{Introduction}
Quantum theory has been extremely successful in explaining results of experiments, ranging from the spectrum of black-body radiation, atomic spectra, molecular chemistry, atomic interferometry, quantum electrodynamics and nuclear physics, to properties of lasers, superconductivity, semi-conductor physics, Bose-Einstein condensation, Josephson junctions, nano-technology, applications in cosmology, and much more. The theory is not contradicted by any experiment. Yet, there is one apparently innocuous observed phenomenon the theory seems unable to explain, and in fact seems to contradict with. This is the observed absence of superposition of different position states in a macroscopic system. Quantum theory, by virtue of the principle of linear superposition,  predicts that a microscopic object such as the electron can be in a superposition of different positions at the same time, and this is of course observed, for example in the famous double-slit interference experiment. Moreover, the theory in principle makes no distinction between microscopic objects and macroscopic ones, and predicts that large objects can also be in more than one place at the same time. But this is not what we observe. A table for example, unlike the electron, is never observed to be `here' and `there' simultaneously. 

Why should this be so? The present review article is devoted to discussing one possible proposed resolution, known as Models of Spontaneous Wave Function  Collapse, which is experimentally falsifiable. Namely that, although quantum theory is extremely successful in the microscopic domain, it is an approximation to a more general theory. This general theory is capable of explaining the absence of macroscopic superpositions. It goes over to quantum mechanics in the microscopic limit, and to classical mechanics in the macroscopic limit, but differs from both quantum and classical mechanics in the intermediate [mesoscopic] regime which marks the transition from the micro- to the macro- world.  A large number of experiments worldwide are operating or are being planned,  to test the validity of linear superposition in the mesoscopic domain, and in this article we will review the proposed modification to quantum mechanics, and the laboratory experiments which can falsify this proposal.

\subsection{The relation between non-relativistic quantum mechanics and classical mechanics}
The classical dynamics of a system of particles having a Hamiltonian $H$ is described in phase space $(q_i,p_i)$ by Hamilton's equations of motion
\begin{equation}
\dot{q_i} = \frac{\partial H}{\partial p_i}, \quad \dot{p_i} = - \frac{\partial H}{\partial q_i}\;
\label{hj}
\end{equation}
or via Poisson brackets
\begin{equation}
\dot{q}_i = \{q_i,H\}\; , \quad \dot{p}_i = \{p_i, H\}\;  .
\end{equation}
The state of the system at an initial time $t_0$ is a point in the phase space, and the equations of motion determine the location of the system point at a later time.  An equivalent description of the dynamics is through the Hamilton-Jacobi equation
\begin{eqnarray}
-\frac{\partial S}{\partial t} = H\left(q_i, \frac{\partial S}{\partial q_i}\right)\;
\end{eqnarray}
where $S$ is the action of the system ~\cite{Landau-Lifshitz:1976}.

In contrast, the quantum dynamics of this system is described by first converting the $q_i$ and $p_i$ to operators $\bf{q}_i,\bf{p}_i$ satisfying the commutation relations $[{\bf{q}_i,\bf{p}_i}]=i\hbar$ and then proposing that the operators evolve via the Heisenberg equations of motion
\begin{equation}
\dot{{\bf q}_i} = -\frac{i}{\hbar}[{\bf q}_i,{\bf H}]\; , \quad \dot{{\bf p}_i} = -\frac{i}{\hbar} [\bf{p}_i, \bf{H}] \; .
\end{equation}
Quantum dynamics is equivalently described by the time evolution of the system's wave-function $\psi$, which is a normalized element of a Hilbert space and obeys the norm-preserving Schr\"{o}dinger equation
\begin{equation}
i\hbar \frac{\partial \psi}{\partial t} = H\psi, \qquad \int dq \psi^{*} \psi = 1\; .
\label{sch}
\end{equation}

In the Heisenberg picture the relation between quantum and classical mechanics is  expressed by 
replacing operators by ordinary functions, 
and the commutators in the equations of motion by Poisson brackets. A more insightful comparison is obtained in the Schr\"{o}dinger picture, and for the purpose of illustration it is adequate to consider the case of a single particle of mass $m$ moving in one dimension, for which the Schr\"{o}dinger equation can be written in the position representation, after defining $\psi\equiv e^{iS/\hbar}$, as
\begin{equation}
-\frac{\partial S}{\partial t} = \frac{1}{2m} \left(\frac{\partial S}{\partial q}\right)^2 + V(q)
-\frac{i\hbar}{2m} \frac{\partial^{2}S}{\partial q^2} \; .
\label{sch-hj}
\end{equation}

In the approximation in which the last term in (\ref{sch-hj}) can be neglected, this equation reduces to the classical Hamilton-Jacobi equation (\ref{hj})
\begin{equation}
-\frac{\partial S}{\partial t} = \frac{1}{2m} \left(\frac{\partial S}{\partial q}\right)^2 + V(q) \;
\label{hj2}
\end{equation}
provided the quantity $S$ is assumed to be real and identified with the action of the system. This essentially corresponds to the limit $S\gg\hbar$. [We will not consider the more precise treatment where $S$ is separated into its real and imaginary parts, as it is not crucial to the present discussion].

There is thus a well-defined sense in which the Schr\"{o}dinger equation goes over to the Hamilton-Jacobi equation in the limit, and a description of the dynamics in Hilbert space gets replaced by a description in terms of evolution of position and momentum coordinates in phase space. Yet, there is a profound aspect which gets lost in the limiting process. The Schr\"{o}dinger equation is linear: if $\psi_1$ and $\psi_2$ are two solutions of (\ref{sch}) then the linear superposition $c_1\psi_1 + c_2\psi_2$ is also a solution, where $c_1$ and $c_2$ are complex coefficients. On the other hand the Hamilton-Jacobi equation (\ref{hj}) is non-linear : if $S_1$ is a solution corresponding to one space-time trajectory, and $S_2$ is a solution corresponding to another space-time trajectory, then clearly $a_1 S_1 + a_2 S_2$ is not a solution of this equation.

In particular, if $\psi_1$ is a wave-packet which is peaked around one classical solution and $\psi_2$ is a wave-packet peaked around another classical solution, quantum mechanics predicts that the sum of these two wave-packets is also a solution, and in principle such solutions should be observed in nature.  However, according to classical mechanics, such a superposition is not a solution of the equations of motion, nor is it observed in the macroscopic world around us. Naively, we believe that classical mechanics, which applies to macroscopic systems, is a limiting case of quantum mechanics, and hence quantum mechanics should apply to large systems as well. Why then we do not observe macroscopic superpositions [such as a table being `here' and `there' at the same time]?

One might argue that even though the Hamilton-Jacobi equation is non-linear, its non-linearity cannot be used to deduce the observed  absence of macroscopic superpositions, because the classical theory is after all an approximation.  The last term in Eqn. (\ref{sch-hj}), howsoever small, is always non-zero and present, and can be used to transform back to the linear Schr\"{o}dinger equation. At a fundamental level, the description of the dynamics, even for a macroscopic classical object, is in terms of the wave-function of the quantum state, and not in terms of the action which appears in the Hamilton-Jacobi equation. Hence superpositions must be there. Nonetheless, one is left with the discomforting feeling that the prediction of the
Hamilton-Jacobi equation regarding position superpositions seems to be at variance with quantum theory, and in accord with what is actually observed. Thus one needs to explain the following\footnote{One should keep in mind the difference between the conceptual issue raised here, and the purely technical fact that performing an experiment which tests macroscopic superpositions (table ``here'' + table ``there'') is practically unfeasible.} : why is it that macroscopic objects which obey the rules of classical mechanics are not found in superposition of different position states, in spite of quantum theory suggesting otherwise? There is no unique universally accepted answer to this question. In this sense this is an unsolved problem.

The absence of macroscopic superpositions is of course at the heart of the so-called quantum measurement problem ~\cite{Bassi2:00}. Suppose a quantum system which is in a superposition of two eigenstates $\psi_1$ and $\psi_2$ of a physical observable ${\bf O}$ interacts with a classical measuring apparatus $A$. Let us say that the state $\psi_1$ of the quantum system corresponds to a pointer position state $A_1$  of the apparatus [meaning that if the system had been in the state $\psi_1$ and interacted with the apparatus, the pointer would result in the position $A_1$ and we would interpret that the observable had the value $O_1$]. Similarly, the pointer position $A_2$ corresponds to the system state $\psi_2$ and a value $O_2$ for the observable ${\bf O}$. Immediately after interaction, the combined state of the system and apparatus is
\begin{equation}
\psi = c_1 \psi_1 A_1 + c_2 \psi_2 A_2
\label{entangled-state}
\end{equation}
where $c_1$ and $c_2$ are complex coefficients proportional to the relative amplitudes for the system to be in the two states $\psi_1$ and $\psi_2$.

According to quantum mechanics, this state $\psi$ of Eqn. (\ref{entangled-state}) should evolve linearly by way of Schr\"{o}dinger evolution, and the linear superposition of the two parts should be preserved. But that is not what is observed in a quantum measurement. The outcome of the measurement is either pointer at position $A_1$ (and hence system is driven to state $\psi_1$) or pointer at position $A_2$ (system is driven to state $\psi_2$). Repeated measurements on the same initial quantum state yield outcome $\psi_1$ or $\psi_2$ with relative probability $|c_1|^2 : |c_2|^2$. This is the Born probability rule. The process of measurement destroys linear superposition of the initial states $\psi_1$ and $\psi_2$. This indeed has to do with the fact that the apparatus [which is a macroscopic object] is never simultaneously observed in a linear superposition of pointer position states $A_1$ and $A_2$.  To the extent that we do not understand why macroscopic objects are not found in superposed states, we do not understand why the measurement process breaks superposition.

Perhaps even more remarkable is the emergence of probabilities. The Schr\"{o}dinger evolution is deterministic, and so is the classical evolution according to the Hamilton-Jacobi equation. In our discussion above, on the transition from the Schr\"{o}dinger equation to the Hamilton-Jacobi equation, nowhere did we encounter probabilities. And for good reason. Because the initial state is always exactly specified [including at the start of the measurement process, as in Eqn. (\ref{entangled-state})], unlike in classical probability theory, where probabilities arise because of uncertainty in our knowledge of the initial state of the system. Thus the status of probabilities in quantum theory is absolutely unique, and besides explaining the absence of macroscopic superpositions one must also explain why during a measurement probabilities arise, in violation of deterministic linear superposition, and the quantum system is driven to one or the other outcome in accordance with the Born rule.

Another important and related unsolved problem is the following: when do we call a physical system a quantum system and when we do call it a classical measuring apparatus? In other words, where is the quantum-classical divide? How much mass, or degrees of freedom [say number of nucleons] should an object have, before it qualifies to be an apparatus? Of course, in order for it to be called an apparatus, different pointer positions should never be simultaneously realized, but one does not know at what mass scale this transition from micro to macro [and the concurrent breakdown of superposition] takes place. Interferometry experiments have shown that quantum theory, and hence linear superposition, holds for molecules at least as large as those having about a thousand atoms [hence a molecular mass of $10^{-21}$ grams]. Efforts are afoot to push this test limit up to objects of about a million atoms  [$10^{-18}$ grams]. On the other end, classical behavior [absence of superpositions of states corresponding to different positions] is known to hold out down to about a microgram
[$10^{18}$ atoms]. There is thus an enormous desert of some fifteen orders of magnitude, where linear quantum superposition yet remains to be tested experimentally. Does quantum mechanics hold at all scales, including macroscopic scales, and is there a way to understand the absence of macroscopic superpositions while staying within the framework of quantum theory? Or is it that somewhere in the grand desert modifications to quantum theory start becoming significant, so that linear superposition becomes more and more of an approximate principle as the size of a system is increased, until for large objects the superposition of states corresponding to different positions is no longer a valid principle? What exactly is the nature of the quantum-to-classical transition? A large number of ongoing and planned experiments worldwide are gearing up to address this question.

The paradoxical issue of deterministic evolution followed by a peculiar probabilistic evolution during a measurement was of course well-appreciated by the founding fathers of quantum mechanics. Over the last eighty-five years or so since the discovery of the Schr\"{o}dinger equation, extraordinary theoretical effort has been invested in trying to find an answer to what is generally known as the measurement problem in quantum mechanics ~\cite{Albert:92, Bell:87, Ghirardi:2005, Maudlin:2011, Wheeler-Zurek:1983, Leggett:2002, Leggett:2005}. In the next sub-section we give a brief overview of a few of the major categories of the explanations, keeping in mind that the modern outlook is to discuss this problem not in isolation, but in conjunction with the question of lack of macroscopic superpositions, and as a part of the much broader investigation of the exact nature of the quantum-to-classical transition.

Our review of the measurement problem will be almost exclusively confined to the context of non-relativistic quantum mechanics, as the relativistic version seems not within reach at the moment (though it does not seem symptomatic of a deep incompatibility of modified quantum mechanics and relativity). Thus we will not discuss issues raised by the instantaneous nature of wave-function collapse, such as the EPR paradox, whether this `violates the spirit of relativity' or whether there is a need for a radical change in our ideas about space-time structure.

\subsection{Proposed resolutions for the quantum measurement problem and for the observed absence of macroscopic superpositions}
\subsubsection{The Copenhagen Interpretation}
The Copenhagen interpretation ~\cite{Bohr:28} [reprinted in ~\cite{Wheeler-Zurek:1983}] postulates an artificial divide between the micro-world and the macro-world, without quantitatively specifying at what mass scale the divide should be. Microscopic objects obey the rules of quantum theory [superposition holds] and macroscopic objects obey the rules of classical mechanics [superposition does not hold]. During a measurement, when a micro-system interacts with a macro-system, the wave-function of the micro-system `collapses' from being in a superposition of the eigenstates of the measured observable, to being in just one of the eigenstates. This collapse is postulated to happen in accordance with the Born probability rule, and no dynamical mechanism is specified to explain how the collapse takes place, in apparent contradiction with the linearity of the Schr\"{o}dinger equation.

von Neumann gave a more precise form to this interpretation by explicitly stating that evolution in quantum theory takes place in two ways : (i) deterministic evolution according to the Schr\"{o}dinger equation before a measurement, as well as {\it after} a measurement, and (ii) non-deterministic, probabilistic evolution [the projection postulate] during  a measurement ~\cite{Neumann:55}.

At a pragmatic level, this can be taken to be a perfectly valid set of rules, in so far as the goal is to apply quantum theory to sub-atomic, atomic and molecular systems, and to compare with experiments the predictions based on theoretical calculations. However, the interpretation bypasses the questions raised in the previous section, by simply raising unresolved issues to the level of postulates. The interpretation creates an ill-defined micro-macro separation, which is being challenged by modern experiments which are verifying superposition for ever larger systems. There is no precise definition as to which systems classify to serve as a `classical measuring apparatus'. Even though there is a sense in which the Hamilton-Jacobi equation is a limit of the Schr\"{o}dinger equation, no attempt is made to explain the apparently different predictions of the two theories with regard to absence of macroscopically different position superpositions. At a fundamental level one should prescribe a physical mechanism which causes the so-called collapse of the wave-function.

The Copenhagen interpretation does not solve the quantum measurement problem, nor does it explain the absence of macroscopic superpositions.

The ``histories'' approach is an observer independent generalization of the Copenhagen interpretation wherein the notions of apparatus and measurement are replaced by the more precise concept of histories. In this approach, the reduction of the state vector appears as a Bayesian statistical rule for relating the density matrix after measurement to the density matrix before measurement ~\cite{Omnes:92, Omnes:94, Omnes:99, Griffiths:02, Hartle:92}.

\subsubsection{Decoherence}
The phenomenon of decoherence, which is observed in laboratory experiments, highlights the role played by the environment when a quantum system interacts with a measuring apparatus, during the process of measurement. By environment is meant the system of particles which surround the apparatus. More precisely one could define the environment as the collection of particles which is present within a radius $cT$ of the apparatus, where $T$ is the duration of the measurement: these are hence particles which can causally interact with and influence the apparatus during a measurement.

To illustrate the effect of decoherence, we assume that the system on which measurement is to be made is a two state system initially in the state
\begin{equation}
\psi(t=0) = c_1 \psi_1 + c_2 \psi_2.
\end{equation}
Denoting the initial state of the apparatus by $\phi_{0A}$ and the initial state of the environment by
$\phi_{0E}$, we can write the net initial state as the direct product
\begin{equation}
\Phi_0 = \psi (t=0) \; \phi_{0A} \; \phi_{0E}.
\end{equation}
Over time, as a result of interaction, this state evolves into the state
\begin{equation}
\Phi(t)=  c_1 \psi_1 \phi(t)_{EA1} +
c_2 \psi_2 \phi(t)_{EA2}.
\label{newstate}
\end{equation}
Here, $\phi(t)_{EA1}$ and $\phi(t)_{EA2}$ denote macroscopically distinguishable entangled states of the apparatus and the environment.

As demonstrated below, during measurement, the process of decoherence operates in such a way that very quickly, the inner product
\begin{equation}
\langle\phi(t)_{EA1}|\phi(t)_{EA2}\rangle\ \rightarrow 0\;
\label{interf}
\end{equation}
starting from the value unity at $t=0$. The final state is reduced to a statistical mixture of states with relative weights $|c_1|^2 : |c_2|^2$.

This by itself does not explain why during a measurement
\begin{equation}
\Phi(t) \rightarrow \psi_1\phi(t)_{EA1} \quad {\rm or}  \quad  \Phi(t) \rightarrow \psi_2\phi(t)_{EA2}\; .
\nonumber
\end{equation}
Decoherence destroys interference amongst alternatives, which is what Eqn. (\ref{interf}) signifies, but because it operates within the framework of linear quantum mechanics, it cannot destroy superposition. Since loss of superposition is what is seen during a measurement, decoherence does not explain the measurement process. What (\ref{interf}) implies is that decoherence forces quantum probability distributions to appear like classical probabilities (weighted sums of alternatives); however this is neither necessary nor sufficient to explain the outcome of an individual measurement. [The issue of non-observation of superposition for macro-systems becomes even more acute in the case of isolated systems as there are no environment degrees of freedom to be traced out. Then the theory seems unable to explain the breakdown of superposition for isolated macroscopic systems, such as the universe as a whole.]

The loss of interference can be understood as a consequence of the interaction of the very large number of particles of the environment with the apparatus. Assuming that the measurement starts at $t=0$, the product $\langle\phi(t)_{EA1}|\phi(t)_{EA2}\rangle$,
which is one at $t=0$, rapidly goes to zero.  To see this, one notes that in general a particle of the environment, say the $i$th particle, will be scattered by the state $A_1$ of the apparatus to a final state different from the one to which it will be scattered from the apparatus state $A_2$. Thus, the product
\begin{equation}
\langle E_1|E_2\rangle(t) = \Pi_i  \ {}_{i}\langle E(t=0)|S_{A1} S_{A2}|E(t=0)\rangle _i \nonumber
\end{equation}
is made up of an ever increasing number of quantities, each of which is smaller than one, $S_{A1}$ and
$S_{A2}$ being scattering matrices describing the action of the apparatus on the environment. Hence this product can be written as $\exp (-\Lambda t)$ and goes to zero for large $t$, $\Lambda$ being the decoherence rate. Because the environment has a very large number of particles, this cross-product between the two environment states is very rapidly suppressed, and is responsible for the emergence of the property described by Eqn. (\ref{interf}). The decoherence time-scale $\Lambda^{-1}$ is much smaller than the duration $T$ of the measurement.

The above discussion is partly based on the article by Adler ~\cite{Adler:2002} where a more detailed description of decoherence in the context of measurement can be found. There is a vast literature on decoherence, including the experiments and models by ~\cite{Brune:96, Harris:81,Gerlich2007kapitza},   books by  ~\cite{Joos:03, Schlosshauer:2007, Breuer:2000} and the seminal
papers ~\cite{Zeh:70, Joos:85, Caldeira:81}  and ~reviews
\cite{Schlosshauer:2005, Vacchini:2009, Zurek:91, Zurek:03}, and ~\cite{Bacciagaluppi:2007}.

\subsubsection{Many-Worlds interpretation}
The Many-worlds interpretation was invented by Everett ~\cite{Everett:57}  to counter the Copenhagen interpretation. According to Everett, evolution during a measurement is also Schr\"{o}dinger evolution, and there is no such thing as a non-deterministic probabilistic evolution during the measurement. Thus in this interpretation the state (\ref{entangled-state}) evolves during a measurement according to the Schr\"{o}dinger equation. Why then does it appear as if only one of the two outcomes has been realized? The answer is that the state continues to be of the form
\begin{equation}
\Psi = c_1 \psi_1 A_1 O_1 + c_2 \psi_2 A_2 O_2
\label{entangled-state-observer}
\end{equation}
where $O_1$ ($O_2$) is the state of the observer where the observer detects the system and apparatus in state one (two). The two parts of this state exist in two different branches of the Universe.

Despite appearances, there is no logical inconsistency in the above interpretation, as has been argued by Everett : it is merely the assertion that Schr\"odinger evolution is universally valid at all scales, and the breakdown of superposition during a measurement is only apparent, not real. The hard part is to explain the origin of probabilities and the Born probability rule. If the evolution is deterministic through and through, why should there be a definite probability associated with an outcome? In our opinion, despite extensive investigation, this problem remains unsolved in the many-worlds
interpretation ~\cite{DeWitt:73, Tegmark:2007, Kent:1990, Wallace:2003, Deutsch:1998, Vaidman:2002,Hsu:2011, Saunders2010, Putnam:05, Barrett:12}.

\subsubsection{Decoherence + Many-Worlds}
Decoherence by itself does not solve the measurement problem, because it does not destroy superposition. However, one could propose that the decohered alternatives both continue to co-exist in different branches of the Universe, in the sense of the many-worlds interpretation, and these branches do not interfere with each other because decoherence is operating. While this merger helps both the decoherence and the many-worlds picture of a measurement, the origin of the Born probability rule continues to be not understood, and as of now is essentially added as a postulate.

This is perhaps today the `establishment view', wherein one believes that one does not need to modify quantum theory in order to explain measurement. Its major weakness though is that it is not experimentally falsifiable. What experiment can one perform in order to find out whether the other branches of the many-worlds exist or not? In the absence of such an experiment, we have at hand another interpretation of the same quantum theory, an interpretation which cannot be experimentally distinguished from the Copenhagen interpretation.

For discussions on decoherence in the context of many-worlds see ~\cite{Bacciagaluppi:2001}.

\subsubsection{Bohmian Mechanics}
Bohmian mechanics is a quantum theory of particles in motion. The positions of the particles of an $N$-particle system are  $Q_k\quad k=1,..., N$ moving in physical space. The role of the wave function, 
being governed by  Schr\"{o}dinger equation, is to direct the motion of the particles. The theory is deterministic, randomness enters like in classical mechanics via typicality.   It is shown that the outcomes in measurement experiments are governed by  Born's statistical  rule. The equation of motion for the particles is given by $v_k = {dQ_{k}}/{dt}$ where
\begin{eqnarray}
\frac{dQ_{k}}{dt} = \frac{\hbar}{m_k} Im \nabla _{Q_{k}} \log \psi (Q_1, Q _2,  ..., Q_N,t)\nonumber .
\end{eqnarray}
Bohmian mechanics is a quantum theory in which the collapse of the wave function is effective, in contrast to collapse models,  so that macroscopic interference is in principle possible.
Predictions of Bohmian mechanics agree with those of orthodox quantum mechanics, whenever the latter are unambiguous.  Bohmian mechanics would be falsified if collapse models were experimentally verified. 

For literature on Bohmian mechanics
see ~\cite{Bohm:52, Bohm2:52, Bub:1997, Duerr:92, Holland, Bohmbook, Duerr, DGZ}.

 \subsubsection{Quantum theory is an approximation to a more general theory}
It is proposed here that the measurement problem and the apparent inability of quantum theory to explain the absence of macroscopic superpositions are a consequence of trying to apply the theory in a domain where it is not valid. It is proposed that there is a universal dynamics, to which quantum theory and classical mechanics both are approximations. In the domain of a quantum measurement, the universal dynamics differs from quantum dynamics, and operates in such a way that interaction of the quantum system with the apparatus causes a collapse of the wave-function from a superposition to one of the eigenstates. The collapse is a physical, dynamical process, and hence the universal dynamics provides a physical explanation for the ad hoc collapse postulated by the Copenhagen interpretation. Furthermore, the collapse is shown to obey the Born probability rule. The universal dynamics is 
stochastic: the outcome of a measurement is random and unpredictable, but the mathematical structure of the dynamics is such that repeated measurements on an ensemble of identically prepared quantum systems are shown to yield different outcomes, in relative frequencies which obey the Born rule.

The universal dynamics must be non-linear, in order to allow for the breakdown of superposition during a measurement. Yet, the non-linearity must be extremely negligible in the microscopic domain, so that the experimentally observed linear superposition in microscopic quantum systems is reproduced. The new dynamics must be stochastic; but once again, stochasticity must be negligible for microscopic systems, so that the deterministic Schr\"{o}dinger evolution prevails. Thirdly, as one progresses from microscopic to macroscopic systems, the universal dynamics must allow for non-unitary [but norm-preserving] evolution :
this is essential so that stochastic evolution can cause all but one outcome to decay exponentially, something which would not be permitted during unitary evolution. Again, non-unitarity must be utterly negligible for microscopic systems. Thus, the universal dynamics possesses a set of parameters, whose effective values are determined for the system under study in such a way that for microscopic systems these parameters take values so that the dynamics is experimentally indistinguishable from quantum dynamics. Similarly, for macroscopic systems, there is an amplification mechanism built in the equations, such that the dynamics coincides with classical dynamics. For systems that are mesoscopic [neither micro nor macro] the dynamics differs from both classical and quantum, and is hence experimentally distinguishable from quantum theory. The properties of non-linearity, stochasticity and non-unitarity also ensure position localization for macroscopic objects and hence dynamically explain the observed absence of macroscopic superpositions. It is clear that the universal dynamics is not tied to or invented for explaining just the measurement process or absence of macroscopic superpositions - these two phenomena just happen to be special cases where the new dynamics plays a vital role in understanding the phenomenon. We say that the universal dynamics, which describes the behaviour of micro, meso and macro objects, is intrinsically non-linear, stochastic, and non-unitary.

Over the last two decades or so there has been a significant progress in developing phenomenological models of such a universal dynamics. At the same time, one would like to know if there are underlying theoretical reasons [new symmetry principles for instance] which compel us to consider a generalization of quantum theory of the kind mentioned above, thereby lending an inevitability to the phenomenological models that have been proposed. There has been important progress on this front too. Thirdly, there have been important technological advances which are now permitting a host of experiments to be carried out to test these phenomenological models, and verify their predictions against those of quantum theory. Needless to add, all these three facets are best described as `work currently in progress'. The purpose of the present review is to present a state-of-the-art description of (i) the phenomenological models for the universal dynamics, (ii) the underlying theories, and (iii) ongoing experiments which aim to test these models and theories. It is our hope that a review of this nature will further stimulate the cross-talk between phenomenologists, theorists and experimentalists in this field, thereby helping the community to sharply focus on those aspects of phenomenology and experimentation which might be most directly accessible and feasible  in the near future.

\smallskip

\noindent {\bf Phenomenological models of modified quantum mechanics}

\noindent From early times, an aspect which has received considerable attention, is possible non-linear modifications of quantum theory, and this is not necessarily because of the measurement problem. Most fundamental differential equations which describe physical phenomena are non-linear, with linearity being a convenient approximation in some appropriate limiting cases. Why then should an equation as fundamental as the Schr\"{o}dinger equation be a singular exception to this rule? [It is of course known that there are very strong bounds on non-linearity in the atomic domain, see for instance the experiment described
in ~\cite{Bollinger:89}].
 Non-linear quantum theories may be classified as deterministic non-linear, and stochastic non-linear. For discussions on deterministic non-linear quantum mechanics the reader is referred to the works by Weinberg ~\cite{Weinberg:89,Weinberg2:89}, by Goldin ~\cite{Goldin:2000, Doebner-Goldin:1992}, and ~\cite{Birula:76}. It has often been suggested, and demonstrated, though perhaps not universally so, that deterministic non-linear modifications result in superluminal
propagation ~\cite{Polchinski:91, Gisin:90,GhirardiGrassi:91}. This, coupled with the fact that stochasticity appears to be an essential ingredient for explaining the origin of probabilities, has meant that investigations of a universal dynamics have tended to focus on stochastic non-linearities; see for instance ~\cite{Diosi:88c, Diosi:88a, Gisin:81,Gisin:84, Gisin:89, Gisin:95, Weinberg:11}.

With regard to the application of stochastic non-linearity to explain measurement, the pioneering paper is due to Pearle ~\cite{Pearle:76} - the paper is aptly titled ``Reduction of the state vector by a non-linear Schr\"{o}dinger equation''.  Pearle proposed to replace the Schr\"{o}dinger equation by a non-linear one, during measurement, and that certain variables which take random values just after the quantum system interacts with the apparatus, drive the system to one or the other outcomes, thus breaking superposition. For the choice of these random variables he suggested the phases of the state vectors immediately after the measurement. An appropriate assignment of the probability distribution of these phases over the allowed parameter space leads to the Born rule. It is noteworthy that this assignment of the probability distribution is something which has to be put in by hand, keeping in mind what probability rule one wants to emerge. This is one aspect where phenomenology and underlying theories need to do better even today: there should be a fundamental reason for the probability distribution over the stochastic variables, which inevitably implies the Born rule. Two important missing pieces, in order to consider the proposed dynamics a {\it universal} dynamics for all physical systems, were the preferred basis on which the wave function should collapse, as well as the trigger mechanism. Both limitations are overcome by the GRW model. Further investigations by Pearle were reported in ~\cite{Pearle:79, Pearle:82, Pearle:84, Pearle:89} and reviewed in ~\cite{Pearle:99}.

The next major advance came from Ghirardi, Rimini and Weber ~\cite{Ghirardi:86} in a seminal paper titled ``Unified dynamics for microscopic and macroscopic systems'' and the model has come to be known as the GRW model.
There were two guiding principles for this dynamical reduction model (also known as QMSL: Quantum Mechanics with Spontaneous Localization):

1. The preferred basis - the basis on which reductions take place - must be chosen in such a way as to guarantee a definite position in space to macroscopic objects.

2. The modified dynamics must have little impact on microscopic objects, but at the same time must reduce the superposition of different macroscopic states of macro-systems. There must then be an amplification mechanism when moving from the micro to the macro level.

The reduction is achieved by making the following set of assumptions:

1. Each particle of a system of $n$ distinguishable particles experiences, with a mean rate $\lambda_{\text{\tiny GRW}}^i$, a sudden spontaneous localization process.

2. In the time interval between two successive spontaneous processes the system evolves according to the usual Schr\"{o}dinger equation.

In their model, GRW introduced two new fundamental constants of nature, assumed to have definite numerical values, so as to reproduce observed features of the microscopic and macroscopic world. The first constant, $\lambda_{\text{\tiny GRW}}^{-1} \sim 10^{16}$ seconds, alluded to above, determines the rate of spontaneous localization (collapse) for a single particle. For a composite object of n particles, the collapse rate is $(\lambda_{\text{\tiny GRW}} n)^{-1}$ seconds. The second fundamental constant is a length scale $r_C \sim 10^{-5}$ cm which is related to the concept that a widely spaced wave-function collapses to a length scale of about $r_C$ during the localization.

A gravity based implementation of the GRW model was studied by Di\'osi ~\cite{Diosi:89} and generalized
by ~\cite{Ghirardi3:90}.

The GRW model has been upgraded into what is known as the CSL (Continuous Spontaneous Localization) model by Ghirardi, Pearle and Rimini~\cite{Ghirardi2:90}. In CSL a randomly fluctuating classical field couples with the particle number density operator of a quantum system to produce collapse towards its spatially localized eigenstates. The collapse process is continuous in time, and this allows to express the dynamics in terms of a single stochastic differential equation, containing both the Schr\"odinger evolution and the collapse of the wave function. The narrowing of the wave function amounts to an increase in the energy of the particle, and actually amounts to a tiny violation of energy conservation.

An outstanding open question with regard to the dynamical reduction models is the origin of the random noise, or the randomly fluctuating classical scalar field, which induces collapse.

The current status of the Spontaneous Collapse models is discussed in detail in Section II.

A modern approach to stochastic reduction is to describe it using a stochastic  non-linear Schr\"{o}dinger equation, an elegant simplified example of which is the following one-particle case [known as QMUPL : Quantum Mechanics with Universal Position Localization ~\cite{Diosi:89}]  [See Section II for details]:
\begin{eqnarray}
d\psi(t) &=& \left[ - \frac{i}{\hbar}  H dt + \sqrt{\lambda} ( q - \langle q\rangle _t)dW_t  -  \right. \nonumber \\
& & \left. \frac{\lambda}{2} (q - \langle q\rangle _t)^2 dt \right] \psi(t) \; .
\label{stocheqn}
\end{eqnarray}
$q$ is the position operator,  $\langle q\rangle _t$ is its expectation value, and $\lambda$ is a constant, characteristic of the model, which sets the strength of the collapse. $W_t$ is a Wiener process which describes the impact of stochasticity on the dynamics.  As for the GRW and CSL models, this equation can be used to explain the collapse of wave-function during a measurement, the emergence of the Born rule, the absence of macroscopic superpositions, and the excellent matching of the linear theory with experiments for microscopic systems.

Various studies and arguments suggest that the structure of this equation is very rigid and tightly controlled, once one assumes [as is true here] that the evolution is norm-preserving, and secondly, superluminal propagation is not possible ~\cite{Gisin:89, Adler:04}.  There is then a unique relation between the coefficient $\sqrt{\lambda}$ of the diffusion [stochastic] term  and the coefficient $-\lambda /2$ of the drift term :
Drift Coefficient = $- 2$ (Diffusion Coefficient)$^2$. This is the well-known martingale structure for a stochastic differential equation.

In QMUPL, stochastic fluctuations take place only in the time direction and hence there is only one free parameter, i.e. $\lambda$. In contrast, in the CSL model the stochastic fluctuations exist over space too, and hence there is a second free parameter $r_C$ (analogous to GRW) which defines the scale of spatial localization. Of course, in the QMUPL and CSL models the stochastic process acts continuously, unlike in GRW, wherein the stochastic jumps are discontinuous and discrete.  In fact the QMUPL model can be understood as a scaling limit of the GRW process ~\cite{Duerr:11} [the collapse frequency goes to infinity and the spread $r_C$ goes to zero in such a way that their product remains a constant].

Part of the experimental effort on testing quantum mechanics, discussed in detail in Section IV, is devoted to testing the validity of equations such as (\ref{stocheqn}) above, and measuring / setting bounds on the rate constant $\lambda$ and the length scale $r_C$.

\smallskip

\centerline{\bf Underlying Theories}
Phenomenological models of dynamical wave-function collapse propose an ad hoc modification of quantum mechanics, albeit retaining certain features such as norm-preservation and no superluminal propagation. In principle, there should be strong underlying theoretical reasons which make a compelling case for a modified quantum theory, rendering the phenomenological models inevitable. Here we mention three different theoretical developments in this connection, two of which arise from attempts to remove one or the other fundamental incompleteness in the formulation of quantum theory, and the third investigates how gravity might play an effective role in wave-vector reduction.

\smallskip

\centerline{\it Trace Dynamics}
Classical mechanics is supposed to be a limiting case of quantum theory. And yet, in its canonical formulation, quantum theory assumes a prior knowledge of classical dynamics! In order to `quantize' a system, one should know the classical configuration variables and their conjugate momenta, and one should first know the Hamiltonian or the action [for a path-integral formulation] of the classical system. This is unsatisfactory. In the canonical formulation, one then proposes canonical commutation relations such as $[{\bf q}, {\bf p}]=i\hbar$ in an ad hoc manner. Why should these be the relations, unless one already knows that they lead to results which match with experiments? It would be desirable to derive quantum theory from a starting point which is not classical mechanics, and then obtain classical mechanics as an approximation [and explain quantum measurement in the process]. The theory of {\it Trace Dynamics} developed by Adler and collaborators does well in progressing towards this goal \cite{Adler:94, Adler:04, Adler-Millard:1996, Adler:06a}.

Trace Dynamics [TD] assumes that the underlying theory is a classical dynamics of Grassmannian matrices, living on a given space-time. However, this classicality does not mean that TD is a `hidden variables' theory - for the eventual description is at an averaged level, where no reference is made to the matrices which have been coarse-grained over. The matrices satisfy the standard Lagrangian and Hamiltonian Dynamics, but as a consequence of global unitary invariance, the theory possesses a remarkable  additional conserved charge, not present in point-particle mechanics. This is the Adler-Millard charge ~\cite{Adler-Millard:1996}
\begin{equation}
\tilde{C} = \sum_i [q_i,p_i] - \sum_j \{ q_j,p_j\}
\end{equation}
where the first sum is over commutators of bosonic matrices, and the second sum is over anti-commutators of fermionic matrices. [See Section III for details]. This conserved charge, which has the dimensions of action, plays a central role in the emergence of quantum theory at a coarse-grained level.

Assuming that these matrix degrees of freedom are at a level sufficiently `microscopic' [e.g. at the Planck scale] that we do not observe them in our routine laboratory experiments, a statistical thermodynamics of this matrix dynamics is constructed. An equipartition theorem for the thermodynamically averaged quantities is derived, which results in the Adler-Millard charge being uniformly distributed across the averaged commutators, each of which is assumed to equal Planck's constant. This is the origin of the quantum commutation relations. As a consequence of the assumed invariance of thermodynamic averages under constant shifts in phase space, a Ward identity is derived, which under suitable assumptions shows that the thermally averaged $q$'s and $p$'s satisfy Heisenberg equations of motion. A relativistic quantum field theory is arrived at, and a non-relativistic Schr\"{o}dinger equation holds in the finite particle limit. Thus quantum theory is shown to emerge as the thermodynamic approximation to an underlying classical dynamics of Grassmann matrices possessing a global unitary invariance.

Perhaps the greatest asset of TD is to be able to go beyond this stage and address the quantum measurement problem in a natural manner. Quantum theory emerges in the thermodynamic approximation of the statistical mechanics of the underlying matrix  mechanics. Next, it is pertinent to consider the impact of Brownian motion fluctuations - remarkably these modify the Schr\"{o}dinger equation and provide the necessary stochastic  element for the collapse process to operate, and for the origin of probabilities. Subject to certain crucial assumptions for which one would eventually like to find a theoretical basis, the modified Schr\"{o}dinger equation is a non-linear and non-unitary [but norm-preserving] stochastic equation of the type used in the CSL model. In this way, Trace Dynamics, through its thermodynamic limit and the associated
statistical fluctuations provides a theoretical underpinning for the phenomenological collapse models.

TD is perhaps the most well-developed underlying theory one has at present for collapse phenomenology. Hence in Section III we give a detailed presentation  of the physics and mathematics of TD, leading to wave-vector reduction, and we also point out the open problems of TD which remain to be addressed.

\smallskip

\centerline{\it Quantum theory without classical spacetime}
Quantum theory requires an external classical time for describing evolution. This is of course so obvious and essential that it is almost never stated explicitly! However this dependence on an external classical time is perhaps the greatest incompleteness of quantum theory. Such a time is part of a classical spacetime geometry which is produced by classical matter fields according to the laws of general relativity. But classical matter fields are a limiting case of quantum fields. If there were no classical fields in the Universe, but only fields subject to quantum fluctuations, there will be no definite metric available to describe the spacetime geometry. An argument due to Einstein, known as the Einstein hole argument ~\cite{Christian:98} then implies that if the metric is subject to quantum fluctuations, there is no longer available an underlying classical spacetime manifold. It is then not possible to describe quantum evolution.

 We see once again that via its dependence on external time, quantum theory depends on its classical limit [the required presence of a Universe dominated by classical matter]. This is unsatisfactory from a fundamental point of view, and hence there must exist an equivalent reformulation of quantum theory which does not refer to classical time. Such a reformulation can be shown to be the limiting case of a non-linear theory, with the non-linearity becoming important at the Planck mass scale. The non-linearity is possibly stochastic, and could have implications for resolution of the quantum measurement problem.   Tentative heuristic discussions towards this investigation have been given in ~\cite{Singh:2006, Singh:2009}.

A detailed systematic program to develop a formulation of quantum theory without classical time and to study its impact on quantum measurement has recently been begun, and is qualitatively described
in ~\cite{Singh:2011}.
The key symmetry principle here is that basic laws should be invariant under coordinate transformations of non-commuting coordinates. The motivation being that if quantum fluctuations destroy a classical spacetime manifold, a possible replacement for ordinary spacetime could be a non-commutative spacetime.
 This approach proposes to generalize Trace Dynamics by raising time, and space, to the level of matrices [operators]. This has been done in ~\cite{Lochan-Singh:2011} and it has been shown that by defining a
non-commutative space-time metric as a Trace over the space-time operators, a
Poincar\'e invariant dynamics can be constructed. We call this a generalized Trace Dynamics. Evolution is described with respect to the scalar constructed by taking Trace over the non-commutative metric - this is the analog of the ordinary proper time.

The next step is to construct, a la TD, a statistical mechanics for this generalized matrix dynamics and obtain the equilibrium thermodynamic approximation - this yields a generalized quantum theory which has an energy-time commutation relation and a generalized Schr\"{o}dinger equation with an operator time as one of the configuration variables. This is the sought for reformulation of quantum theory which does not refer to an external classical time ~\cite{Lochan:2012}. If the Universe is dominated by macroscopic objects, the consideration of Brownian motion fluctuations should yield position localization {\it and} the concurrent emergence of a classical space-time. This is the classical Universe, dominated by classical macroscopic objects and in possession of a classical spacetime. This Universe has a `sprinkling' of quantum fields and non-relativistic quantum systems. On the backdrop of this classical Universe one can postulate standard quantum theory [now that an external time is given] and then proceed to implement the program of Trace Dynamics to derive quantum dynamics from matrix mechanics, for this `sprinkling' of quantum matter fields on the classical spacetime background, and to resolve the attendant measurement problem.

The program described here aims to address a limitation of Trace Dynamics - a matrix treatment for matter fields while leaving spacetime as having point structure, thus leaving spacetime untouched. We regard such a limitation as one which should be addressed - in the process we see that removing time from quantum theory drives us to a starting point [generalized Trace Dynamics] whose eventual outcome is a possible resolution for the measurement problem. We thus wish to assert that there is a deep connection between the problem of time in quantum theory and the measurement problem in quantum theory ~\cite{Singh:2012}. Addressing the former will possibly compel us to consider a modification of quantum theory and that modification will have a bearing on the measurement problem.

Since this work is at present in an early stage of development, we will not discuss it any further in the remainder of the review.

\smallskip

\centerline{\it Gravity induced wave-function collapse}

\noindent The fact that the fundamental mass scale, Planck mass $M_{Pl} = (\hbar c/G)^{1/2} \sim 10^{-5}$ grams is not far from the scale where the micro to macro transition takes place has often intrigued some physicists. Mass seems to have something to do with deciding which objects are quantum, and which are classical, and mass also produces gravity. Could gravity thus play some role in causing wave-function collapse, and in localization of macro- objects? The idea that gravity might somehow be responsible for wave-function collapse has been seriously pursued by Karolyhazy and collaborators ~\cite{Karolyhazi:66, Karolyhazi:86}, by ~\cite{Diosi:87}, and by Penrose ~\cite{Penrose:96}. Penrose's proposal is also the subject of an important ongoing experiment aimed at testing it \cite{Marshall:03}. These issues will be discussed in Section III.

\bigskip

\centerline{\bf Experimental Tests}
The Copenhagen interpretation was a need of the times when it was proposed: pioneering experiments were being carried out for atomic systems. The measuring apparatus was a classical object, and the Born probability rule had to be invoked to explain the random outcomes of measurements. For some, this dual aspect of quantum theory - unitary evolution followed by wave-packet reduction - was the `truth' in quantum theory; this is how nature is. For others, this was completely unacceptable, and reinterpretations and new mathematical formulations such as many-worlds, Bohmian mechanics, and decoherent histories, were developed. However, the idea that quantum theory may be an approximation to a holistic theory which better explains both the unitary and reductionist aspects as limits of a unified mathematical description has taken shape only over the last three decades or so. And yet, none other than Einstein
himself ~\cite{Einstein:1949}  saw it this way early on, and had this to say about quantum theory:

{\it  ``... it would,  within the framework of future physics, take an approximately analogous position to that of statistical mechanics within the framework of classical mechanics''}.

In the light of the theory of Trace Dynamics and models of spontaneous wave function collapse, these words are prophetic. These modern ideas suggest the emergence of probabilities as a consequence of thermodynamic averaging in a deterministic theory, and the related significance of stochastic fluctuations. Above all, their predictions for results of experiments differ from the predictions of quantum theory. The difference will be far too small to be detectable for an atomic system, but starts becoming significant as the size of the system is increased. The best example of an experiment which could detect such a difference is double slit interference. If an object of mass $m$ is directed at a suitably prepared double slit, with appropriate slit width and separation between the slits, quantum theory predicts that an interference pattern will be seen on the screen no matter what the value of $m$. Not so, say collapse models. According to these models, the superposition state created after the object has passed the slits lasts only for a finite time $\tau$, where $\tau$ decreases with increasing $m$, and its value can be calculated precisely from a given theoretical model. Thus, according to these models, if the time of travel from the slits to the screen is greater than $\tau$, superposition will break down before the screen is reached, and no interference pattern will be seen. This is perhaps the cleanest confrontation that spontaneous collapse and gravity collapse models make with experiment. A successful diffraction experiment in the right mass domain will irrefutably confirm or rule out these models.

One should of course stay cautioned against assuming that quantum theory will be successful through and through, and that interference will be seen for all values of $m$. The fact that a theory is extremely successful in one part of the parameter space should not be taken as guarantee that it will continue to be successful in a different part of the parameter space - in the present instance the absence of macroscopic superpositions already provides reason for caution. And there are historical examples of long-standing successful theories eventually turning into approximations to more general theories when their extrapolation into a new part of the parameter space failed to be confirmed by experiment: (i) classical mechanics became an approximation to special relativity at speeds close to the speed of light, (ii) quantum dynamics took over from classical dynamics in the atomic domain, and (iii) Newton's inverse square law of gravitation was replaced by the laws of general relativity for strong gravitational fields.

Interference experiments with matter have a fascinating history, with  a quest developing over decades to test
superposition
using larger and larger objects. Over eighty years have passed since the classic experiment by Davisson and Germer in 1927 where interference was demonstrated with electrons.  Landmarks on the way included confirmation of interference for Helium [1930] and neutrons [1988]. A great modern breakthrough came in 1999 when interference was demonstrated for $C_{60}$ in the famous fullerene
experiment ~\cite{Arndt:99}. [There seems to be prevalent a popular belief in some quarters that the discovery of quantum superposition in a molecule as `large' as fullerene means the end of all theories which
predict breakdown of superposition for large systems. This of course is not true - breakdown of superposition and the quantum-to-classical transition is expected around $10^{6}$ amu to $10^{9}$ amu.]
This opened the door for larger molecules, and today, a decade later, interference has been demonstrated for molecules with 7,000 nucleons ~\cite{Gerlich2011}. Proposed future interferometry experiments plan to push the limit to macromolecules with a million nucleons and beyond, going up to molecules with a 100 million nucleons. Doing so involves overcoming great technological challenges ~\cite{Hornberger2011review}, and there are many orders of magnitudes in the mass scale yet to be covered. But we certainly live in exciting times where predictions of collapse models and gravity based models are being tested by these experiments, and constraints are being put on model parameters. See Section IV for details.

Also, this is perhaps a good place to clear another misconception regarding the domain over which quantum mechanics has been tested. Various remarkable macroscopic internal states have been achieved experimentally, in which an enormous collection of internal degrees of freedom behave as a collective one-particle coherent state. We have in mind of course systems such as superconductors, superfluids and Bose-Einstein condensates. The existence of such states however does not explain why macroscopic objects are not found in superposition of {\it position} states. Quantum mechanics may yet have to be modified so that the modified theory can explain the absence of position superpositions, but the modified theory will certainly continue to successfully explain a collective phenomenon such as superconductivity. In other words, the discovery of superconductivity does not solve / trivialize the Schr\"{o}dinger cat paradox!

Apart from direct laboratory experiments, collapse model parameters are also constrained by their effect on known measurements.  Section IV discusses the various experimental tests of the phenomenological models.

\subsection{Plan and outline of the article}

Sections II, III and IV are the main parts of the review. Sec. II reviews phenomenological models of spontaneous wave function collapse, which explain the absence of macroscopic superpositions, via a stochastic non-linear modification of the Schr\"{o}dinger equation. Sec. III gives a review of Trace Dynamics and gravity-induced-collapse as possible underlying theories for the phenomenology discussed in Sec. II. Sec. IV is a review of the techniques and results of ongoing and planned experiments which are testing the proposed phenomenological models. Sec. V provides a critique of the current understanding on the theoretical and experimental front, and lists open problems. 


Sec. II begins by introducing spontaneous collapse, and recalls the various collapse models that have been proposed. The original GRW model is then introduced. This is then followed by a detailed review of the QMUPL model, which is applied to show how stochasticity induces collapse, and how the Born probability rule is derived. The possible origin of the noise field is discussed.  The next sub-section discusses the most widely used (but physically equivalent to GRW) collapse model, i.e. the CSL model, and its generalizations. Lastly, the current understanding of the numerical values of the two parameters of the collapse model is reviewed.

Sec. III reviews Adler's Trace Dynamics as a candidate fundamental theory for Spontaneous Localization. The fundamental matrix degrees of freedom of the theory are introduced, and their dynamics described. The conserved charges of the theory, including the all-important Adler-Millard charge, are derived. This is followed by the construction of the statistical mechanics and the canonical ensemble for thermodynamic equilibrium for the theory. Following this, an important Ward identity, which is an analog of the equipartition theorem, is proved. It is shown how the commutation relations for quantum theory, and the Schr\"{o}dinger equation, emerge at this coarse-grained level, from the microscopic theory. Finally, consideration of fluctuations described by Brownian motion leads to generalization from the Schr\"{o}dinger equation to the stochastic non-linear Schr\"odinger equation, which makes contact with the CSL model. Subsequent sections describe the gravity based models for collapse, based on the work of Karolyhazy et al., Di\'osi, and the work of Penrose.

Sec. IV on Experimental Tests starts by discussing the basics of the collapse theory necessary for performing and interpreting the diffraction experiments with macromolecules. Matter-wave interferometry and optomechanics experiments with mechanical cantilevers are reviewed in detail. Cavity optomechanics with micro-spheres and nanoparticles is discussed, followed by a review of new developments which combine optical tweezing techniques with near-field matter-wave interferometry. The challenges proposed to these experiments by various kinds of decoherence are considered. The current bounds on collapse model parameters coming from the diffraction experiments and from other measurement processes are summarized.

\section{Spontaneous Collapse Models}

\subsection{Introducing spontaneous collapses}

Quantum Mechanics, in its standard textbook formulation, refers only to the
outcomes of measurements, but it has nothing to say about the
world as it is, independently of any measurement or act of
observation. This is a source of serious difficulties, which have
been clearly elucidated e.g. by J. Bell~\cite{Bell:00}: {\it It
would seem that the theory is exclusively concerned about `results
of measurements', and has nothing to say about anything else. What
exactly qualifies some physical systems to play the role of
`measurer'? Was the wave function  of the world waiting to jump for
thousands of millions of years until a single-celled living
creature appeared? Or did it have to wait a little bit longer, for
some better qualified system ... with a Ph.D.?}

Measuring devices, like photographic plates and bubble chambers, are
very sophisticated and highly structured physical systems, which
anyhow are made of atoms; we then expect them to be ultimately
described in quantum mechanical terms by means of the Schr\"odinger
equation. What else should we expect, taking into account that physicists
are trying to describe even the entire universe quantum
mechanically? But if we describe measurements in this way, then the theory does not predict any definite outcome, at the end of the process. The
Schr\"odinger equation is linear, the superposition principle holds, and it does so in such a way that all possible outcomes
are there simultaneously in the wave function, but none of them is
selected as the one which actually occurs. Yet, if we perform a
measurement, we always get a definite outcome. So we have a problem
with Quantum Mechanics.

Continuing quoting Bell: {\it If the theory is to apply to anything
but highly idealized laboratory operations, are we not obliged to
admit that more or less `measurement-like' processes are going on
more or less all the time, more or less everywhere? Do we not have
jumping then all the time?}

The basic idea behind the dynamical
reduction program is precisely this: spontaneous and random
collapses of the wave function occur all the time, for all particles,
whether isolated or interacting, whether they form just an atom
or a complex measuring device. Of course, such collapses must be rare
and mild for microscopic systems, in order not to alter their
quantum behavior as predicted by the Schr\"odinger equation. At the
same time, their effect must add up in such a way that, when
thousands of millions of particles are glued together to form a
macroscopic system, a single collapse occurring to one of the
particles affects the global system. We then have thousands of
millions of such collapses acting very frequently on the
macro-system, which together force its wave function to be very rapidly well-localized in space.

On the mathematical level, the program is accomplished by modifying the
Schr\"odinger evolution, introducing new terms having the
following properties:
\begin{itemize}
\item They must be {\it non-linear}: The new dynamics must break the
superposition principle at the macroscopic level and guarantee the
localization of the wave function of macro-objects.
\item They must be {\it stochastic}: When describing
measurement-like situations, the dynamics must explain why the outcomes
occur randomly; more than this, it must explain why they are
distributed according to the Born probability rule. On top of this, stochasticity is necessary because otherwise the non-linear terms would allow for faster than light communication. 
\item There must be an {\it amplification mechanism} according to
which the new terms have negligible effects on the dynamics of
microscopic systems but, at the same time, their effect becomes
very strong for large many-particle systems such as macroscopic
objects, in order to recover their classical-like behavior.
\item They must not allow for {\it superluminal signaling}, as one would like to preserve the causal structure of spacetime.
\end{itemize}

Looking carefully at these requirements, one soon realizes that they are
very demanding: there is no reason beforehand, that they
can be consistently fulfilled. One of the greatest
merits of collapse models is to have shown that this program can
be implemented in a consistent and satisfactory way.

\subsection{The zoo of collapse models}

In the literature, different collapse models have been proposed. A first characterization depends on the choice of the {\it collapse operators}, i.e. on the basis on which the wave function is localized. Some models induce the collapse in the energy basis~\cite{Adler:02,Adler:04,Adler:01,Adler2:01,Adler:00}, ~\cite
{Hughston:96, Brody:02, Milburn:91}, others in the momentum basis \cite{Benatti:88}, or the spin basis~\cite{Bassi:04, PEARLE1}. However, only models which collapse in the {\it position basis} make sure that different macroscopic superpositions rapidly collapse towards localized states. To understand this, one can think of a superposition of two spatially separated states of a macroscopic object, which have the same (or very similar) energy. In this case, an energy-based collapse model would not be able to collapse the superposition fast enough, because the superposition in energy is null or negligible. Such a model would not be able to guarantee that macro-objects always occupy a definite position in space. Only {\it space collapse models} make sure that macroscopic objects always behave classically, and therefore we will consider only them in the following.

Space collapse models can be conveniently grouped depending on the properties of the noise, which is responsible for the collapse. A first distinction is between {\it white} and {\it non-white} models. In white-noise models, the collapse-noise is assumed to be a Wiener process in time, and the resulting evolution is Markovian. All frequencies of the noise contribute to the collapse with the same weight. Examples of collapse models of this type are the GRW model~\cite{Ghirardi:86}, the CSL model ~\cite{PEARLE2}, ~\cite{Ghirardi2:90}, the QMUPL model~\cite{Diosi:89,Bassi2:05}. In non-white noise models, the collapse-noise is taken to be a generic Gaussian noise, with mean equal to zero, and a generic correlation function. The corresponding dynamics turns out to be non-Markovian, and they are more difficult to analyze. A model of this kind is the non-Markovian QMUPL model~\cite{Bassi:09,Bassi2:09}, while the non-Markovian CSL model is still under development~\cite{Adler:07,Adler:08}. General non-Markovian collapse models have been discussed in~\cite{Pearle:93,Pearle:96,Diosi:98,Bassi:02}.

A second distinction is between {\it infinite temperature} and {\it finite temperature} models. In the first type of models, the collapse-noise acts like a reservoir at infinite temperature. The wave function collapses, but at the same time the energy of the quantum system increases steadily; no dissipative effects are taken into account. This is a well-known feature of collapse models. Mathematically, these models are characterized by the fact that the wave function and the collapse-noise are coupled through the position operator only. The GRW model, the (Markovian and non-Markovian) CSL model, the (Markovian and non-Markovian) QMUPL model all belong to this group. In the finite temperature models instead, the collapse-noise behaves like a reservoir at finite temperature. The wave function still collapses, but now dissipative terms are included (through a position and momentum coupling between the wave function and the noise), which thermalize any quantum systems to the temperature of the noise. The only such model so far available is the non-dissipative QMUPL model~\cite{Bassi3:05}, though also the other models can be generalized in this sense. Recently, the QMUPL model has been generalized in order to include both non-Markovian and dissipative effects~\cite{Ferialdi:11}.

A final distinction is between {\it first quantized} models and {\it second quantized} models. Models of the first type consider only system of distinguishable particles; the GRW model, the QMUPL model and its non-Markovian and/or dissipative generalization belong to this group. Models of the second type are formulated in the language of quantum field theory and include systems of identical particles. The Tumulka-GRW model~\cite{Tumulka:06} and the CSL model belong to this group.

We also mention the earlier contributions of L. Diosi ~\cite{Diosi:88a,Diosi2:88}, N. Gisin ~\cite{Gisin:84, Gisin:89} and J. Percival ~\cite{PERCIVAL1,PERCIVAL2} to developing stochastically modified Schr\"odinger equations for describing the process of wave function collapse.  

Some comments are in order. The first one is that all space-collapse model are qualitatively equivalent: they all induce the collapse of the wave function in space, and the collapse is faster, the larger the system. Of course, they can differ also in a significant way in the technical details, as we will see.  The second comment refers to the nature of the stochastic character of the collapse process. One way to look at it---which corresponds to the original attitude towards these models---is that nature is intrinsically stochastic, therefore stochastic differential equations are the natural type of equations for describing the dynamics of physical systems. A new way to look at it is to assume that there is a random field, filling space, which couples to quantum matter in a non-standard way, and is responsible for the collapse of the wave function. The new terms in the modified Schr\"odinger equation are meant to describe such a coupling. Since this noise fills the whole space, most likely it has a cosmological origin. According to this scenario, the physically most reasonable collapse model is a model, where the collapsing field is ``cosmologically reasonable'', e.g. it has a typical cosmological correlation function and a typical cosmological temperature. This could be the case for the colored-noise and dissipative CSL model, which however has not been formulated yet. What one can do, is to extrapolate predictions from the other models already available. Therefore, in the following we will focus our attention on two of the above mentioned models: the CSL model, the one that more closely resembles the physically most reasonable model; and the QMUPL model, which is less physical, but has already been generalized in order to include dissipation, as well as colored noises, and is relatively easy to analyze mathematically.

A third comment is about the origin of the noise field. The important thing to bear in mind, is that this field cannot be a standard quantum field, otherwise we would fall back in the realm of standard quantum mechanics, with the superposition principle and the measurement problem. This field couples to quantum matter through an anti-hermitian and non-linear coupling. The most intriguing guess is that this noise has a gravitational origin. In fact, a gravitational background is part of the standard cosmological scenario; gravity is non-linear; gravity has not been successfully quantized yet, and we do not know today what shape a quantum theory of gravity will eventually take. Gravity-induced collapse models have been formulated in the literature, and we will discuss them in Sec. III.B.

A fourth comment is about the relativistic extension of collapse models. All models previously listed are non-relativistic. Their generalization to relativistic quantum field theories has not been successful. The reason is very simple to understand: the collapse of the wave function is an instantaneous process, or at least faster than light ~\cite{Maudlin:2011}. This is a necessary requirement, in order to reproduce non-local quantum correlations encoded in Bell inequalities~\cite{Bell:87}, which have been verified experimentally. An instantaneous collapse process is not welcome in a relativistic framework, hence the difficulty in formulating relativistic collapse models. We will discuss this issue in Sec. II.I

\subsection{The GRW model}

In order to appreciate how collapse models work, and what they are able to achieve, we briefly review the GRW model, the first consistent model proposed. Though it is not expressed in terms of a compact stochastic differential equation, it has the advantage of being physically very intuitive. In presenting the model, we follow the exposition of J. Bell ~\cite{BELL1} [reprinted in ~\cite{Bell:87}] in terms of discrete jumps of the wave function.

Let us consider a system of $N$ particles which, only for the sake of
simplicity, we take to be scalar and spinless; the GRW model is defined
by the following postulates: \\

\noindent {\bf States.} The state of the system is represented by a wave
function $\psi({\bf x}_{1}, {\bf x}_{2}, \ldots {\bf x}_{N})$
belonging to the Hilbert space ${\mathcal L}^2({\bf R}^{3N})$. Spin and other internal degrees of freedom
are ignored for the sake of simplicity. \\

\noindent {\bf Dynamics.} At  random times, the wave function experiences a sudden jump of
the form:
\begin{eqnarray}
\lefteqn{\psi_{t}({\bf x}_{1}, {\bf x}_{2}, \ldots {\bf x}_{N}) \quad
\longrightarrow \quad} \qquad\qquad \nonumber \\
& & \frac{L_{n}({\bf x}) \psi_{t}({\bf x}_{1},
{\bf x}_{2}, \ldots {\bf x}_{N})}{\|L_{n}({\bf x}) \psi_{t}({\bf
x}_{1}, {\bf x}_{2}, \ldots {\bf x}_{N})\|},
\end{eqnarray}
where $\psi_{t}({\bf x}_{1}, {\bf x}_{2}, \ldots {\bf x}_{N})$ is
the state vector of the whole system at time $t$, immediately prior
to the jump process. $L_{n}({\bf x})$ is a linear operator which is
conventionally chosen equal to:
\begin{equation}
L_{n}({\bf x}) =
\frac{1}{(\pi r_C^2)^{3/4}} e^{- ({\bf
q}_{n} - {\bf x})^2/2r_C^2},
\end{equation}
where $r_C$ is a new parameter of the model which sets the
width of the localization process, and ${\bf q}_{n}$ is the position
operator associated to the $n$-th particle of the system; the random variable
${\bf x}$ corresponds to the place where the jump occurs. Between
two consecutive jumps, the state vector evolves according to the
standard Schr\"odinger equation.

The probability density for a jump taking place at the position
${\bf x}$ for the $n$-th particle is given by:
\begin{equation}
p_{n}({\bf x}) \quad \equiv \quad \|L_{n}({\bf x}) \psi_{t}({\bf
x}_{1}, {\bf x}_{2}, \ldots {\bf x}_{N})\|^2,
\end{equation}
and the probability densities for the different particles are
independent.

Finally, it is assumed that the jumps are distributed in time like
a Poissonian process with frequency $\lambda_{\text{\tiny GRW}}$; this is the second
new parameter of the model.

The standard numerical values for $r_C$ and $\lambda_{\text{\tiny GRW}}$ are:
\begin{equation} \label{eq:num}
\lambda_{\text{\tiny GRW}} \; \simeq \; 10^{-16} \, \makebox{sec$^{-1}$}, \qquad
r_C \; \simeq \; 10^{-7} \, \makebox{m}.
\end{equation}
We will come back to the issue of numerical value of these parameters in Sec. II.J.

\noindent {\bf Ontology.} In order to connect the mathematical formalism with the physical world, one needs to provide an ontology, which is rather straightforward for collapse models. Let $m_{n}$ be the mass associated to
the $n$-th ``particle'' of the system (one should say: to what is
called ``a particle'', according to the standard terminology); then
the function:
\begin{eqnarray}
\rho^{(n)}_{t}({\bf x}_{n}) & \equiv & m_{n} \int d^3 x_{1} \ldots
d^3 x_{n-1} d^3 x_{n+1} \nonumber \\
& & \ldots d^3 x_{N} \, | \psi_{t}({\bf x}_{1},
{\bf x}_{2}, \ldots {\bf x}_{N}) |^2 \quad
\end{eqnarray}
represents the {\it density of mass}~\cite{Ghirardi:95} of that ``particle''
in space, at time $t$.

These are the axioms of the GRW model: as we see, words such as
`measurement', `observation', `macroscopic', `environment' do not
appear. There is only a {\it universal} dynamics governing all
physical processes, and an ontology which tells how the physical
world is, according to the model, independently of any act of
observation.

The GRW model, as well as the other dynamical reduction models
which have appeared in the literature, has been extensively
studied (see~\cite{Bassi:03} and~\cite{Pearle:99} for a review on this
topic); in particular---with the numerical choice for $\lambda_{\text{\tiny GRW}}$
and $r_C$ given in~(\ref{eq:num})---the following three
important properties have been proved, which we will state in more
quantitative terms in the following section:
\begin{itemize}
\item At the microscopic level, quantum systems behave almost
exactly as predicted by standard Quantum Mechanics, the
differences being so tiny that they can hardly be detected with
present-day technology.
\item At the macroscopic level, wave functions of macro-objects
are almost always very well localized in space, so well localized
that their centers of mass behave, for all practical purposes,
like point-particles moving according to Newton's laws.
\item In a measurement-like situation, e.g. of the von Neumann type,
the GRW model reproduces---as a consequence of the modified dynamics---both
the Born probability rule and the standard postulate of wave-packet
reduction.
\end{itemize}
In this way, models of spontaneous wave function collapse provide
a unified description of all physical phenomena, at least at the
non-relativistic level, and a consistent solution to the
measurement problem of Quantum Mechanics.

It may be helpful to stress some points about the world-view
provided by the GRW model, and collapse models in general. According to the ontology given by
the third axiom, there are no particles at all in the theory! There
are only distributions of masses which, at the microscopic level,
are in general quite spread out in space. An electron, for example, is not a
point following a trajectory---as it would be in Bohmian
Mechanics---but a wavy object diffusing in space. When, in a
double-slit experiment, it is sent through the two apertures, it literally
goes through both of them, as a classical wave would do. The
peculiarity of the electron, which qualifies it as a quantum system,
is that when we try to locate it in space, by making it
interact with a measuring device, e.g. a photographic film,
then, according to the collapse dynamics, its wave function very rapidly shrinks in space till
it gets localized to a spot, the spot where the film is exposed
and which represents the outcome of the position measurement. Such a behavior,
which is added {\it ad hoc} in the standard formulation of Quantum
Mechanics, is a direct consequence of the universal dynamics of
the GRW model.

Also macroscopic objects are waves; their centers of mass are not
mathematical points, rather they are represented by some function
defined throughout space. But macro-objects have a nice property:
according to the GRW dynamics, each of them is always almost
perfectly located in space, which means that the wave functions
associated with their centers of mass are appreciably different from
zero only within a very tiny region of space (whose linear extension
is of order $10^{-14}$ m or smaller, as we shall see), so tiny that
they can be considered point-like for all practical purposes. This
is the reason why Newton's mechanics of point particles is such a
satisfactory theory for macroscopic classical systems.

Even though the GRW model contains no particles at all, we will
keep referring to micro-systems as `particles', just as a matter
of convenience.

Though the collapse dynamics is expressed entirely in terms of the wave function, not of the density matrix, in order to eliminate any possible ambiguity about the nature of the collapse, it is nevertheless convenient to look at the collapse dynamics for the density matrix, to analyze specific features of the model. The 1-particle master equation of the GRW model takes the form~\cite{Ghirardi:86}:
\begin{equation} \label{eq:asdui}
\frac{d}{dt} \rho(t) = - \frac{i}{\hbar} [ H, \rho(t) ] - T[\rho(t)],
\end{equation}
where $H$ is the standard quantum Hamiltonian of the particle, and $T[\cdot]$ represents the effect of the spontaneous collapses on the particle's wave function. In the position representation, this operator becomes:
\begin{eqnarray} \label{eq:fsgdffpq}
\lefteqn{\langle {\bf x} | T[\rho(t)] | {\bf y} \rangle =} \qquad\\
& &  \lambda_{\text{\tiny GRW}} [ 1 - e^{-({\bf x} - {\bf y})^2/4r_C^2} ] \langle {\bf x} | \rho(t) | {\bf y} \rangle. \nonumber
\end{eqnarray}
As expected, the effect of the spontaneous collapse is to suppress the off-diagonal elements of the density matrix, with a rate proportional to $\lambda_{\text{\tiny GRW}}$, depending also on the distance between the off-diagonal elements: distant superpositions are suppressed faster than closer ones.

The many particle master equation is the generalization of Eqn.~(\ref{eq:asdui}), where an operator $T_i[\cdot]$, $i = 1,2, \ldots N$ appears for each particle. For ordinary matter---and with a good approximation---one can separate the center-of-mass motion from the internal motion~\cite{Ghirardi:86}. The reduced density matrix for the internal motion obeys the standard Schr\"odinger equation, while that for the center of mass is equivalent to Eqn.~(\ref{eq:asdui}), where now the collapse rate entering the definition of the operator $T[\cdot]$ is $N \lambda_{\text{\tiny GRW}}$, with $N$ the total number of particles making up the object. This is a manifestation of the {\it amplification mechanics}, perhaps the most important feature of collapse models: the wave function of an object collapses with a rate which is proportional to the size of the system. This is the mathematical reason why collapse models can accommodate both the quantum dynamics of microscopic systems (negligible collapse rate) and the classical dynamics of macroscopic systems (fast collapse) within one unified dynamical principle.

\subsection{The QMUPL model} \label{sec:qmupl}

We now focus our attention on the QMUPL (Quantum Mechanics with Universal Position Localizations) model. As previously anticipated, the reason is that this model has the virtue of being both physically realistic, though very simplified compared to the more realistic GRW and CSL models, and mathematically simple enough to be analyzed in great detail. The axioms defining this model are the same as those of the GRW model, with the only difference that the dynamics is described by a stochastic differential equation.
The one-particle equation takes the form (for simplicity, we will work only in one dimension in space):
\begin{eqnarray} \label{eq:qmupl1}
d \psi_t & = &  \left[ -\frac{i}{\hbar} H dt + \sqrt{\lambda} (q - \langle q \rangle_t) dW_t \right. \nonumber \\
& & - \left. \frac{\lambda}{2} (q - \langle q \rangle_t)^2 dt \right] \psi_t,
\end{eqnarray}
where $q$ is the position operator of the particle, $\langle q \rangle_t \equiv \langle \psi_t | q | \psi_t \rangle$ is the quantum expectation, and $W_t$ is a standard Wiener process. For simplicity, we work in only one spatial dimension, the generalization to three dimensions being straightforward. The collapse constant $\lambda$ sets the strength of the collapse mechanics, and it is chosen proportional to the mass $m$ of the particle according to the formula\footnote{One should keep in mind that the collapse strength depends on the type of model. For the GRW model, $\lambda_{\text{\tiny GRW}}$ is a rate. For the QMUPL model, $\lambda$ has the dimensions of an inverse time, times an inverse square length. The two constants are related, by requiring that the collapse strengths according to different models coincide in the appropriate limit ~\cite{BassiDuerr:09}}:
\begin{equation} \label{eq:sfjdsutar}
\lambda = \frac{m}{m_0}\; \lambda_0,
\end{equation}
where $m_0$ is the nucleon's mass and $\lambda_0$ measures the collapse strength~\cite{Bassi2:05}. If we set $\lambda_0 \simeq 10^{-2} \text{m}^{-2} \text{s}^{-1}$, then the strength of the collapse mechanism according to the QMUPL model corresponds to that of the GRW and CSL models in the appropriate limit ~\cite{BassiDuerr:09}. Note also that the QMUPL model is defined in terms on only one parameter ($\lambda$), while the GRW model (and similarly the CSL model) is defined in terms of two parameters ($\lambda_{\text{\tiny GRW}}$ and $r_C$).

We will come back to the numerical values of the collapse parameter in Sec. II.J. The generalization for a many-particle system can be easily obtained by considering the position operator $q_i$ of every particle, each coupled to a different Wiener process $W^{(i)}_t$. The structure remains the same, with a sum to include the contribution to the collapse coming from each particle.

As we expect, Eqn.~(\ref{eq:qmupl1}) contains both non-linear and stochastic terms, which are necessary to induce the collapse of the wave function. In order to see this, let us consider a free particle ($H = p^2/2m$), and a Gaussian state:
\begin{equation} \label{gsol}
\psi_{t}(x) = \makebox{exp}\left[ - a_{t} (x -
\overline{x}_{t})^2 + i \overline{k}_{t}x + \gamma_{t}\right].
\end{equation}
It is not too difficult to show that $\psi_{t}(x)$ is solution of Eqn.~(\ref{eq:qmupl1}), provided that the time dependent functions in the exponent solve appropriate stochastic differential equations~\cite{Bassi2:05}. In particular, the equations for $a_t$ which controls the spread both in position and momentum, for the mean position $\overline{x}_{t}$ and the mean momentum $\overline{k}_{t}$ are\footnote{The superscripts ``R'' and ``I''
denote, respectively, the real and imaginary parts of the
corresponding quantities.}:
\begin{eqnarray}
d a_{t} & = & \left[ \lambda - \frac{2i\hbar}{m}\,
\left( a_{t} \right)^{2}\right] dt, \label{efa1} \\
d \overline{x}_{t} & = &\frac{\hbar}{m}\, \overline{k}_{t}\, dt +
\frac{\sqrt{\lambda}}{2a_{t}^{\makebox{\tiny R}}} \, dW_{ t},  \label{efab} \\
d \overline{k}_{t} & = & - \sqrt{\lambda}\,\,
\frac{a_{t}^{\makebox{\tiny I}}}{a_{t}^{\makebox{\tiny R}}}  \label{efac} \,
dW_{t}.
\end{eqnarray}
Eqn.~(\ref{efa1}) is deterministic and easy to solve. The spreads in position and momentum:
\begin{eqnarray}
\sigma_{q}(t) & = & \frac{1}{2}\sqrt{\frac{1}{a_{t}^{\makebox{\tiny
R}}}}, \nonumber \\
\sigma_{p}(t) & = & \hbar\,\sqrt{\frac{(a_{t}^{\makebox{\tiny R}})^2 +
(a_{t}^{\makebox{\tiny I}})^2}{a_{t}^{\makebox{\tiny R}}}},
\end{eqnarray}
are given by the following analytical expressions:
\begin{eqnarray}
\sigma_{q} & = &
\sqrt{\frac{\hbar}{m\omega}\frac{\cosh (\omega t + \varphi_{1}) + \cos (\omega t +
\varphi_{2})}{\sinh (\omega t + \varphi_{1}) + \sin (\omega t +
\varphi_{2})}},  \nonumber \\
& & \label{sx} \\
\sigma_{p} & = & \sqrt{\frac{\hbar m\omega}{2}\frac{\cosh (\omega t + \varphi_{1}) - \cos (\omega t +
\varphi_{2})}{\sinh (\omega t + \varphi_{1}) + \sin (\omega t +
\varphi_{2})}}, \nonumber \\
\label{sp}
\end{eqnarray}
with:
\begin{equation}
\omega \; = \; 2\,\sqrt{\frac{\hbar \lambda_{0}}{m_{0}}} \; \simeq
\; 10^{-5} \; \makebox{s$^{-1}$}.
\end{equation}
The two parameters $\varphi_{1}$ and $\varphi_{2}$ are functions of the
initial condition. Also, note that setting $\lambda_{0}=0$ will give the same results as one will get by using the Schr\"{o}dinger equation instead of Eqn. (\ref{eq:qmupl1}).

Eqns.~(\ref{sx}) and~(\ref{sp}) tell that the spreads in position and momentum do not increase in time, but reach an asymptotic final value given by:
\begin{equation} \label{aval1}
\sigma_{q}(\infty) = \sqrt{\frac{\hbar}{m\omega}} \simeq
\left( 10^{-15} \sqrt{\frac{\makebox{Kg}}{m}}\, \right)\,
\makebox{m},
\end{equation}
and:
\begin{equation} \label{aval2}
\sigma_{p}(\infty) = \sqrt{\frac{\hbar m\omega}{2}}
\simeq \left( 10^{-19} \sqrt{\frac{m}{\makebox{Kg}}}\,
\right)\, \frac{\makebox{Kg m}}{\makebox{sec}},
\end{equation}
such that:
\begin{equation}
\sigma_{q}(\infty)\, \sigma_{p}(\infty) \; = \;
\frac{\hbar}{\sqrt{2}}
\end{equation}
which corresponds to almost the minimum allowed by Heisenberg's
uncertainty relations. As we see, the spread in position does not increase indefinitely, but stabilizes to a finite value, which is a compromise between the Schr\"odinger's dynamics, which spreads the wave function out in space, and the collapse dynamics, which shrinks it in space. For microscopic systems, this value is still relatively large ($\sigma_{q}(\infty) \sim 1$m, for an electron, and $\sim 1$mm, for a buckyball containing some 1000 nucleons), such as to guarantee that in all standard experiments---in particular, diffraction experiments---one observes interference effects. For macroscopic objects instead, the spread is very small ($\sigma_{q}(\infty) \sim 3 \times 10^{-14}$m, for a 1g object), so small that for all practical purposes the wave function behaves like a point-like system. Once again, this is how collapse models are able to accommodate both the ``wavy'' nature of quantum systems and the ``particle'' nature of classical objects, within one single dynamical framework. One should also note that, as a byproduct of the collapse in position, one has an almost perfect collapse in momentum, compatibly with Heisenberg's uncertainty relations.

Eqn.~(\ref{efac}) says that the mean momentum undergoes a diffusion process. For microscopic systems, such a diffusion is appreciably large: the wave function is kicked back and forth by the collapse noise. For larger objects instead, the diffusion becomes weaker and weaker, to the point that at the macroscopic level it is almost entirely negligible. The same is true for the mean in position, according to Eqn.~(\ref{efab}). In this way, collapse models can explain both the stochastic nature of quantum phenomena and the (apparently) deterministic nature of classical ones. Moreover, the average momentum ${\mathbb E}[\langle p \rangle_t]$ is constant ($\langle p \rangle_t = \hbar k_t$), while the average position is given by ${\mathbb E}[\langle q \rangle_t] = {\mathbb E}[\langle p \rangle_t]/m$: the particle, on the average, moves along a straight line, depending on its initial momentum.

Two comments are in order. The first one is that the above results refer only to the special case of Gaussian wave functions, like that of Eqn.~(\ref{gsol}). However, in~\cite{Bassi5:08,Bassi:10} a remarkable result has been proven: with probability one, {\it any} initial state converges asymptotically to a {\it Gaussian} wave function, having a fixed spread both in position and in momentum, given by Eqn.~(\ref{aval1}) and~(\ref{aval2}) respectively. The collapse process not only localizes wave functions, but also smoothes all their bumps and eventually shapes them as Gaussian functions. The second comment is that the above results refer only to a free particle. Also the harmonic oscillator can be treated in a fully analytical way, but more general potentials require perturbative approaches, which have not been explored so far.

To conclude this section, let us consider the many-particle equation:
\begin{eqnarray} \label{nlemp}
d\,\psi_{t} & = & \left[ -\frac{i}{\hbar}
H dt + \!\sum_{i=1}^{N}\!\sqrt{\lambda_{i}} (
q _{i} - \langle q _{i} \rangle_{t}) d W_{t}^{(i)}\right. \nonumber \\
& & \left. - \frac{1}{2}\,\sum_{i=1}^{N} \lambda_{i} ( q _{i} -
\langle  q_{i} \rangle_{t})^2 dt \right] \psi_{t},
\end{eqnarray}
where $H$ is the  quantum Hamiltonian
of the composite system, the operators $ q _{i}$ ($i = 1, \ldots
N$) are the position operators of the particles of the system, and
$W_{t}^{(i)}$ ($i = 1, \ldots N$) are $N$ independent standard
Wiener processes.

As often in these cases, it is convenient to switch from the particles' coordinates ($x_1, x_2, \ldots x_N$) to
the center--of--mass ($R$) and relative ($\tilde{x}_{1},
\tilde{x}_{2}, \ldots \tilde{x}_{N}$) coordinates:
\begin{equation}
\left\{
\begin{array}{lcl}
R & = & \displaystyle \frac{1}{M}\, \sum_{i=1}^{N}
m_{i}\, x_{i} \qquad M \; = \; \sum_{i=1}^{N} m_{i},\\
& & \\
x_{i} & = & R + \tilde{x}_{i};
\end{array}
\right.
\end{equation}
let $Q$ be the position operator for the center of mass and
$\tilde q _{i}$ ($i = 1 \ldots N$) the position operators
associated to the relative coordinates.
It is not difficult to show that---under the assumption
$H = H_{\makebox{\tiny CM}} +
H_{\makebox{\tiny rel}}$---the dynamics for the center of mass
and that for the relative motion decouple; in other words,
$\psi_{t}(\{ x \}) = \psi_{t}^{\makebox{\tiny CM}}(R) \otimes
\psi_{t}^{\makebox{\tiny rel}}(\{\tilde{x}\})$ solves Eqn.
(\ref{nlemp}) whenever $\psi_{t}^{\makebox{\tiny CM}}(R)$ and
$\psi_{t}^{\makebox{\tiny rel}}(\{\tilde{x}\})$ satisfy the
following equations:
\begin{eqnarray}
d\psi_{t}^{\makebox{\tiny rel}} \!& = &\! \left[
-\frac{i}{\hbar} H_{\makebox{\tiny rel}} dt +
\!\sum_{i=1}^{N}\!\sqrt{\lambda_{i}}\, ({\tilde{q}}_{i} - \langle
{\tilde{q}}_{i} \rangle_{t}) d W_{t}^{(i)}\right. \nonumber \\
& & \left. - \frac{1}{2}\,\sum_{i=1}^{N} \lambda_{i}
({\tilde{q}}_{i} - \langle {\tilde{q}}_{i} \rangle_{t})^2 dt
\right] \psi_{t}^{\makebox{\tiny rel}},
\end{eqnarray}
and:
\begin{eqnarray}
d\,\psi_{t}^{\makebox{\tiny CM}}\! & = & \! \left[
-\frac{i}{\hbar} H_{\makebox{\tiny CM}} dt +
\sqrt{\lambda_{\makebox{\tiny CM}}} (Q - \langle
Q \rangle_{t}) d W_{t}\right. \nonumber \\
& & \left. - \frac{\lambda_{\makebox{\tiny CM}}}{2}\, (Q - \langle
Q \rangle_{t})^2 dt \right] \psi_{t}^{\makebox{\tiny CM}},
\label{eqcm}
\end{eqnarray}
with:
\begin{equation}
\lambda_{\makebox{\tiny CM}} \; = \; \sum_{n=1}^{N} \lambda_{n} \;
= \; \frac{M}{m_{0}}\, \lambda_{0}.
\end{equation}
The first of the above equations describes the internal motion of
the composite system: it basically says that the internal structure of the system behaves quantum mechanically, modulo tiny modifications given by the collapse process. The second equation describes the center-of-mass evolution, and here we can see once again the most important feature of collapse models: the {\it amplification mechanism}. The collapse strength of the center of mass is proportional to the size (i.e. the number of constituents) of the system. For microscopic systems, $\lambda_{\makebox{\tiny CM}}$ is similar to $\lambda_0$, i.e. very weak; in these cases, the collapse is almost negligible. For macroscopic objects, $\lambda_{\makebox{\tiny CM}}$  ($\sim N \lambda_0$, with $N \sim 10^{24}$) can be very strong, implying a rapid and efficient collapse of the wave function. It is precisely because of the amplification mechanism that, with a single choice of $\lambda_0$, one can describe both quantum and classical phenomena at the same time.

\subsubsection{The quantum formalism derived}

Collapse models contain a unique and universal dynamics, which applies to all physical situations. Measurements play no special role in collapse models. It then becomes interesting and important to show how the entire phenomenology of quantum measurements emerges from the universal dynamics of collapse models.  To do this, we use the QMUPL model, because of its relatively simple mathematical structure.
We will show that measurements always have a definite outcome, are randomly distributed according to the Born rule, and that at the end of the measurement process, the wave function of the micro-system collapses according to the von Neumann projection postulate. All these features are included in Eqn.~(\ref{eq:qmupl1}), without any need for extra axioms.

The
measurement setup we consider consists of a microscopic system~${\mathcal S}$ interacting
with a macroscopic system~${\mathcal A}$, which acts like a measuring
apparatus; both systems are described in quantum mechanical terms. We assume that the measurement includes a
{\it finite} set of outcomes.  Accordingly, we assume that the
microscopic system~${\mathcal S}$ can be described by a finite-dimensional complex
Hilbert space. For the sake of simplicity, and without loss of
generality, we can consider the simplest case:
$\mathcal{H}_{{\mathcal S}} = \mathbb{C}^{2}$, because the
generalization of what follows to~$\mathbb{C}^{n}$ is quite
straightforward.  Since the most general self-adjoint operator~$O$
acting on~$\mathbb{C}^{2}$ can be written as
\begin{equation}
  \label{eq:1}
  O = o_{+} | + \rangle\langle +| + o_{-} | - \rangle\langle -|,
\end{equation}
where~$| + \rangle$ and~$| - \rangle$ are the eigenstates of $O$,
while $o_{+}$ and $o_{-}$ are its two real eigenvalues, for
definiteness and with no loss of generality, in what follows we will
take $o_{\pm} = \pm \hbar / 2$ and $O$ to be the $z$-component of
the spin, $S_{z}$, of a spin $1/2$  particle.

We take the following model for the
measuring apparatus~${\mathcal A}$, which is general enough to
describe all interesting physical situations: we assume that the
apparatus consists of a fixed part plus a pointer moving along a
graduated scale, in such a way that different positions of the
pointer along the scale correspond to different possible outcomes of
the measurement.  To simplify the analysis, we study the evolution
of the center of mass of the pointer only, and disregard all other
macroscopic and microscopic degrees of freedom; accordingly, the
pointer will be treated like a macroscopic quantum particle of
mass~$m$ moving in one dimension only, whose state space is
described by the Hilbert space $\mathcal{H}_{{\mathcal A}} =
\mathrm{L}^{2} (\mathbb{R})$.

We assume that the wave function of the pointer of~${\mathcal A}$
is subject to a spontaneous collapse process according to Eqn.~(\ref{eq:qmupl1}), while the wave function of the microscopic
system~${\mathcal S }$ evolves according to the standard
Schr\"odinger equation, since, as typical of dynamical reduction
models, the stochastic collapse terms have negligible effects on
microscopic quantum systems. For definiteness, let us consider a pointer of mass $m
= 1$ g (i.e., a pointer made of an Avogadro number of nucleons).

We take the total Hamiltonian~$H$ to be of the form: $H =
H_{{\mathcal S}} + H_{{\mathcal A}} + H_{\mathrm{INT}}$.  The first
term is the quantum Hamiltonian for the microscopic system: we
assume that the time scale of the free evolution of the microscopic
system is much larger than the characteristic time scale of the
experiment (``instantaneous measurement'' assumption); accordingly
we take~$H_{{\mathcal S}}$ to be the null operator. The second term
is the quantum Hamiltonian of the pointer, which we take equal to
that of a non-relativistic free quantum particle of mass~$m$:
$H_{{\mathcal A}} = p^{2} / (2m)$, where~$p$ is the momentum
operator.
Finally, we assume the interaction term~$H_{\mathrm{INT}}$ between
the two systems to be of the von~Neumann type, devised in such a
way as to measure the spin $S_{z}$:
\begin{equation} \label{eq:5}
H_{\mathrm{INT}}(t) = \kappa\, \Delta^{T}_{t}\, S_{z} \otimes p,
\end{equation}
where~$\kappa$ is a coupling constant and $\Delta^{T} \colon t
\mapsto \Delta^{T}_{t}$ is a $T$-normalized,\footnote{\label{fn:4}By
a $T$-normalized function, we simply mean
\[
\int_{-\infty}^{+\infty} \Delta^{T}_{t}\, dt = \int_{t_{0}}^{t_{0} +
 T} \Delta^{T}_{t}\, dt = T.
\]
Note that $\Delta^{T}_{t}$ depends also on the initial time $t_{0}$;
we will omit indicating this explicitly, when no confusion arises.}
non-negative, real valued, function of time, identically equal to
zero outside a given interval of the form~$(t_{0}, t_{0} + T)$,
i.e., outside the time interval of length~$T$, say $T = 1$s,
during which the experiment takes place; we choose the time origin
in such a way that the experiment begins at $t_{0} = 0$. As
is well known in standard Quantum Mechanics, $H_{\mathrm{INT}}$ generates the following type of
evolution, depending on the initial state of the micro-system
${\mathcal S}$:
\begin{eqnarray} \label{eq:6}
\lefteqn{\left[ c_{+} |+\rangle + c_{-} |-\rangle \right] \otimes \phi_{0}} \qquad \nonumber \\
& \mapsto &  c_{+} |+\rangle \otimes \phi_{+}  + c_{-} |-\rangle
\otimes \phi_{-},
\end{eqnarray}
where $\phi_{\pm}$ are final pointer states spatially
translated with respect to the initial state~$\phi_{0}$ by the
quantity $\pm(\hbar/2)\, \kappa\, T$. We will see how collapse models modify this linear evolution.

The strength of the coupling constant~$\kappa$ has to be chosen in
such a way that the distance $\hbar\, \kappa\, T$ between the
initial state $\phi_{0}$ of the pointer and any of the two
final states $\phi_{\pm}$ is macroscopic; for definiteness,
let us choose $\hbar\,\kappa = 1$ cm s$^{-1}$, so that
$\hbar\,\kappa\,T = 1$ cm.

We take the initial states of the
microscopic system~${\mathcal S}$ and of the macroscopic
apparatus~${\mathcal A}$ to be completely uncorrelated, as it is
customary and appropriate for the description of a measurement
process. Accordingly, we assume the initial state of the total
system ${\mathcal S} + {\mathcal A}$ to be:
\begin{equation} \label{eq:7}
\left[ c_{+} |+\rangle + c_{-} |-\rangle \right] \otimes \phi_{0},
\end{equation}
where~$\phi_{0}$ describes the ``ready'' state of the
macroscopic apparatus~${\mathcal A}$.

Some considerations are in order, regarding the initial state~$\phi_{0}$ of the
pointer.  In the previous section we
showed that, according to Eqn.~(\ref{eq:qmupl1}), the wave
function for the centre of mass of an isolated quantum system
reaches asymptotically (and very rapidly, for a macro-object) a
Gaussian state of the form
\begin{equation} \label{eq:8}
\phi^{\mathrm{G}}_{t} (x) = \sqrt[4]{\frac{1}{2 \pi
\sigma_{q}^{2}}}\! \exp\! \left[ \!- \frac{1 - i}{4 \sigma_{q}^2} (x -
\bar{x}_{t})^2 \!+ i \bar{k}_{t} x \right]
\end{equation}
(modulo a time-dependent global phase factor) with $\sigma_{q}$
defined as in Eqn.~(\ref{aval1}), taking the value: $\sigma_q \simeq 4.6 \times 10^{-14} \text{m}$ for $m = 1 \text{g}$.
The dispersion of the Gaussian function of Eqn.~(\ref{eq:8}) in
momentum space is $\sigma_{p} \simeq  1.6
\times 10^{-21}\, \makebox{kg m s$^{-1}$}$, as described in Eqn.~(\ref{aval2}).

In our measurement model, we assume that the pointer is isolated for
the time prior to the experiment; during this time its wave function converges rapidly towards a state
close to~(\ref{eq:8}), which we therefore assume to be the initial
state of the pointer. To summarize, we take as the initial state of
the composite system ${\mathcal S} + {\mathcal A}$ the vector:
\begin{equation} \label{eq:10}
\Psi_{0} = \left[ c_{+} |+\rangle + c_{-} |+\rangle \right]
\otimes \phi^{\mathrm{G}}.
\end{equation}
We
choose the natural reference frame where the pointer is initially at
rest, so that $\bar{k}_{0} = 0$ m$^{-1}$, with the origin set up in
such a way that $\bar{x}_{0} = 0$ m.

We are now ready to solve the collapse equation. It is not difficult to show that, for the given initial condition, the solution takes the form
\begin{equation}
  \label{eq:30}
  \psi_{t} = |+\rangle \otimes \phi^{+}_{t}
  + |-\rangle \otimes \phi^{-}_{t},
\end{equation}
where~$\phi^{-}_{t}$ have the
form:
\begin{equation} \label{eq:15}
\phi^{\pm}_{t} (x) = \exp\! \left[ \!-  \alpha_{t} (x -
\bar{x}^{\pm}_{t})^{2} \!+ \!i \bar{k}^{\pm}_{t} x \!  + \!
\gamma^{\pm}_{t} \! + \! i \theta^{\pm}_{t} \right]
\end{equation}
whose parameters $\alpha_{t} \in \mathbb{C}$, and
$\bar{x}^{\pm}_{t}$, $\bar{k}^{\pm}_{t}$, $\gamma^{\pm}_{t}$,
$\theta^{\pm}_{t} \in \mathbb{R}$ satisfy a complicated set of non-linear stochastic differential equations~\cite{Bassi:07}, with given initial conditions. In particular:
\begin{equation} \label{eq:orygzf}
\gamma^{\pm}_{0} \; = \; \ln | c_{\pm} |
\end{equation}
(of course we now assume that $c_{\pm} \neq 0$).

In order to extract the relevant physical information, let us consider the
differences $X_{t} := \bar{x}^{+}_{t} - \bar{x}^{-}_{t}$ and $K_{t}
:= \bar{k}^{+}_{t} - \bar{k}^{-}_{t}$, which represent
the distance in position and (modulo $\hbar$) momentum space
between the centres of the two Gaussian
functions~$\phi^{+}_{t}$ and~$\phi^{-}_{t}$.  One can easily prove that~$X_{t}$ and~$K_{t}$ satisfy a set of linear and deterministic equations~\cite{Bassi:07}:
\begin{equation}
  \label{eq:31}
  \frac{d}{dt}
  \begin{bmatrix}
    X_{t} \\ K_{t}
  \end{bmatrix} =
  \begin{bmatrix}
    - \omega                  & \hbar/m\\
    - 2 \lambda & 0
  \end{bmatrix}
  \begin{bmatrix}
    X_{t} \\ K_{t}
  \end{bmatrix} +
  \begin{bmatrix}
    \hbar\, \kappa\, \Delta^{T}_{t}\\ 0
  \end{bmatrix};
\end{equation}
where both the non-linear and the stochastic terms cancel out.
The solution of the above system depends of course on the specific
choice for the function $\Delta^{T}_{t}$; a simple reasonable choice
is the following:
\begin{equation}
\Delta^{T}_{t} \quad = \quad \left\{
\begin{array}{ll}
1 \quad & t \, \in \, [0,T] \\
0 & \makebox{else},
\end{array}
\right.
\end{equation}
which---in a standard quantum scenario---means that during the
measurement each term of the superposition moves with a constant speed towards the
left and towards the right, respectively. According to this choice, $X_{t}$, given the initial
condition $X_{0} = 0$ m, evolves in time as follows:
\begin{equation}
  \label{eq:n2}
  X_{t} \!=\!
  \left\{
  \begin{array}{l}
  \displaystyle
    \!\frac{2\hbar\kappa}{\omega}\,
    e^{-\omega t/2}
    \sin\frac{\omega}{2} t \qquad
    \text{for $0\leq t\leq T$},  \\
   \\
  \displaystyle
  \!\frac{2\hbar\kappa}{\omega}
    e^{-\omega t/2}\!\!\left[
    \sin\frac{\omega}{2} t
     - e^{\omega T/2}\!
    \sin\frac{\omega}{2} (t\!-\!T)
    \right] \\
    \qquad\qquad\qquad\qquad\qquad\quad\text{for $t \geq T$}.
 \end{array}
 \right.
\end{equation}
Since $\omega^{-1} \simeq 2.0 \times 10^{4}$s is a very long time
compared to the measurement-time, we can meaningfully expand
Eqn.~\eqref{eq:n2} to first order in $\omega t$:
\begin{equation}
  \label{eq:n2bis}
  X_{t} \!\simeq\!
  \left\{
  \begin{array}{ll}
  \!\hbar \kappa t \quad &
    \text{for $0\leq t\leq T = (\hbar\kappa)^{-1}$} = 1\, \text{s},  \\
  \!1 \makebox{cm} \quad &
    \text{for $T \leq t \ll \omega^{-1} \simeq 2.0 \!\times\! 10^{4}$ s}
 \end{array}
 \right.
\end{equation}
As we see, the distance between the two peaks increases almost
linearly in time, reaching its maximum (1 cm) at the end of the
measurement process, as predicted by the standard Schr\"odinger
equation; after this time, their distance remains practically
unaltered for extremely long times, and only for $t \simeq 2.0
\times 10^{4}$ s it starts slowly to decrease, eventually going to
0. Note that such a behavior, being determined by $\omega$, does
{\it not} depend on the mass of the pointer, thus a larger pointer
will not change the situation. The moral is that $X_{t}$ behaves as
if the reduction mechanism were not present (as if $\lambda_{0} =
0$) so we have to look for the collapse somewhere else.

As we shall see now, the collapse occurs
because, in a very short time, the measure of one of the two Gaussian
wave functions ($\phi^{+}_{t}$ or $\phi^{-}_{t}$)
becomes much smaller than the measure of the other component; this
implies that one of the two components practically disappears, and
only the other one survives, the one which determines the outcome of the
experiment. Of course, this process is random and, as we shall
prove, it occurs with a probability almost equivalent to the Born
probability rule.

The relative damping between the two Gaussian components of
Eqn.~(\ref{eq:30}) is measured by the stochastic process: 
\begin{equation} \label{eq:dfhgpwwm}
\Gamma_{t}
= \gamma^{+}_{t} - \gamma^{-}_{t}.
\end{equation} 
Note that, according to Eqn.~(\ref{eq:orygzf}), $e^{\Gamma_0} = |c_+|/|c_-|$. If, at the end of the measurement process, it
occurs that $\Gamma_{t} \gg 1$, it means that~$\phi^{-}_{t}$ is suppressed with respect
to~$\phi^{+}_{t}$, so that the initial state~(\ref{eq:10})
practically evolves to~$|+\rangle \otimes \phi^{+}_{t}$; the opposite
happens if $\Gamma_{t} \ll -1$.

In~\cite{Bassi:07} it is shown that $\Gamma_t$ satisfies the nonlinear stochastic differential equation:
\begin{equation}
  \label{eq:37b}
  d \Gamma_{t} \; = \; \lambda X_{t}^{2} \tanh \Gamma_{t}\, dt
  \, + \, \sqrt{\lambda} X_{t}\, dW_{t},
\end{equation}
to be solved with initial condition $\Gamma_{0} = \ln |c_{+}/c_{-}|$.
To proceed
further with the analysis, it is convenient to perform the following
time change:
\begin{equation}
  \label{eq:45}
  t \quad \longrightarrow \quad s_{t} \; := \; \lambda \int_{0}^{t}
  X_{t}^{2}\, dt',
\end{equation}
which allows us to describe the collapse process in terms of the
dimensionless quantity $s$ that measures its effectiveness. Using
Eqn.~\eqref{eq:n2}, one can solve exactly the above integral and
compute $s$ as a function of $t$. Such a function however cannot be
inverted analytically in order to get $t$ from $s$. Therefore, we use the
simplified expression~\eqref{eq:n2bis} in place of the exact
formula~\eqref{eq:n2} to compute the integral, an expression
which, as we have seen, represents a very good approximation to the
time evolution of $X_{t}$ throughout the whole time during which the
experiment takes place. Accordingly,
we have:
\begin{eqnarray} \label{eq:n45}
s & \equiv & s_{t} \; \simeq \; \frac{\lambda \hbar^2 \kappa^2}{3}\,
t^3 \; \simeq \; 2.0 \times 10^{17}\, (t/\text{s})^3 \nonumber \\
& & \quad\quad\qquad \text{for}\; 0
\, \leq t \, \leq T = 1 \; \text{s}, \\
t & \equiv & t_{s} \; \simeq \; \sqrt[3]{\frac{3}{\lambda \hbar^2
\kappa^2}\,s} \; \simeq \; (1.7 \times 10^{-6}\, \sqrt[3]{s})\;
\text{s} \nonumber \\
& & \text{for}\; 0 \, \leq s \, \leq \lambda \hbar^2 \kappa^2/3 =
2.0 \times 10^{17}.
\end{eqnarray}
Note that, according to the above equations, the physical time $t$
depends on $s$ through the inverse cubic-root of $\lambda$, i.e. on
the inverse cubic-root of the mass of the pointer; this time
dependence of $t$ on $\lambda$ is important since, as we shall see,
it will affect the collapse time. We do not study the functional
dependence between $s$ and $t$ for $t \geq T$ since, as we shall
soon see and as we expect it to be, the collapse occurs at times
much smaller than $T$.

Written in terms of the new variable $s$, Eqn.~\eqref{eq:37b} reduces
to:
\begin{equation} \label{eq:n14} d \Gamma_{s} \; = \; \tanh
\Gamma_{s} \, ds \, + \, dW_{s};
\end{equation}
this equation belongs to a general class of stochastic differential equations  whose properties are known in detail~\cite{Gikhman:72}. Here we report the main results.

{\it Collapse time.} The collapse time is the time when $|\Gamma_s| \geq b$ (where $b$ is some fixed number much larger than 1), i.e. the time when one of the two terms of the superposition becomes dominant with respect to the other:
\begin{equation}
S_{\text{\tiny COL}} \; \equiv \; \inf \{ s: \;\; |\Gamma_{s}| \geq b \}.
\end{equation}
This is a random variable (for each run of the experiment, the collapse time slightly changes), whose mean and variance can be exactly computed~\cite{Gikhman:72}. In particular, if we start with an equal-weight superposition ($\Gamma_0 = 0$), then: ${\mathbb
E}_{\mathbb P}[S_{\text{\tiny COL}}] \simeq b$ and ${\mathbb V}_{\mathbb
P}[S_{\text{\tiny COL}}] \simeq b$, where ${\mathbb
E}_{\mathbb P}[\cdot]$ and ${\mathbb V}_{\mathbb
P}[\cdot]$ denote the mean and variance, respectively. If we transform back from $s$ to the physical time $t$, we have the following estimate for the collapse time for the 1-g pointer~\cite{Bassi:07}:
\begin{equation} \label{eq:n13}
T_{\text{\tiny COL}} \; \simeq \; 1.5 \times 10^{-4} \, \makebox{s}.
\end{equation}
(This value refers to $b = 35$.) The collapse occurs within a time interval smaller than the
perception time of a human observer. Moreover, as proven in~~\cite{Bassi:07}, $T_{\text{\tiny COL}}$ is proportional to the inverse cubic-root
of the mass of the pointer: therefore, the bigger the pointer, the shorter the collapse time. With
our choice for $\lambda_{0}$, even for a 1-g pointer the reduction
occurs practically instantaneously.

It is important to note that, at time $T_{\text{\tiny COL}} \simeq 1.5 \times
10^{-4}$ s, the distance between the two Gaussian components, according to Eqn.~(\ref{eq:n2bis}), is
approximately $X_{T_{\text{\tiny COL}}} \simeq 1.5 \times 10^{-4}$ cm: this means
that, with very high probability, the collapse occurs {\it before}
the two components have enough time to spread out in space to form a
macroscopic superposition. This means that, from the physical point
of view, there is {\it no} collapse of the wave function at all,
since it always remains perfectly localized in space at any stage of
the experiment.

{\it Collapse probability.} Let us call $P_{+}$ the probability that $\Gamma_{s}$ hits the point
$+b$ before the point $-b$, i.e. the probability that
$\phi^{+}_{s}$ survives during the collapse process so that
the outcome of the measurement is ``$+\hbar/2$''. Such a probability
turns out to be equal to~\cite{Bassi:07}:
\begin{equation}
P_{+} \; = \; \frac{1}{2}\, \frac{\tanh b + \tanh \Gamma_{0}}{\tanh
b};
\end{equation}
while the probability $P_{-}$ that $\Gamma_{s}$ hits the point $-b$
before the point $+b$, i.e. that the outcome of the experiment is
``$-\hbar/2$'', is:
\begin{equation}
P_{-} \; = \; \frac{1}{2}\, \frac{\tanh b - \tanh \Gamma_{0}}{\tanh
b}.
\end{equation}
By taking into account that $\tanh b \simeq 1$, since we have assumed that $b \gg 1$, and resorting to Eqns.~(\ref{eq:orygzf}) and~(\ref{eq:dfhgpwwm}), we can write, 
with extremely good approximation~\cite{Bassi:07}:
\begin{eqnarray}
P_{+} & \simeq & \frac{1}{2}\, \left[ 1 + \tanh \Gamma_{0} \right]
\; = \; \frac{e^{\Gamma_{0}}}{e^{\Gamma_{0}} + e^{-\Gamma_{0}}} \nonumber \\
& = & \frac{e^{2\gamma^{+}_{0}}}{e^{2\gamma^{+}_{0}} +
e^{2\gamma^{-}_{0}}} \; = \; |c_{+}|^2,
\\
P_{-} & \simeq & \frac{1}{2}\, \left[ 1 - \tanh \Gamma_{0} \right]
\; = \; \frac{e^{-\Gamma_{0}}}{e^{\Gamma_{0}} + e^{-\Gamma_{0}}} \nonumber \\
& = & \frac{e^{2\gamma^{-}_{0}}}{e^{2\gamma^{+}_{0}} +
e^{2\gamma^{-}_{0}}} \; = \; |c_{-}|^2.
\end{eqnarray}
We see that the probability of getting one of the two possible
outcomes is practically equivalent to the {\it Born probability
rule.} On the one hand, this is an entirely expected result, since
collapse models have been designed precisely in order to solve the
measurement problem and in particular to reproduce quantum
probabilities; on the other hand, it is striking that a very general
equation like Eqn.~(\ref{eq:qmupl1}), which is meant to describe both
quantum systems as well as macroscopic classical objects (i.e. all
physical situations, at the non-relativistic level), when applied to
a measurement situation, provides not only a consistent description
of the measurement process, but also reproduces quantum
probabilities with such a good precision.

{\it State vector after the collapse.} At time $t \geq T_{\text{\tiny COL}}$ the state of the composite system is:
\begin{equation}
\Psi_{t} \quad = \quad \frac{|+\rangle \otimes \tilde{\phi}^+_t
\; + \; \epsilon_{t} |-\rangle \otimes \tilde{\phi}^-_t}{\sqrt{1 + \epsilon_{t}^2}},
\end{equation}
where $\epsilon_{t} \; \equiv \; e^{-(\gamma^{+}_{t} -
\gamma^{-}_{t})} $ and the normalized Gaussian states $\tilde{\phi}^{\pm}_t$ are defined as follows:
\begin{eqnarray}
\tilde{\phi}^{\pm}_t & = &
\sqrt[4]{\frac{1}{2\pi\sigma_{q}^{2}}}  \exp \left[ -
\frac{1-i}{4\sigma_{q}^{2}} (x - \bar{x}^{\pm}_{t})^{2}  \right. \nonumber \\
& & \qquad\qquad\qquad\left. + i
\bar{k}^{\pm}_{t} x + i\, \theta^{\pm}_{t} \right].
\end{eqnarray}
Let us assume that the collapse occurred in favor of the
positive eigenvalue, i.e. in such a way that $\Gamma_{t} \geq
b$ for $t \geq T_{\text{\tiny COL}}$; it follows that:
\begin{equation}
\epsilon_{t} \; \leq \; e^{-b} \; \simeq \; 0 \quad \text{if $b \gg 1$},
\end{equation}
and we can write, with excellent accuracy:
\begin{equation}
\Psi_{t} \quad \simeq \quad |+\rangle \otimes \tilde{\phi}^+_t.
\end{equation}
We recover in this way the {\it postulate of wave packet reduction}
of standard quantum mechanics: at the end of the measurement
process, the state of the micro-system reduces to the eigenstate
corresponding to the eigenvalue which has been obtained as the
outcome of the measurement, the outcome being defined by the
surviving Gaussian component ($\tilde{\phi}^+_t$ in this case). Note
the important fact that, according to our model, the collapse acts
directly only on the pointer of the measuring apparatus, not on the
micro-system; however, the combined effect of the collapse plus the
von Neumann type of interaction is that the microscopic
superposition of the spin states of the micro-system is rapidly
reduced right after the measurement.

Note finally that, after the collapse, the states of the
micro-system and of the pointer are de facto factorized: as such,
after the measurement process one can, for all practical purposes,
disregard the pointer and focus only on the micro-system for future
experiments or interactions with other systems, as is the custom in
laboratory experiments.

To conclude, we have seen how collapse models can describe quantum measurements in a precise way, without ambiguities and paradoxes. We have seen that the standard recipe for quantum measurements (definite outcomes, the Born rule, the postulate of wave function collapse) derives from the dynamical equation~(\ref{eq:qmupl1}) and need not be postulated in an ad hoc way. But there is something more: it can be shown~\cite{Bassi2:07} that also the Hilbert space operator formalism---according to which observable quantities are represented by self-adjoint operators, whose eigenstates and eigenvalues have the role ascribed to them by standard quantum mechanics---can also be derived from Eqn.~(\ref{eq:qmupl1}). In other words, in collapse models there is only the wave function and a collapse equation like Eqn.~(\ref{eq:qmupl1}): everything else can be derived from it.

\subsection{On the origin of the noise field}

As we mentioned earlier, the QMUPL model can be generalized in order to include dissipative effects~\cite{Bassi3:05} and non-white noises~\cite{Bassi:09,Bassi2:09,Ferialdi:11}. According to these models, the noise field acquires a more physical character: it can be assigned a finite temperature, and its spectrum is arbitrary; moreover, this noise is assumed to fill space. It becomes then natural to consider whether it can have a cosmological origin. At present it is too early to answer such a question, though some work has already
been done~\cite{Adler:08} and some people have already suggested that it could have a
gravitational~\cite{Penrose:96,Karolyhazi:86,Feynman:95,Diosi:89} or
pre-quantum~\cite{Adler:04} nature. Moreover, it is still not clear why it has
an anti-Hermitian coupling to matter, which is necessary to ensure the
collapse\footnote{With an Hermitian coupling, one would have a standard quantum Hamiltonian with a random potential; the equation would be linear and no suppression of quantum superpositions would occur.}.
However, one can meaningfully ask whether a noise with `typical' cosmological properties (in terms of temperature and correlation function) can induce an efficient collapse of the wave function, where by `efficient' we mean that the collapse is fast enough to avoid the occurrence of macroscopic superpositions.

Let us consider a Gaussian state as in Eqn.~(\ref{gsol}), whose time evolution can be analytically unfolded in all models so far described. The quantity we are interested in, is its spread in space $\sigma_t = (2\sqrt{\alpha_{t}^{\text{\tiny R}}})^{-1/2}$.
This is plotted in Fig. 1. The top row shows the difference in the evolution of the spread as given by the
dissipative QMUPL model and by the original QMUPL model.
\begin{figure}[!htb]
\begin{center}
\includegraphics[width=8.0cm]{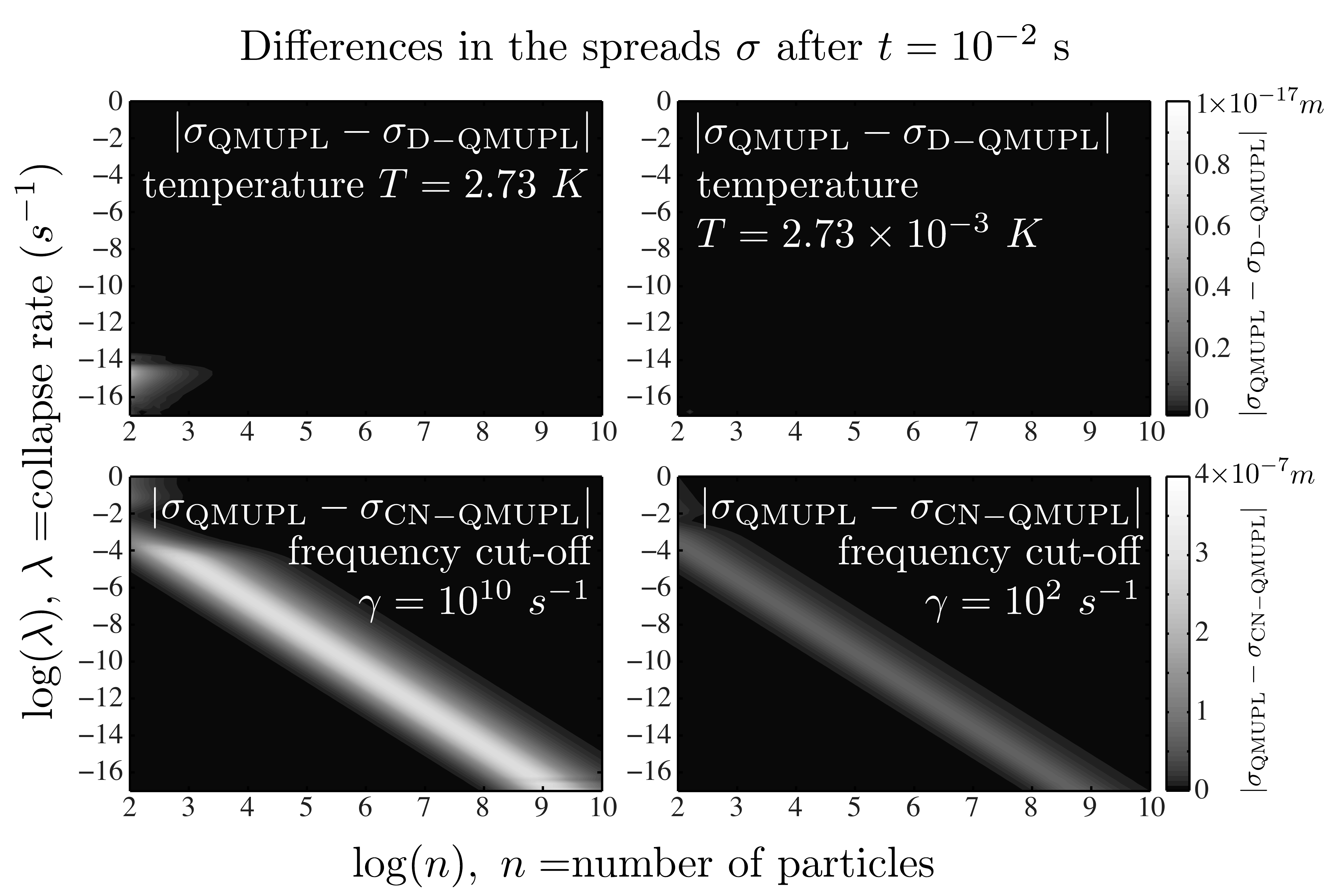}
\caption{Difference between the spread $\sigma$ predicted by the QMUPL model with that given by the dissipative QMUPL model (D-QMUPL), for T = 2.73K (top row left) and T = 2.73 $\times 10^{-3}$K (top row right); and by the colored-noise model (CN-QMUPL), with cutoff at $10^{10}$Hz (bottom row left) and $10^2$Hz (bottom row right). As the color bars on the right show, the whiter the region, the greater the difference in the spreads.  The initial value $\sigma_0 = 5 \times 10^{-7}$m and the elapsed time $t = 10^{-2}$s reproduce the typical geometry of the macromolecule diffraction experiments. At lower temperatures or lower cutoffs the wave function tends to collapse more slowly, which results in a bigger difference with respect to the QMUPL model. Regarding the plots in the top row, the discrepancy manifests in the lower left corner of the plot for $T=2.73\times 10^{-3}$K which disappears for $T=2.73$K. Regarding the plots in the bottom row, it is manifest in the diagonal strip for $\gamma=10^{2}$Hz which decreases for $\gamma=10^{10}$Hz. This diagonal feature exactly lies on the ridge between the quantum and the classical regime. The large discrepancy there is due to the missing high frequencies in the noise spectrum of the colored-noise model. Without these high frequencies the colored-noise model is not able to reproduce the sharpness of the ridge predicted by the QMUPL model.
  }\label{fig:2}
\end{center}
\end{figure}
Two temperatures have been considered: $T = 2.73$K (top row left), as for the CMBR and $T =
2.73 \times 10^{-3}$K (top row right). In both cases the difference is negligible. This means
that even a rather cold thermal field, according to cosmological standards, can
induce an efficient collapse of the wave function, as efficient as with the
standard QMUPL model.
The bottom row shows the difference in the evolution of the spread as given by the
QMUPL model, and by the colored-noise model, with a noise having a frequency cutoff.
Neither high-frequency cutoff affects the collapsing properties of the model in an
appreciable way. These results can be compared with the behavior of typical
cosmological fields such as the CMBR, the relic neutrino background and the
relic gravitational background. The spectrum of the first two have a cutoff
(measured or expected) at $\sim 10^{11}$ Hz, while the spectrum of the third
one probably lies at $\sim 10^{10}$ Hz~\cite{Grishchuk:10}. All these cutoffs
as well as that of $\sim 10^{15}$ Hz proposed in~\cite{Bassi:03} for the collapse noise
 ensure a rapid collapse of the wave function. While the collapse is robust
over a large range of cutoffs, other effects, such as the emission of
radiation from charged particles, highly depend on the spectrum of the noise
correlator~\cite{Adler2:07}.

Therefore, the message which can be drawn is that a cosmological field with `typical'
properties can induce an efficient collapse. A great challenge is to test the
existence of such a field.

\subsection{The CSL model}

The QMUPL model has the advantage of allowing for quite a rigorous mathematical analysis of the main features of collapse models, as was shown in the previous sections. However it does not seem physically realistic, for two main reasons. The first one is that it is built for systems of distinguishable particles, and its generalization to identical particles does not seem straightforward. The second reason is that the noise field depends only on time, not on space, thus it cannot be immediately identified with a random field of Nature. The CSL model ~\cite{PEARLE2}, ~\cite{Ghirardi2:90} overcomes the above difficulties, and so far remains the most advanced collapse model. In its {\it mass proportional} version~\cite{Pearle:94}, it is defined by the following stochastic differential equation in the Fock space:
\begin{eqnarray} \label{eq:csl-massa}
d\psi_t & = &  \left[-\frac{i}{\hbar}Hdt \right. \\
& + &\frac{\sqrt{\gamma}}{m_{0}}\int d\mathbf{x} (M(\mathbf{x}) - \langle M(\mathbf{x}) \rangle_t)
dW_{t}(\mathbf{x}) \nonumber \\
& - & \left. \frac{\gamma}{2m_{0}^{2}} \int d\mathbf{x}\,
(M(\mathbf{x}) - \langle M(\mathbf{x}) \rangle_t)^2 dt\right] \psi_t;  \nonumber
\end{eqnarray}
as usual, $H$ is the standard quantum Hamiltonian of the system and the other two
terms induce the collapse of the wave function in space. The mass $m_0$ is a reference mass, which as usual is taken
equal to that of a nucleon. The parameter $\gamma$ is a positive coupling
constant which sets the strength of the collapse process, while $M({\bf x})$ is
a smeared {\it mass density} operator:
\begin{eqnarray}
M(\mathbf{x})
& = & \underset{j}{\sum}m_{j}N_{j}(\mathbf{x}), \label{eq:dsfjdhz}\\
N_{j}(\mathbf{x})
& = & \int d\mathbf{y}g(\mathbf{y-x})
\psi_{j}^{\dagger}(\mathbf{y})\psi_{j}(\mathbf{y}), \qquad
\end{eqnarray}
$\psi_{j}^{\dagger}(\mathbf{y})$,
$\psi_{j}(\mathbf{y})$ being, respectively, the creation and
annihilation operators of a particle of type $j$ in the space point
$\mathbf{y}$. The smearing function $g({\bf x})$ is taken equal to
\begin{equation} \label{eq:nnbnm}
g(\mathbf{x}) \; = \; \frac{1}{\left(\sqrt{2\pi}r_{c}\right)^{3}}\;
e^{-\mathbf{x}^{2}/2r_{C}^{2}},
\end{equation}
where $r_C$ is the second new phenomenological constant of the model. $W_{t}\left(\mathbf{x}\right)$ is an
ensemble of independent Wiener processes, one for each point in space. [In the original, i.e. `not mass proportional', CSL model, the integrals are proportional to the number density operator, instead of the mass density operator].

As one can see from Eqn.~(\ref{eq:csl-massa}), in the CSL model the collapse operators are the {\it density number} operator $\psi_{j}^{\dagger}\left(\mathbf{y}\right)\psi_{j}\left(\mathbf{y}\right)$, which means that superpositions containing different numbers of particles in different points of space are suppressed. This is equivalent to collapsing the wave function in space, in a second-quantized language.

The collapse occurs more or less as in the QUMPL model, though it is more difficult to unfold: an easier and more handy way to look at the collapse is through the density matrix, in particular how its off-diagonal elements decay in time. Since ordinary matter is made just of electrons and nucleons and, according to Eqns.~(\ref{eq:csl-massa}) and~(\ref{eq:dsfjdhz}), the collapse effect on electrons is negligible in comparison to the effect on nucleons, we focus our attention only on nucleons.

According to Eqn.~(\ref{eq:csl-massa}), the decay of the
off-diagonal elements of the density matrix $\rho_t \equiv {\mathbb E}[|\psi_t\rangle\langle\psi_t|]$ (where ${\mathbb E}[\cdot]$ denotes the stochastic average) of a many-nucleons system, in the {\it position basis}, is~\cite{Ghirardi2:90}:
\begin{equation}
\frac{\partial}{\partial t} \langle \bar{{\bf x}}' |\rho_t| \bar{{\bf x}}'' \rangle
= - \Gamma({\bar{{\bf x}}',\bar{{\bf x}}''}) \, \langle \bar{{\bf x}}' |\rho_t| \bar{{\bf x}}'' \rangle,
\end{equation}
where $\bar{{\bf x}}' \equiv {\bf x}'_1, {\bf x}'_2, \ldots {\bf x}'_N$ (and
similarly for $\bar{{\bf x}}''$). In the above equation, we have neglected the
standard quantum evolution. The decay function $\Gamma$ is:
\begin{eqnarray}\label{eq:blacsl}
\Gamma & = & \frac{\gamma}{2}\sum_{i,j}\left[
G({\bf x}'_i-{\bf x}'_j) + G({\bf x}''_i-{\bf x}''_j) \right. \nonumber \\
& - & \left. 2G({\bf x}'_i-{\bf x}''_j)
\right],
\end{eqnarray}
where the indices $i,j$ run over the $N$ nucleons of the system, and:
\begin{equation} \label{eq:oudssvc}
G({\bf x}) = \frac{1}{(4\pi r_C^2)^{3/2}}e^{-{\bf x}^2/4r_C^2}.
\end{equation}

As a first observation, in the case of a single nucleon $\Gamma$ reduces to:
\begin{equation}
\Gamma({\bf x}', {\bf x}'') = \frac{\gamma}{(4\pi r_C^2)^{3/2}}\left[1 - e^{-|{\bf x}' - {\bf x}''|^2/4r_C^2} \right],
\end{equation}
which is precisely the GRW 1-particle collapse term (see Eqn.~(\ref{eq:fsgdffpq})). 
Accordingly, the two models give similar predictions regarding the collapse effects on systems containing just a few particles, while for many-particle systems important differences emerge, as we will see soon. 
The collapse rate is defined in terms of $\gamma$ as follows:
\begin{equation}
\lambda_{\text{\tiny CSL}} \; = \; \frac{\gamma}{(4\pi r_C^2)^{3/2}}.
\end{equation}
In~\cite{Ghirardi2:90} the following choice for $\gamma$ was made:
$ \gamma \sim 10^{-30} \text{cm}^3\text{s}^{-1}$,
corresponding to:
\begin{equation} \label{eq:ffgfkgyi}
\lambda_{\text{\tiny CSL}} \sim 2.2 \times 10^{-17} \text{s}^{-1}.
\end{equation}
Note the difference of about one order of magnitude between $\lambda_{\text{\tiny GRW}}$ and $\lambda_{\text{\tiny CSL}}$.

Several useful approximate formulas can be obtained from Eqn.~(\ref{eq:blacsl}). The first one is for {\it large distances}. When the particles in a superposition are displaced by a distance $\ell = | {\bf x}' - {\bf x}''| \ll r_C$, then according to Eqn.~(\ref{eq:oudssvc}) their contribution to $\Gamma$ is negligibly small. Thus,
only superpositions with $\ell \geq r_C$ contribute to $\Gamma$ and trigger the
collapse of the wave function. In such a case, the formula shows that for
groups of particles separated (in each term of the superposition) by less than
$r_C$, the rate $\Gamma$ increases quadratically with the number of particles
while for groups of particles separated by more than $r_C$ it increases
linearly. Thus we have the following simplified formula for the collapse rate~\cite{Adler3:07}:
\begin{equation} \label{eq:ridert}
\Gamma \; = \; \lambda_{\text{\tiny CSL}} n^2 N,
\end{equation}
where $n$ is the number of particles within a distance $r_C$, and $N$ is the
number of such clusters. We note that the quadratic dependence of $\Gamma$ on the number of particles---which is absent in the original GRW model---is a direct effect of the identity of particles. This means that the identity of particles works in favor of the collapse.

An estimate for {\it small distances} can be obtained by Taylor expanding $G({\bf x})$ as follows:
\begin{equation}
G({\bf x}) \simeq \frac{1}{(4\pi r_C^2)^{3/2}}\left[1 - \frac{{\bf x}^2}{4r_C^2} \right],
\end{equation}
which leads to:
\begin{equation}
\Gamma({\bf x}', {\bf x}'') \; \simeq \; \frac{\lambda}{4 r_C^2}\left( \sum_i ({\bf x}'_i-{\bf x}''_i) \right)^2.
\label{eq:lambda}
\end{equation}
As we can see---and as expected---the collapse strength grows quadratically with the superposition distance for small distances, like in the GRW and QMUPL models, the important difference here being that there is a quadratic dependence also on the number of particles. In both cases (large- and small-distance approximation), we see the amplification mechanism: the collapse rate increases with the size of the system.

Another useful formula can be obtained for macroscopic rigid systems, for which the mass distribution can be expressed by a density function D({\bf x}), averaging the contributions of the single nucleons. In such a case the decay function $\Gamma$ takes the simpler expression~\cite{Ghirardi2:90}:
\begin{eqnarray} \label{eq:sdfgerd}
\Gamma({\bf X}', {\bf X}'') & = & \gamma \int d {\bf x} [D^2({\bf x}) \nonumber \\
& - & D({\bf x})D({\bf x} + {\bf X}' - {\bf X}'')], \quad
\end{eqnarray}
where ${\bf X}'$ and ${\bf X}''$ are the positions of the center of mass of the object, in the two terms of the superposition. The physical meaning of Eqn.~(\ref{eq:sdfgerd}) can be understood by making reference to a homogeneous macroscopic body of constant density $D$. Then the decay rate becomes:
\begin{equation}
\Gamma = \gamma D n_{\text{\tiny OUT}},
\end{equation}
where $n_{\text{\tiny OUT}}$ is the number of particles of the body when the center-of-mass position is ${\bf X}'$, which do not lie in the volume occupied by the body when the center of mass position is ${\bf X}''$.

Further properties of the CSL model are discussed in~\cite{Ghirardi2:90} and~\cite{Bassi:03}.

\subsubsection{On the nature of the noise field of the CSL model}

As anticipated, contrary to the QMUPL model, in the CSL model the noise field can be given a straightforward physical interpretation. In order to see this, it is convenient to rewrite the CSL dynamical equation~(\ref{eq:csl-massa}) in the following equivalent form:
\begin{eqnarray} \label{eq:csl-massa2}
d\psi_t & = &  \left[-\frac{i}{\hbar}Hdt \right. \\
& + &\frac{\sqrt{\gamma}}{m_{0}}\int d\mathbf{x} (M(\mathbf{x}) - \langle M(\mathbf{x}) \rangle_t)
d\overline{W}_{t}(\mathbf{x}) \nonumber \\
& - & \frac{\gamma}{2m_{0}^{2}} \int d\mathbf{x} d\mathbf{y}
(M(\mathbf{x}) - \langle M(\mathbf{x}) \rangle_t) \cdot \nonumber \\
& & \left. \phantom{\frac{1}{2}} G({\bf x} - {\bf y}) (M(\mathbf{y}) - \langle M(\mathbf{y}) \rangle_t) dt\right] \psi_t,  \nonumber
\end{eqnarray}
where $M({\bf x})$ is still the mass density operator defined in Eqn.~(\ref{eq:dsfjdhz}), while $N_j({\bf x})$ now is the standard number density operator: $N_j({\bf x}) = \psi_{j}^{\dagger}\left(\mathbf{x}\right)\psi_{j}\left(\mathbf{x}\right)$, and $G({\bf x})$ is the same Gaussian function defined in Eqn.~(\ref{eq:oudssvc}). The Wiener processes $\overline{W}_{t}(\mathbf{x})$ are not independent anymore, instead they are Gaussianly correlated in space, the correlator being $G({\bf x})$. The ``white'' noise field $w(t, {\bf x}) \equiv d \overline{W}_{t}(\mathbf{x}) / dt$ corresponds to a {\it Gaussian} field with zero mean and correlation function:
\begin{equation}
{\mathbb E}[w(t, {\bf x})w(s, {\bf y})] = \delta(t-s) G({\bf x} - {\bf y}).
\end{equation}
As we see, the noise field of the CSL model can be interpreted as a (classical) random field filling space. It is white in time, for the very simple reason that white noises are easier to analyze mathematically. It is Gaussianly correlated in space, with correlation length equal to $r_C$. As discussed in connection with the QMUPL model, it is tempting to suggest that such a field has a cosmological nature, and preliminary calculations show that a noise with typical cosmological features yields a satisfactory collapse of the wave function~\cite{Bassi2:10}. However at this stage this is only a (very tempting) speculation. Moreover, one has to justify the non-Hermitian coupling and the non-linear character of the collapse equations, which are necessary in order to obtain the effective collapse dynamics.

\subsubsection{Generalizations of the CSL model and relativistic models}

Like the QMUPL model, the CSL model can also be generalized in several directions. The first generalization one would make, is to include {\it dissipative} terms, in order to cure the problem of energy
non-conservation. In fact, in the CSL model also the energy increases at a steady rate, though such an increase is negligible for all practical purposes. This can be easily seen by noting that the 1-particle master equation of the CSL model coincides with the 1-particle equation of the GRW model. This type of generalization should not present particular problems, however it has not been worked out so far.

The second generalization consists in replacing the white noise field with a more {\it general Gaussian noise}. As for the QMUPL model, the equations become non-Markovian, therefore difficult to analyze mathematically. The analysis has been carried out to the leading perturbative order---with respect to the collapse parameters $\gamma$---in~\cite{Adler:07,Adler:08}. The result of the analysis is the expected one: the collapse qualitatively occurs with the same modalities as in the white noise case, the rate depending on the correlation function of the noise. In particular, the rate is robust against changes of the correlation functions, while other predictions are very sensitive to the form of the time correlator~\cite{Adler2:07}. Much more work is needed in order to understand the properties of the non-Markovian  CSL model.

The great challenge of the dynamical reduction program is to
formulate a consistent model of spontaneous wave function collapse
for {\it relativistic} quantum field theories; many attempts have been
proposed so far, none of which is as satisfactory as the
non-relativistic GRW and CSL models.

The first attempt~\cite{Ghirardi:90, Pearle:90} aimed at making the CSL model
relativistically invariant by replacing Eqn.~(\ref{eq:csl-massa}) with a
Tomonaga-Schwinger equation of the type:
\begin{eqnarray} \label{eq:ts}
\frac{\delta\psi(\sigma)}{\delta \sigma(x)}
& = &
\left[ -\frac{i}{\hbar} {\mathcal H}(x) \right. \\
& + &
\sqrt{\gamma} \left( {\mathcal
L}(x) - \langle {\mathcal L}(x) \rangle \right) w(x) \nonumber \\
& - & \left.
\frac{\gamma}{2} \left( {\mathcal L}(x) - \langle {\mathcal L}(x)
\rangle \right)^2 \right]\psi(\sigma), \nonumber
\end{eqnarray}
where now the wave function is defined on an arbitrary space-like
hypersurface $\sigma$ of space-time. The operator ${\mathcal
H}(x)$ is the Hamiltonian density of the system ($x$ now denotes a
point in space-time), and ${\mathcal L}(x)$ is a {\it local}
density of the fields, on whose eigenmanifolds one decides to
localize the wave function. The $c$-number function $w(x)$ is a random field on space-time with mean equal to zero, while
the correlation function---in order for the theory to be Lorentz
invariant in the appropriate stochastic sense~\cite{Ghirardi:90}---must
be a Lorentz scalar. And here the problems arise.

The simplest Lorentz invariant choice for the correlation function
is:
\begin{equation} \label{eq:cf}
{\mathbb E}[w(x) w(y)] = \delta^{(4)}(x - y),
\end{equation}
which however is not physically acceptable as it causes an infinite
production of energy per unit time and unit volume. The reason is
that in Eqn.~(\ref{eq:ts}) the fields are {\it locally} coupled to
the noise which, when it is assumed to be {\it white}, is too
violent, so to speak, and causes too many particles to come out of
the vacuum. To better understand the situation, let us go back to
the non-relativistic Eqn.~(\ref{eq:csl-massa}): also there we basically
have a white-noise process, which however is not coupled locally to
the quantum field $a^{\dagger}(s, {\bf y}) a(s, {\bf y})$, the
coupling being mediated by the smearing Gaussian function appearing
in the definition of $N({\bf x})$. One can compute the energy
increase due to the collapse mechanism, which turns out to be {\it
proportional} to $r_C$. Now, if we want to have a local coupling
between the quantum field and the noise, we must set $r_C
\rightarrow + 0$, in which case the energy automatically
diverges also for finite times.

The simplest way out one would think of, in order to cure this
problem of Eqn.~(\ref{eq:ts}), is to replace the local coupling
between the noise and the quantum field by a non-local one, as in
the CSL equation~(\ref{eq:csl-massa}); this procedure would essentially
amount to replacing the white noise field with a non-white one. In
both cases we need to find a Lorentz invariant function which either
smears out the coupling or replaces the Dirac-$\delta$ in the
definition of the correlation function~(\ref{eq:cf}). This however
is not a straightforward task, for the following reason.

One of the reasons why the third term $(\gamma/2) \left( {\mathcal
L}(x) - \langle {\mathcal L}(x) \rangle \right)^2$ appears in
Eqn.~(\ref{eq:ts}) is to guarantee that the collapse mechanism occurs
with the correct quantum probabilities (for those who are experts in
stochastic processes, the third term is such that the equation
embodies an appropriate martingale structure); if we change the
noise, we then have to change also the third term, and it turns out
that we have to replace it with a {\it non-local} function of the
fields~\cite{Nicrosini:03,Myrvold:09}. But, having a non-local function of the
fields jeopardizes the entire (somehow formal) construction of the
theory based on the Tomanaga-Schwinger formalism, as the
integrability conditions are not automatically satisfied. More analysis is required, however it is
very likely that the model will turn out to be inconsistent.

What we have briefly described is the major obstacle to finding a
relativistic dynamical reduction model. We want to briefly mention
three research programs which try to overcome such an impasse.

P. Pearle has spent many years in trying to avoid the infinite
energy increase of relativistic spontaneous collapse models, e.g.
by considering a tachyonic noise in place of a white noise as the
agent of the collapse process~\cite{Pearle2:99}, obtaining suggestive
results. Unfortunately, as he has recently admitted~\cite{Myrvold:09},
this program was not successful. A promising new way to tackle this problem has been recently proposed by D. Bedingham~\cite{Bedingham:10, Bedingham:11}.

Dowker and Henson have proposed a spontaneous collapse model for a
quantum field theory defined on a 1+1 null lattice~\cite{Dowker:04,Dowker2:04}, studying issues like the non-locality of the model and the
no-faster-than-light constraint. More work needs to be done in
trying to apply it to more realistic field theories; in particular,
it would be important to understand if, in the continuum limit, one
can get rid of the divergences which plague the relativistic CSL
model.

In a recent paper~\cite{Tumulka2:06}, generalizing a previous idea of
Bell~\cite{Bell:87}, Tumulka has proposed a discrete, GRW-like,
relativistic model, for a system of $N$ non-interacting particles,
based on the multi-time formalism with $N$ Dirac equations, one per
particle; the model fulfills all the necessary requirements, thus it
represents a promising step forward in the search for a relativistic
theory of dynamical reduction. Now it is important to understand
whether it can be generalized in order to include also interactions.

Recently, a completely different perspective towards relativistic collapse models emerged~\cite{Adler:04}. If one assumes that the random field causing the collapse of the wave function is a {\it physical} field filling space and possibly having a cosmological origin, then there is no need to make the equations relativistically invariant. The noise would select a privileged reference frame---in pretty much the same way in which the CMBR identifies a preferred frame. Then the collapse equation, being a phenomenological equation, need not be relativistic invariant, being dependent on the noise. The underlying theory, out of which these equations would emerge at an appropriate coarse-grained level, should respect the appropriate symmetries (Lorentz invariance or a possible generalization of it). The theory would explain the origin of the noise field which, because of the initial conditions, would break the relevant symmetry. This is only speculation at the moment. However it is a reasonable program, though difficult to carry out.

\subsection{Choice of the parameters}

The choice~(\ref{eq:num}) for $\lambda_{\text{\tiny GRW}}$ (or~(\ref{eq:ffgfkgyi}) for $\lambda_{\text{\tiny CSL}}$) and $r_C$ makes sure that collapse models agree with all observational evidence about the quantum nature of microscopic systems, while macro-objects are always localized in space and behave according to Newton's laws (within experimentally testable limits).  It sets a quantum-classical threshold at
$\sim 10^{13}$ nucleons~\cite{Ghirardi2:90,Bassi:03}.

Recently, a much stronger bound on
the collapse rate has been proposed~\cite{Adler3:07}, namely:
\begin{equation} \label{eq:sdiydsgf}
\lambda_{\text{\tiny Adler}} \simeq 2.2 \times
10^{-8 \pm 2} \text{s$^{-1}$},
\end{equation}
corresponding to a threshold of  $\sim 10^5$ nucleons.
The underlying motivation is that the
collapse should be effective in all measurement processes, also those involving
only a small number of particles as happens in the process of latent image
formation in photography, where only $\sim 10^5$ particles are displaced more
than $r_C$. In order for the collapse to be already effective at this scale,
one has to increase the conventional CSL value $\lambda_{\text{\tiny CSL}}$ by $\sim 10^{9 \pm 2}$ orders of magnitude. Eqn.~(\ref{eq:sdiydsgf}) is also the value one has to take in order to make sure that a superposition of 6 photons\footnote{6 photons corresponds to the threshold of vision.} reaching the human eye, collapses within the eye itself~\cite{Bassi2:10}. However, the value $\lambda_{\text{\tiny GRW}}$ makes sure that the collapse occurs before the signal reaches the brain and turns into a perception ~\cite{Aicardi:91}. 

Both values $\lambda_{\text{\tiny CSL}} \simeq 2.2 \times
10^{-17} \text{s$^{-1}$}$ and $\lambda_{\text{\tiny Adler}} \simeq 2.2 \times
10^{-8 \pm 2} \text{s$^{-1}$}$ are compatible with known experimental data~\cite{Adler:09}.
However, such a large discrepancy of $\sim 9$ orders of magnitude shows that there is no
general consensus on the strength of the collapse process and consequently on
the scale at which violations of quantum linearity can be expected to manifest. Experiments will decide which value---if any---is the correct one.
\section{Underlying Theories}
It is beyond doubt that if Continuous Spontaneous Collapse models are the right way to resolve the problems of quantum theory,
 these models must derive from an underlying physical theory based on new symmetry principles. Here, we review two such possibilities. 
Trace Dynamics is such a theory due to Stephen Adler and collaborators, which bears the same relation to quantum theory that 
thermodynamics bears to statistical mechanics and the kinetic theory of gases. The other, which is not quite a theory yet, is the idea 
that gravity has a role to play in bringing about collapse of the wave-function.
\subsection{Trace Dynamics, Quantum Theory, and Spontaneous Collapse}
The commutation rules in canonical quantization are obtained starting from the Poisson brackets
 in the classical theory. This may be somewhat unsatisfactory, since the classical theory itself is supposed to be a limiting case of the fundamental
quantum theory. In order to know the fundamental theory we  should not have to know its limit. The quantum commutations should be achieved in a more
 fundamental manner and this is what Adler's scheme of Trace Dynamics sets out to do. However, eventually the theory goes beyond deriving quantum
 theory from a more fundamental framework, and provides a plausible resolution of the quantum measurement problem which is experimentally testable.

The physics of Trace Dynamics 
can be described in the following three well-laid
 out steps :

(i) The classical theory, which is the Newtonian dynamics of Grassmann-valued non-commuting matrices; and which as a consequence of global unitary
 invariance possesses a unique non-trivial conserved charge of great significance [Sections 1.1.1 - 1.1.4 below].

(ii) The statistical thermodynamics of this theory, the construction of the canonical ensemble and the derivation of equilibrium. The derivation of an
all important Ward identity, as a consequence of assumed invariance under constant shifts in phase space, from which there emerge, as thermodynamic averages, the canonical commutation relations of quantum theory, the Heisenberg
 equations of motion and the equivalent Schr\"{o}dinger equation of quantum theory [Sections 1.1.5 - 1.1.7].

(iii) The consideration of Brownian motion fluctuations around the above thermodynamic approximation, the consequent non-linear stochastic modification
 of the Schr\"{o}dinger equation, the resolution of the quantum measurement problem and derivation of the Born probability rule [Section 1.1.8].

The review below is based on the book by Adler ~\cite{Adler:04}. The interested reader should consult Adler's book and the references therein for further details.

\subsubsection{The fundamental degrees of freedom}

 In this scheme the fundamental degrees of freedom are {\bf matrices} living on a background spacetime with complex Grassmann numbers (or more precisely
 elements
 of a graded algebra
${\cal{G}_C}$ of complex Grassmann numbers) as elements. Grassmann numbers have the following properties:
\begin{eqnarray}
 &\theta_i\theta_j&+\theta_j\theta_i=0, \hspace{0.5 in} \theta_i^2=0;\nonumber\\
&\chi&=\theta_R+i\theta_{I},\hspace{0.5 in} \{\chi_r,\chi_s\}=0. \nonumber
\end{eqnarray}
Thus we have a matrix field with the help of which to each spacetime point we can associate a matrix
$$M(x)= \left(
\begin{array}{llll}
\chi_{11}(x) & \chi_{12}(x) & .. & ...\\
\chi_{21}(x) & \chi_{22}(x) & .. & ...\\
.. & .. & .. & ..\\
.. & .. & .. & ..
\end{array}
\right).$$
Some further important properties of these Grassmann (matrix-) elements are  the following: \\
(i) Product of an even number of Grassmann elements commutes with all the elements of Grassmann algebra.\\
(ii) Product of an odd number of Grassmann elements anti-commutes with any other odd number product.
Therefore we have two disjoint sectors:\\
{\bf Bosonic Sector- B}: Consists of the identity and the Even Grade elements of the algebra:
$B\equiv\{I,\chi_a\chi_b,\chi_a\chi_b\chi_c\chi_d, ....\}$\\
{\bf Fermionic Sector-F}: Consists of the Odd Grade elements of the algebra:\\
$F\equiv\{\chi_a,\chi_a\chi_b\chi_c, ....\}.$\\
Therefore, the fundamental degrees of freedom of the trace dynamics theory are the matrices made out of elements from these sectors
      $$B_I\in \{M;M_{ij}\in B \}$$
      $$\chi_I\in \{M;M_{ij}\in F \}$$
      $${B_I,\chi_I}\in {\cal{G}}_M$$
where ${\cal G}_M$ is the graded algebra of complex Grassmann matrices.

A general matrix can be decomposed as
$$M={\cal A}_1(\in B_I)+{\cal A}_2(\in \chi_I)$$
into Bosonic and Fermionic sectors. We note that being matrices the degrees of freedom of this scheme are non-commutative in nature. Dimensionality of the matrices can be arbitrary;
however we work with finite dimensional matrices with the assumption that everything done subsequently can be extended to infinite dimensional matrices as
well.\\
Next, we define an operation {\it Trace} on this matrix field as follows
$$ Tr:{\cal{G}}_M \rightarrow {\cal{G}}_C,$$
which is a map from the space of matrices [${\cal{G}}_M$] to the field of complex Grassmann numbers [${\cal{G}}_C$] and is given by
the sum of the diagonal elements of a given matrix.
There are some nice {\it Trace properties} satisfied by the degrees of freedom of the theory
\begin{eqnarray}
TrB_1B_2 &=&
TrB_2B_1 \label{T1}\\
Tr\chi_1\chi_2 &=& 
-Tr\chi_2\chi_1 \label{T2}\\
TrB\chi &=& Tr\chi B \label{T3}
\end{eqnarray}
One also has some interesting Trace tri-linear cyclic identities
\begin{eqnarray}
TrB_1\langle\langle B_2,B_3 \rangle\rangle=TrB_2\langle\langle B_3,B_1\rangle\rangle\nonumber\\
=TrB_3 \langle\langle B_1,B_2\rangle\rangle \\
TrB[\chi_1,\chi_2]=Tr\chi_1[\chi_2,B]\nonumber\\
=Tr\chi_2[\chi_1,B]\\
Tr\chi_1\{B,\chi_2\}=Tr\{\chi_1,B\}\chi_2\nonumber\\
=Tr[\chi_1,\chi_2]B\\
Tr\chi\langle\langle B_1,B_2\rangle\rangle=TrB_2\langle\langle\chi,B_1\rangle\rangle\nonumber\\
=TrB_1\langle\langle B_2,\chi \rangle\rangle \\
Tr\chi_1\langle\langle\chi_2,\chi_3\rangle\rangle=Tr\chi_2\langle\langle\chi_3,\chi_1\rangle\rangle\nonumber\\
=Tr\chi_3\langle\langle\chi_1,\chi_2\rangle\rangle, \label{T4}
\end{eqnarray}
where $\langle\langle\hspace{0.25 in} \rangle\rangle $ can be commutator or anti-commutator.
Next, we can verify that the matrices have the following adjoint rule
$$(O_1^{g_1}...O_n^{g_n})^{\dag}=(-1)^{\sum_{i<j}g_ig_j} {O_n^{g_n}}^{\dag}.....{O_1^{g_1}}^{\dag},$$
where $g_i$ is the grade (odd/even) of the matrix $O$.
The anti-commutative feature of matrix elements induces non-triviality in the adjointness properties, as seen above.

We now examine the dynamics of these degrees of freedom and later will construct the statistical
mechanics of a gas of such `particles' to 
find that the equations of quantum theory are 
identities, valid in the thermodynamic limit of this underlying theory.
\subsubsection{Classical dynamics}
We can construct a polynomial $P$ from these non-commuting matrices (say $O$) and obtain the {\it trace} (indicated in bold) of the polynomial 
$${\bf P} =TrP.$$
Trace derivative of ${\bf P}$ with respect to the variable $O$ is defined as 
$$\delta{\bf P} =Tr\frac{\delta{\bf P}}{\delta O}\delta O, $$
i.e. $\delta$- variation in {\bf P} should be written in a way that resulting $\delta$- variation of $O$ in each monomial sits
on the right. Then terms coming on the left of $\delta O$ are defined as the trace derivative. It should be mentioned that one always constructs 
 $P$ to be an even graded element of Grassmann algebra. Moreover, $\delta O$ and $\delta {\bf P}$ are also taken to be of same type
(Bosonic/Fermionic) as $O$ and ${\bf P}$ respectively. Thus $\frac{\delta{\bf P}}{\delta O}$ will be of the same type as $O$. \\
For example, let 
$$P=AOBOC,$$
be a polynomial, where $A,B,C$ are operators which in general do not commute with each other or with the variable $O$. Then,
\begin{eqnarray}
\delta Tr P =\delta{\bf P} = Tr\left[\epsilon_{AO}BOCA(\delta O)\right.\nonumber\\
\left.+\epsilon_{OC}CAOB(\delta O)\right], \nonumber
\end{eqnarray}
using trace properties (\ref{T1}), (\ref{T2}), (\ref{T3}) and (\ref{T4}).
Hence, the trace derivative will be
$$\frac{\delta {\bf P}}{\delta O}=\epsilon_{AO}BOCA+\epsilon_{OC}CAOB.$$
Above, $\epsilon_{XY}=+1$ if either $X$ or $Y$ is bosonic, and it is equal to $-1$ if both $X$ and $Y$ are fermionic.
\subsubsection{Lagrangian and Hamiltonian Dynamics} 
Armed with these tools we write {the \it Lagrangian} of a theory as a Grassmann even polynomial function of bosonic/fermionic operators
$\{q_r\}$ and their
time derivatives $\{\dot{q_r}\}$ with $\{q_r,\dot{q_r}\in {\cal{G}}_M\}$.
We define the Trace Lagrangian
$${\bf L}[\{q_r\},\{\dot{q_r}\}]= TrL[\{q_r\},\{\dot{q_r}\}],$$
and subsequently the Trace Action
$${\bf S}=\int dt {\bf L}.$$
Using the trace derivative we obtain the equation of motion by extremizing the action with respect to variation in ${q_r}$ using the differentiation technique
described above
\begin{equation}
\frac{\delta {\bf L}}{\delta q_r} = \frac{d}{dt}\frac{\delta {\bf L}}{\delta \dot{q_r}}.
\end{equation}
Above is a matrix equation; in component form we can write down  $N^2$ Euler-Lagrange equations of motion 
$$\left(\frac{\delta {\bf L}}{\delta q_r}\right)_{ij} =\frac{\partial{\bf L}}{\partial (q_r)_{ji}} . $$
Further we define conjugate momenta as
$$ p_r\equiv\frac{\delta {\bf L}}{\delta \dot{q_r}}.$$
Since the Lagrangian is Grassmann even, momentum will be of same type (Bosonic/Fermionic) as $ q_r$.
In general the coordinates or the momenta do not commute among each other, 
for these are all arbitrary matrices.
The {\it Trace Hamiltonian} is obtained as
$${\bf H}=\sum_rp_r\dot{q_r}-{\bf L}.$$
Therefore, the Hamiltonian equations of motion are
\begin{equation}
\frac{\delta {\bf H}}{\delta q_r}=-\dot{p_r},\qquad
\frac{\delta {\bf H}}{\delta p_r}=\epsilon_r \dot{q_r}, \label{HEV}
 \end{equation}
where $\epsilon_r=\pm 1,$ depending upon whether $r$ is a bosonic or fermionic degree of freedom.

We define a generalized Poisson Bracket over the {\it phase space} $\{q_r,p_r\}$
\begin{equation}
\{{\bf A},{\bf B}\}=Tr\sum_r\epsilon_r\left(\frac{\delta {\bf A}}{\delta q_r}\frac{\delta {\bf B}}{\delta p_r}-
\frac{\delta {\bf B}}{\delta q_r}\frac{\delta {\bf A}}{\delta p_r}\right),
\end{equation}
which satisfies the Jacobi identity
$$\{{\bf A},\{{\bf B},{\bf C}\}\}+\{{\bf C},\{{\bf A},{\bf B}\}\}+\{{\bf B},\{{\bf C},{\bf A}\}\}=0.$$
For ${\bf A}[\{q_r\},\{p_r\},t]$ one can easily verify that,
\begin{equation}
\dot{{\bf A}}=\frac{\partial {\bf A}}{\partial t}+\{{\bf A},{\bf H}\}. \label{TE}
\end{equation}
We observe that the matrix dynamics obtained above is non-unitary in general.
For operators that do not have explicit time dependence, (\ref{TE}) does not show a unitary evolution of the type
$$\dot{x}_r(t)=i[G,x(t)].$$
However, for {\it Weyl-ordered} Hamiltonians, trace-dynamics evolution and Heisenberg unitary time evolution can be shown to be equivalent, on an initial time slice on which the phase space variables are canonical.
\subsubsection{Conserved parameters}
As an obvious result of (\ref{TE}), the trace Hamiltonian itself is conserved,
\begin{equation}
\dot{{\bf H}}=\{{\bf H},{\bf H}\}=0.
\end{equation}
Moreover, for a Trace Hamiltonian restricted to a bilinear form in the Fermionic sector with a self-adjoint kinetic part 
\begin{eqnarray}
{\bf H} = Tr\sum_{r,s\in F}(p_rq_sB_{1rs}+
p_rB_{2rs}q_s) \nonumber\\
+ bosonic,  \label{H4N}
\end{eqnarray}
the quantity {\it Trace Fermion number}
$${\bf N}=\frac{1}{2}iTr\sum_{r\in F}[q_r,p_r]$$
is conserved, i.e. $\dot{{\bf N}}=0$.\\
This conserved charge corresponds to $U(1)$ gauge transformations of Fermionic degrees of freedom:
$$q_r\rightarrow \exp{\{i\alpha\}}q_r: p_r\rightarrow \exp{\{i\alpha\}}p_r,$$
for real and constant  $\alpha$ and $r\in F$.\\
Requirement of bilinear Fermionic sector of $H$ and self adjoint kinetic part forces,
$$B_{1rs}=-B_{1rs}^{\dagger}; B_{2rs}=-B_{2rs}^{\dagger}, $$
and $p_r=q_r^{\dagger}$, resulting in
$${\bf N}=-iTr\sum_{r\in F}q_r^{\dagger}q_r,$$
which resembles the number operator for Fermionic degrees of freedom.

Consider the restriction to matrix models in which the only non-commuting matrix quantities are the dynamical variables. Thus the trace Lagrangian and Hamiltonian are constructed from the dynamical  
variables using only $\it c$-number complex coefficients, excluding the more general case in which fixed matrix coefficients are used. 
Then there is a N\"{o}ether charge corresponding to a global unitary invariance possessed by the trace Hamiltonian (or equivalently Lagrangian) i.e.
\begin{equation}
 {\bf H}[\{U^{\dagger}q_rU\},\{U^{\dagger}\dot{q_r}U\}]={\bf H}[{q_r},{\dot{q_r}}], \label{UT2}
\end{equation}
for some constant unitary matrix $U$. From (\ref{UT2}) and (\ref{HEV}), it can be shown that this N\"{o}ether charge is
\begin{equation}
\tilde{C}=\sum_{r\in B}[q_r,p_r]-\sum_{r\in F}\{q_r,p_r\}, \label{adler-millard-charge}
\end{equation}
which we call the Adler-Millard charge. 
This charge 
 having the dimensions of action and being trivially zero in point particle mechanics, makes all the difference between Trace Dynamics, and ordinary 
classical mechanics of point 
particles where all position and momenta commute with each other. Note that in
(\ref{adler-millard-charge}) the individual \{anti-\}commutators take arbitrary values in time;  yet the particular combination shown in this
 equation remains conserved.

Now, if the Fermionic degrees of freedom have the adjointness property of $p_r=q_r^{\dagger}$ and the Bosonic degrees of freedom are self-adjoint
 (or anti-self adjoint) then the
conserved charge is anti-self adjoint 
and traceless
\begin{equation}
 \tilde{C}= - \tilde{C}^{\dagger}, \hspace{0.25 in} Tr \tilde{C}= 0. \label {CTP}
\end{equation}
Since $\tilde{C}$  is the N\"{o}ether charge corresponding to global unitary invariance of the matrix model, it
can be used to construct the generator of the global unitary transformation
\begin{eqnarray}
{\bf G_{\Lambda}}&=&Tr\Lambda\tilde{C} \text{,  with}\nonumber\\
\{{\bf G_{\Lambda}}, {\bf G_{\Sigma}}\}&=&{\bf G_{[\Lambda, \Sigma]}}
\end{eqnarray}
as the algebra of the generators.
Now we consider those canonical transformations which have global unitary invariant generator ${\bf G}$.
For those canonical transformations, clearly,
\begin{equation}
\{{\bf G},{\bf G_{\Lambda}}\}=0. \label{GUCT}
\end{equation}
Alternatively we can interpret (\ref{GUCT}) by saying that ${\bf G_{\Lambda}}$ is invariant under the action of ${\bf G}$.
Then, along these lines it can be shown that $\tilde{C}$ is Poincar\'{e} invariant  when the trace Lagrangian is Poincar\'{e} invariant, where Poincar\'{e} transformations are generated by trace functional
Poincar\'{e} generators. These generators are global unitary invariant when the trace Lagrangian is Poincar\'{e} invariant.
Hence, we can make use of this charge $\tilde{C}$ in Poincar\'{e} invariant theories.\\
If we consider a Lagrangian which has a fermionic Kinetic part given by
\begin{equation}
 {\bf L}_{kin}=Tr \sum_{r,s\in F}q_r^{\dagger}A_{rs}\dot{q_s} \label{New_L}
\end{equation}
where $A_{rs}$ is a constant matrix having the property (for real trace Lagrangian) $A_{rs}=A_{sr}^{\dagger},$
then $\tilde{C}$ is still conserved but now it can have a self-adjoint part as well,
\begin{equation}
\tilde{C}+\tilde{C}^{\dagger}=-\sum_{r,s\in F}[q_sq_r^{\dagger},A_{rs}].
\end{equation}
Generically, for a continuous space-time based trace Lagrangian written in terms of trace Lagrangian density
${\cal L}(\{q_l(x)\}, \{\partial_{\mu} q_l(x)\})$
which is invariant under the following symmetry transformations
\begin{eqnarray}
 q_l(x)\rightarrow q_l(x)+\alpha(x)\Delta_l(x) \nonumber\\
\partial_{\mu} q_l(x)\rightarrow\partial_{\mu} q_l(x)+\alpha(x)\partial_{\mu}\Delta_l(x)\nonumber\\
+\partial_{\mu}\alpha(x)\Delta_l(x),
\end{eqnarray}
there is a local trace current
$${\bf J}^{\mu}=Tr\sum_l\frac{\delta{\cal L}}{\delta\partial_{\mu}q_l(x)}\Delta_l(x),$$
for which
$$\partial_{\mu}{\bf J}^{\mu}=0,$$
suggesting that there is a conserved charge
$$Q=\int d^3x{\bf J}^0(x).$$
For a global unitary and Poincar\'{e} invariant theory the conserved charges are the following
\begin{eqnarray}
 \tilde{C}=\int d^3x\sum_l(\epsilon_lq_lp_l-p_lq_l),\\
{\bf P}^{\mu}=\int d^3x {\cal T}^{0\mu}; \\
{\bf M}^{\mu \nu}=\int d^3x{\cal M}^{0\mu\nu}
\end{eqnarray}
where the momentum conjugate to $q_l(x)$ is
\begin{equation}
p_l(x)= \frac{\delta {\cal L}}{\delta \partial_{0}q_l(x)};
\end{equation}
trace energy momentum density is
\begin{equation}
 {\cal T}^{\mu \nu}=
\eta^{\mu \nu}{\cal L}-Tr \sum_l\frac{\delta {\cal L}}{\delta \partial_{\mu}q_l}\partial^{\nu}q_l;
\end{equation}
$\eta^{\mu \nu}=diag(-1,1,1,1)$ is the Minkowski metric and
\begin{eqnarray}
 {\cal M}^{\lambda \nu \mu}=x^{\mu}{\cal T}^{\lambda \nu}-x^{\nu}{\cal T}^{\lambda \mu}\nonumber\\
+Tr\sum_{lm}\frac{\delta {\cal L}}{\delta \partial_{\lambda}q_l}\chi_{lm}^{\nu \mu}q_m
\end{eqnarray}
is the trace angular momentum density, with $\chi_{lm}$ being the matrix characterizing the intrinsic spin structure of the field $q_l$ such that under four space rotation
\begin{eqnarray}
 x_{\mu}\rightarrow x_{\mu}^{'}= x_{\mu}+\theta_{\mu\nu}x^{\nu}\nonumber\\
q_l(x')=q_l(x)+(1/2)\theta_{\mu\nu}\sum_m\chi _{lm}^{\mu\nu}q_m(x),\nonumber
\end{eqnarray}
for the antisymmetric infinitesimal rotation parameter $\theta_{\mu\nu}$.
${\bf P}^0$ is the conserved trace Hamiltonian. $ {\bf P}^{\mu}$ and ${\bf M}^{\mu \nu} $ together form a complete set of Poincar\'{e} generators.

The matrix operator phase space  is well behaved. We can define on this phase space a measure
\begin{eqnarray}
 d\mu = \prod_{r,m,n} d(x_r)^A_{mn},
\end{eqnarray}
where $A=0,1$ for
$$(x_r)_{mn}=(x_r)_{mn}^0+ i (x_r)_{mn}^1.$$
This measure is invariant under canonical transformations (Louiville's theorem), hence under dynamic evolution  of the system
as time evolution is a canonical transformation generated by $dt{\bf H}$.
$$d\mu[\{x_r+\delta x_r\}]=d\mu[\{x_r\}].$$
Further, bosonic and fermionic measures can be separated
$$d\mu=d\mu_Bd\mu_F,$$
which are separately invariant under the adjointness properties assumed above. 
\subsubsection{Canonical Ensemble}
With the matrix equations of motion (\ref{HEV}) for time evolution in trace dynamics we study the evolution of phase space distribution.
We assume that a large enough system rapidly forgets its initial distribution and the time averages of physical quantities are equal to the
statistical averages over an equilibrium ensemble which is determined by maximizing combinatoric probability subject to conservation laws.
If $$dP=d\mu[\{x_r\}]\rho[\{x_r\}]$$ is the probability of finding the system in operator phase space volume element $d\mu[\{x_r\}]$, then,
$$\int dP=1.$$
For a system in statistical equilibrium, phase space density distribution is constant
$$\dot{\rho}[\{x_r\}]=0.$$
Hence, $\rho$ depends only upon conserved operators, conserved trace functionals and constant
parameters. By going to a frame where the system is not translating, accelerating or rotating the charges associated with Poincar\'{e} symmetry
can be put to zero. In that case
$$\rho=\rho(\tilde{C}, {\bf H}, {\bf N}).$$
In addition the  distribution function of dynamical variables can depend on constant parameters
$$\rho=\rho(Tr\tilde{\lambda}\tilde{C};{\bf H},\tau; {\bf N}, \eta)$$
where $\tilde{\lambda},\tau$ and $\eta$ are the Lagrange multipliers conjugate to $\tilde{C},{\bf H}$ and ${\bf N}$ respectively.
One important aspect to note is that while ${\bf H}$ and ${\bf N}$ belong to ${\cal{G}}_C$, $\tilde{C}\in{\cal {G}}_M$. Hence,
$\tau,\eta \in {\cal{G}}_C$ while $\tilde{\lambda}\in{\cal {G}}_M$.  The dependence of $\rho$ on $Tr\tilde{\lambda}\tilde{C}$ is motivated from
 global unitary invariance.\\
If $\tilde{C}$ has a self-adjoint part as well, one can break it into its self-adjoint ($sa$) and anti-self adjoint ($asa$) parts,
$$\rho=\rho(Tr\tilde{\lambda}^{sa}\tilde{C}^{sa},Tr\tilde{\lambda}^{asa}\tilde{C}^{asa};{\bf H},\tau; {\bf N}, \eta).$$
Next, we define ensemble average of an operator $O$ as
$$\langle O \rangle = \int d\mu \rho O,$$
which is a map from ${\cal {G}}_M$ to ${\cal {G}}_M$.\\
This ensemble average has nice properties\\
(i) When $O$ is constructed only from phase space variables $\{x_r\}$ then this ensemble average depends only on the constant parameters
 $\bar{\lambda}=\{\tilde{\lambda}, \tau, \eta \}$
$$\langle O \rangle = F_O(\bar{\lambda}).$$
(ii) Since the integration measure is unitary invariant and $O$ is made up of only $\{x_r\}$ using $c$- number
coefficients, under a global unitary transformation
$$F_O(\bar{\lambda})=UF_O(U^{-1}\bar{\lambda}U)U^{-1}.$$
(iii) As a consequence $F_O(\bar{\lambda})$ depends explicitly on $\bar{\lambda}$ only and hence commutes with $\tilde{\lambda}$
$$[\bar{\lambda},\langle O \rangle] =0.$$
Taking a specific case when $O=\tilde{C}$
$$[\bar{\lambda},\langle \tilde{C} \rangle] =0.$$
Being \{anti-\} self-adjoint $\bar{\lambda}$  can be diagonalized by a unitary
transformation, hence so will be $\langle \tilde{C} \rangle.$
Again specializing to anti-self-adjoint $\bar{\lambda}$
and $\langle \tilde{C} \rangle$ we get\\
(i) for a real, non-negative, diagonal {\it magnitude} operator $D_{eff}$ and unitary diagonal {\it phase}-operator $i_{eff}$,
$$\langle \tilde{C} \rangle = i_{eff}D_{eff}.$$
(ii) Since, $\tilde{C}$ is traceless,
$$ Tr(i_{eff}D_{eff})=0.$$
(iii) anti-self adjointness of $\tilde{C}$ is ensured with
$$i_{eff}=-i_{eff}^{\dagger}.$$
and $$[i_{eff},D_{eff}]=0.$$
(iv) As a consequence of this decomposition
$$i_{eff}^2 =-I.$$
For an ensemble symmetric in Hilbert space basis, i.e. the ensemble does not prefer any state, the averaged operator should have identical
entries as eigenvalues. Therefore,
$$ D_{eff} = k{I}.$$
Clearly, $ D_{eff}$ is determined by a single real number with dimension of action
\begin{eqnarray}
 \langle \tilde{C} \rangle =i_{eff}\hbar. \label{CanAvg}
\end{eqnarray}
The constant $\hbar$ will eventually be identified with Planck's constant.

The traceless $\langle \tilde{C} \rangle$ implies
$$Tri_{eff}=0.$$
The above mentioned properties of $i_{eff}$ along-with property (iv) forces the dimension of Hilbert
space to be even and uniquely fixes $i_{eff}$ to
$$ i_{eff}=i[ diag(1,-1,1,-1,...,1,-1)].$$ Next, we obtain the functional form of $\rho$ through maximizing the {\it entropy} defined as
\begin{equation}
S=\int d\mu\rho \log{\rho}, \label{entropy}
\end{equation}
subject to the following constraints
\begin{eqnarray}
 \int d\mu\rho=1,\hspace{0.3 in} \int d\mu\rho\tilde{C}=\langle \tilde{C} \rangle\hspace {0.1 in} \nonumber\\
\int d\mu\rho{\bf H}=\langle {\bf H} \rangle,\hspace{0.25in} \int d\mu\rho{\bf N}=\langle {\bf N} \rangle, \hspace {0.1 in} \nonumber
\end{eqnarray}
which gives
\begin{eqnarray}
\rho_j = Z_j^{-1}\exp(-Tr\tilde{\lambda}\tilde{C}
-\nonumber \\ \tau{\bf H}- \eta{\bf N}- \sum_rTrj_{r}x_r)\nonumber \\
Z_j=\int d\mu\exp(-Tr\tilde{\lambda}\tilde{C}-\nonumber \\
\tau{\bf H}-\eta{\bf N} -\sum_rTrj_rx_r),
\end{eqnarray}
where we have introduced a book-keeping matrix source term $j_r$ for each matrix
variable $x_r$ of same type ($B/F$) and adjointness that can be varied and set to zero.
This helps us in obtaining the ensemble properties of functions made explicitly of the dynamic variables $\{x_r\}.$\\
In that case,
\begin{equation}
 \langle O \rangle_j = \int d\mu \rho_j O, \label{AvO}
\end{equation}
with $$\langle O \rangle_{AV,j}=\int d \mu\rho_j O.$$
Using this distribution and partition function we can evaluate ensemble averages
\begin{eqnarray}
 \langle \tilde{\lambda} \rangle =-\frac{\delta \log{Z_j}}{\delta \tilde{\lambda}}\\
 \langle {\bf H}\rangle =-\frac{\partial \log{Z_j}}{\partial \tau} \\
 \langle {\bf N}\rangle = -\frac{\partial \log {Z_j}}{\partial \eta}
\end{eqnarray}
and the mean square fluctuations
\begin{eqnarray}
 \Delta_{\tilde{P}\tilde{C}}^2\equiv (Tr\tilde{P}\frac{\delta}{\delta \tilde{\lambda}})^2\log{Z_j}\\
 \Delta_{\bf H}^2\equiv \frac{\partial^2 \log {Z_j}}{(\partial \tau)^2}\\
 \Delta_{\bf N}^2\equiv \frac{\partial^2 \log {Z_j}}{(\partial \eta)^2}
\end{eqnarray}
with $\tilde{P}$ being  any arbitrary fixed anti-self-adjoint operator.\\
We further study the structure of averages of dynamical variables in the canonical ensemble. This study is essential for subsequently making connection with the emergent quantum theory.  According to the previous discussion,
$$\langle \tilde{C} \rangle =i_{eff}D_{eff},$$
and $\tilde{\lambda}$ is related to $\langle \tilde{C} \rangle$ only using $c$-number coefficients. Since $ D_{eff}$ is a constant times identity,
\begin{equation}
 \tilde{\lambda}=\lambda i_{eff},
\end{equation}
for a real c-number $\lambda$. Now for a unitary matrix $U_{eff}$ that commutes with $i_{eff}$
\begin{equation}
 [U_{eff}, \tilde{\lambda}]=0 \Rightarrow U_{eff} \tilde{\lambda}  U_{eff}^{\dagger} =\tilde{\lambda}.  \label{unitary}
\end{equation}
Clearly, the presence of $Tr \tilde{\lambda}\tilde{C} $ term in the partition function breaks the global unitary invariance since under,
$$q_r \rightarrow U^{\dagger} q_r U,\hspace{0.05 in} p_r \rightarrow U^{\dagger} p_r U,\hspace{0.05 in} \tilde{C} \rightarrow U^{\dagger} \tilde{C} U,$$
\begin{equation}
 Tr \tilde{\lambda}\tilde{C} =\lambda Tr i_{eff}\tilde{C}\rightarrow \lambda Tr Ui_{eff}U^{\dagger}\tilde{C}.         \label{LC}
\end{equation}
Thus, presence of this term breaks the global unitary invariance to $\{U_{eff}\}.$ With this residual invariance in canonical ensemble we define an
integration measure as
$$d\mu=d[U_{eff}]d\hat{\mu},$$
with $d[U_{eff}]$ as the Haar measure on the residual symmetry group and $ d\hat{\mu}$ is the measure over operator phase space when  overall
global unitary transformation $U_{eff}$ is kept fixed.\\
In this case, for a polynomial $R_{eff}$ which is a function of $i_{eff}$ and dynamical variables $\{x_r\}$,
\begin{equation}
 R_{eff AV} \equiv\frac{\int d[U_{eff}]d\hat{\mu}\rho R_{eff}}{\int d[U_{eff}]d\hat{\mu}\rho}.
\end{equation}
Now if we fix the $U_{eff}$ rotation
$$q_r = U_{eff}^{\dagger} \hat{q}_r U_{eff}, \quad p_r = U_{eff}^{\dagger} \hat{p}_r U_{eff}, $$
it results in, $$ R_{eff} = U_{eff}^{\dagger} \hat{ R}_{eff} U_{eff}$$
and hence,
$$R_{eff AV} \equiv\frac{\int d[U_{eff}]U^{\dagger} R_{eff \hat{AV}}U_{eff}}{\int d[U_{eff}]}.$$
In the above equation,
$$R_{eff \hat{AV}} \equiv \frac{\int d\hat{\mu}\hat{\rho}\hat{R}_{eff}}{\int d\hat{\mu}\hat{\rho}},$$
where $\hat{R}_{eff}$ and $\hat{\rho}$ are obtained from $R_{eff}$ and $\rho$ by replacing $q, p$ therein by $\hat{q},\hat{p}$ as defined above.\\
Writing $i_{eff}=i\sigma_{3}{\bf 1}_{K}$, where ${\bf 1}_{K}$ is a unit $K\times K$ matrix, a general 
$N\times N$ matrix $M$ can be decomposed in the form (with $N=2K$)
\begin{equation}
  M = M_{eff}+M_{12} 
\end{equation}
with
$$M_{eff}=\frac{1}{2}(\sigma_0+\sigma_3)M_++\frac{1}{2}(\sigma_0-\sigma_3)M_-$$ and
$$M_{12}=\sigma_1M_1+\sigma_2M_2.$$
Here $\sigma_i, i=1,2,3$  are  $2\times 2$ Pauli matrices and $\sigma_0=1_2$. The
$M_{+,-,1,2}$ are four $K\times K$ matrices.
These new matrices satisfy
\begin{eqnarray}
 [i_{eff},M_{eff}]=0;\hspace{0.2 in}
\{i_{eff},M_{12}\}=0\hspace{0.2 in}\\
2i_{eff}M_{eff}=\{i_{eff},M_{eff}\}.\hspace{0.2 in} \label{iCom}
\end{eqnarray}
In irreducible systems unitary fixing can be done by fixing the global unitary rotation of one canonical pair of dynamical variables.
With the restricted measure we need to know the restricted canonical average $\tilde{C}_{\hat{AV}}$. Since unitary fixing does not
disturb $i_{eff}$, for any operator $O$ made up of $\{x_r\}$ using $c$-number coefficients,
$$[\tilde{\lambda},O_{\hat{AV}}]=0.$$
When $O=\tilde{C}$,
$$[\tilde{\lambda},\tilde{C}_{\hat{AV}}]=0.$$
Using properties (\ref{iCom}), the most general $\tilde{C}_{\hat{AV}}$ commuting with $\tilde{\lambda}$ is
\begin{equation}
 \tilde{C}_{\hat{AV}}=i_{eff}\hbar+\frac{1}{2}(\sigma_0+\sigma_3)\Delta_++\frac{1}{2}(\sigma_0-\sigma_3)\Delta_-.
\end{equation}
Thus, the ensemble average in the unitary fixed system is different from the canonical ensemble average (\ref{CanAvg}). However, since the unitary
fixing has been done by restricting only one canonical pair, in systems involving large number of canonical pairs, the restricted average should
be close to unrestricted average. Therefore, $\Delta_\pm$ should be small.
\subsubsection{General Ward Identity and Emergence of Quantum Theory}
Thus far we have progressed from the classical theory of Trace Dynamics, to developing a statistical thermodynamics of this theory.
 We now have in hand the tools necessary to describe the emergence of quantum theory in this thermodynamic approximation. The first step 
is the derivation of a crucial general Ward identity. This identity should be thought of as an analog of the equipartition theorem in
 statistical mechanics, and its implications in the present context are deeply connected with the existence of the Adler-Millard charge and
 its canonical average (\ref{CanAvg}). Averages now are defined with respect to the restricted measure.
Under a constant shift of any matrix variable apart from the restricted pair
$$x_r\rightarrow x_r+\delta x_r, r\neq R,R+1,$$
\begin{equation}
 \int d\hat{\mu} \delta_{x_r}(\rho_j O)=0.  \label{shift}
\end{equation}
Using this and the definition of average given above for an $O= \{ \tilde{C},i_{eff}\}W$
we get
\begin{eqnarray}
 \int d\hat{\mu}\delta_{x_s} [\exp{(-Tr\tilde{\lambda}\tilde{C}-\tau{\bf H}
   -\eta{\bf N}}\hspace{0.2 in}\nonumber\\
-\sum_rTrj_rx_r)Tr\{\tilde{C},i_{eff}\}W ]=0.\hspace{0.2 in}
\end{eqnarray}
Using the chain rule we have
\begin{eqnarray}
\int d\hat{\mu}\exp{(-Tr\tilde{\lambda}\tilde{C}-\tau{\bf H}
-\eta{\bf N}} \hspace{0.2 in}\nonumber\\
-\sum_rTrj_rx_r)[(-Tr\tilde{\lambda}\delta_{x_s}\tilde{C}
-\tau \delta_{x_s}{\bf H} \hspace{0.2 in} \nonumber\\
-\eta \delta_{x_s} {\bf N}-Tr j_s\delta x_s)\{\tilde{C},i_{eff}\}W \hspace{0.2 in}\nonumber\\
+\delta_{x_s}\{\tilde{C},i_{eff}\}W] =0. \hspace{0.2 in} \label{WD1}
\end{eqnarray}
Evaluating term by term in (\ref{WD1}) we have\footnote{$\omega =diag(\Omega_B,...,\Omega_B,\Omega_F,..., \Omega_F),$ with $$\Omega_B=\left(
\begin{array}{ll}
0 & 1\\
-1 & 0
\end{array}
\right),
\Omega_F=-\left(
\begin{array}{ll}
0 & 1\\
1 & 0
\end{array}
\right) $$ for bosonic and fermionic d.o.f.},
\begin{eqnarray}
Tr\tilde{\lambda}\delta_{x_s}\tilde{C}=Tr\left[\tilde{\lambda},\sum_r\omega_{rs}x_r\right]\delta x_s \nonumber\\
\delta_{x_s}{\bf H}=\sum_r \omega_{rs}Tr\dot{x_r}\delta x_s, \nonumber\\
\delta_{x_s}{\bf N}=i\sum_r \tilde{\omega}_{rs}Tr x_r\delta x_s, \nonumber\\
\delta_{x_s}Tr\{\tilde{C},i_{eff}\}W=Tr(\{i_{eff}, W\}\delta_{x_s}\tilde{C} + \nonumber\\
                 \{\tilde{C}, i_{eff}\}\delta_{x_s}W), \nonumber
\end{eqnarray}
with,
further, for a polynomial $W$
$$\delta_{x_s}W=\sum_lW_s^{Ll}\delta x_s W_s^{Rl},$$
where $l$ labels each monomial in the polynomial. $W_s^{Ll/Rl}$ is the left (right) fraction of a monomial.
Collecting above terms and plugging back in (\ref{WD1})
with some manipulations leads to the generalized {\bf Ward identity}
\begin{eqnarray}
\langle \Lambda_{u_{eff}} \rangle_j=\langle(-\tau\dot{x}_{ueff}+i\eta\xi_ux_{ueff}-
\nonumber\\
\Sigma_s\omega_{us}j_{seff})Tr\tilde{C}i_{eff}W_{eff} +[i_{eff}W_{eff},x_{ueff}]\nonumber\\
+\sum_{s,l}\omega_{us}\epsilon_l\left(W_s^{Rl}\frac{1}{2}
\{\tilde{C},i_{eff}\}W_s^{Ll}\right)_{eff}\rangle_j=0\hspace{0.1in}\nonumber 
\end{eqnarray}
with 
$\xi_{u}=1(-1)$ for fermionic $q(p)$, zero for bosonic $x_u$ and
\begin{equation}
 \sum_s\omega_{us}\omega_{rs}=\delta_{ur}.
\end{equation}
The Ward identity can be written more compactly as
\begin{equation}
\langle {\cal{D}}x_{ueff}\rangle_j-\sum_s\omega_{us}j_{seff}\langle Tr\tilde{C}i_{eff}W_{eff}\rangle_j=0 \label{WD2}
\end{equation}
where,
\begin{eqnarray}
{\cal{D}}x_{ueff}=(-\tau\dot{x}_{ueff} + i\eta\xi_ux_{ueff})\hspace{0.5 in}\nonumber\\
\times Tr\tilde{C}i_{eff}W_{eff}
+[i_{eff}W_{eff},x_{ueff}]+\hspace{0.2 in}\nonumber\\
\sum_{s,l}\omega_{us}\epsilon_l\left(W_s^{Rl}
\frac{1}{2}
\{\tilde{C},i_{eff}\}W_s^{Ll}\right)_{eff} \hspace{0.3 in}
 \label{WD3}
\end{eqnarray}
From (\ref{WD2}) we see that for a polynomial $S$ made up of $x_{reff}$ and $c$-number coefficients,
$$\langle S_L(x_{teff})({\cal D}S(x_{reff}))S_R(x_{teff})\rangle_{0}=0,$$
for left and right decompositions of the polynomial $S$.\\
We now make the following realistic assumptions \\
(i) Support properties of $\dot{x}_{ueff}$ and $\tilde{C}_{eff}$ are such that
$$-\tau\dot{x}_{ueff}Tr\tilde{C}i_{eff}W_{eff}$$
in (\ref{WD3}) can be neglected.\\
(ii) Chemical potential $\eta$ is very small, such that the term
$$i\eta\xi_ux_{ueff}Tr\tilde{C}i_{eff}W_{eff}$$
in (\ref{WD3}) can be neglected. In fact, for bosonic degrees of freedom this term vanishes and it is taken to be small for fermionic degrees of freedom.\\
(iii) When the number of degrees of freedom is large, $\tilde{C}$ can be replaced by its zero-source ensemble average
$$\langle \tilde{C}_{eff} \rangle_{\hat{AV}}=i_{eff}\hbar.$$
With these assumptions the RHS of the identity (\ref{WD3}) simplifies to
$$i_{eff}[W_{eff},x_{ueff}]-\hbar\sum_s\omega_{us}\left(\frac{\delta{\bf W}}{\delta x_s}\right)_{eff}$$
and (\ref{WD2}) implies $\langle {\cal D} x_{ueff}\rangle_{0}=0$.
 If we consider $ W=H$ in the Ward identity, we obtain
$${\cal{D}}x_{ueff}=i_{eff}[H_{eff},x_{ueff}]-\hbar\dot{x}_{eff}$$
{\it which gives the effective Heisenberg equations of motion for the dynamics} when sandwiched between $S_L(x_{teff})$ and $S_R(x_{teff})$ and averaged over zero-source ensemble.\\
For an arbitrary polynomial function $P_{eff}$ made up of $x_{reff}$
\begin{eqnarray}
& & \langle S_L(x_{teff}) \dot{P}_{eff} S_R(x_{teff})\rangle_{0}=\nonumber\\
& &\langle S_L(x_{teff}) i_{eff}\hbar^{-1}[H_{eff},P_{eff}]  S_R(x_{teff})\rangle_{0},\nonumber
\end{eqnarray}
suggesting that within our assumptions, $H_{eff}$ is a constant of motion.
Next, for $W=\tilde{\sigma}_vx_v$ for some $c$-number parameter $\tilde{\sigma}_v$ 
we get 
\begin{equation}
i_{eff}{\cal{D}}x_{ueff}=[x_{ueff},\tilde{\sigma_v}x_{veff}]-i_{eff}\hbar\omega_{uv}\tilde{\sigma}_v. \label{Com1}
\end{equation}
Thus, when multiplied with  $S_L(x_{teff})$ and $S_R(x_{teff})$ and averaged over zero source ensemble EFFECTIVE CANONICAL COMMUTATORS EMERGE
\begin{eqnarray}
& &\langle\langle q_{ueff},q_{veff}\rangle\rangle=\langle\langle p_{ueff},p_{veff}\rangle\rangle=0, \nonumber\\
& & \langle\langle q_{ueff},p_{veff}\rangle\rangle=i_{eff}\hbar\delta_{uv}, \label{Com2}
\end{eqnarray}
with $\langle\langle \hspace{ 0.1 in} \rangle\rangle$ being anti-commutator (commutator) for $u,v$ being fermionic (bosonic).
It is important to emphasize that these commutation relations emerge only upon statistical averaging, as a consequence of there
 being the conserved Adler-Millard charge. At the level of the underlying theory of Trace Dynamics, the commutators/anti-commutators
 amongst the above operators are arbitrary.

Next, let $W=G$ be a self-adjoint operator such that ${\bf G}$ generates canonical transformation
\begin{equation}
 \hbar^{-1}{\cal{D}}x_{ueff}=i_{eff}\hbar^{-1}[G_{eff},x_{eff}] - \delta x_{eff} \label{CanUnit}
\end{equation}
Thus, on sandwiching between $S_L(x_{teff})$ and $S_R(x_{teff})$ that do not contain $x_u$, and averaging, we see that
infinitesimal canonical transformations at the ensemble level and within the above-mentioned  assumptions are generated
by unitary transformations
$$ U_{can \hspace{0.1 cm} eff} = \exp{(i_{eff}\hbar^{-1}G_{eff})}. $$
Therefore, we have at hand the essential features of quantum field theory. The \{anti-\}commutator structure, time evolution in Heisenberg picture and
unitary generation of canonical transformations emerge when we carry out the statistical thermodynamics of the matrix variables.\\
Now we make the following correspondences between operator polynomials in trace dynamics and operator polynomials in quantum field theory
$$S(\{x_{reff}\})\Leftrightarrow S(\{X_{reff}\}),$$
with $X_{reff}$ being quantized operators in quantum field theory. Here, $i_{eff}$ acts as two blocks of $i$ and $-i$. With the assumptions\\
(a) in the continuum limit the trace Lagrangian is Poincar\'{e} invariant,\\
(b) the Hamiltonian $H_{eff}$ is bounded from below by the magnitude of the corresponding effective three-momentum  operator $\vec{P}_{eff}$, and there is
a unique eigenvector $\psi_0$ with lowest eigenvalue of  $H_{eff}$ and zero eigenvalue of $\vec{P}_{eff}$,\\
we also have a correspondence between trace dynamics canonical averages and Wightman functions in emergent quantum field theory,
$$\psi^{\dagger}_0\langle S(\{x_{reff}\}) \rangle_{\hat{AV}}\psi_0=\langle vac | S(\{X_{reff}\})|vac \rangle.$$

Using (\ref{Com1}), (\ref{Com2}) and the  proposed correspondence we have at the quantum level
\begin{equation}
 [X_{ueff}, \tilde{\sigma}_vX_{veff}]=i_{eff}\hbar\omega_{uv}\tilde{\sigma}_v,
\end{equation}
which gives the appropriate commutators at the quantum level, both for bosonic and fermionic degrees.

Time evolution is given by
\begin{equation}
 \dot{X}_{ueff}=i_{eff}\hbar^{-1}[H_{eff},X_{ueff}], \label{HEOM1}
\end{equation}
or equivalently for a polynomial $S_{eff}$ of $\{X_{eff}\}$
\begin{equation}
 \dot{S}_{ueff}=i_{eff}\hbar^{-1}[H_{eff},S_{ueff}].\label{HEOM2}
\end{equation}
For the fermionic anti-commutator to be an operator equation, the following adjointness  assignment is required,
\begin{eqnarray}
 \psi_{reff}=q_{reff}\leftrightarrow Q_{reff}=\Psi_{reff}, \nonumber\\
 \psi^{\dagger}_{reff}=p_{reff}\leftrightarrow P_{reff}=i_{eff}\Psi^{\dagger}_{reff},
\end{eqnarray}
 such that
\begin{equation}
 \{\Psi_{ueff},\Psi^{\dagger}_{veff}\}=\hbar\delta_{uv}.
\end{equation}
For the bosonic sector we obtain creation and annihilation operators $A_{reff},A^{\dagger}_{reff}$
such that,
\begin{eqnarray}
 Q_{ueff}=\frac{1}{\sqrt{2}}(A_{reff}+A^{\dagger}_{reff})\nonumber\\
 P_{ueff}=\frac{1}{\sqrt{2}i_{eff}}(A_{reff}-A^{\dagger}_{reff})
\end{eqnarray}
and,
 $$[A_{ueff},A_{veff}]=[A^{\dagger}_{ueff},A^{\dagger}_{veff}]=0,$$
\begin{equation}
[A_{ueff},A^{\dagger}_{veff}]=\hbar\delta_{uv}.
\end{equation}
Thus, we have the correct commutation/anti-commutation rules for bosonic/fermionic degrees of freedom in both the $i_{eff}=\pm i$ sectors.

Once we have the Heisenberg equations of motion (\ref{HEOM1}) and (\ref{HEOM2}), we can make the transition to the Schr\"{o}dinger picture as usual,
without making any reference to the background trace dynamics theory. 
When the effective Hamiltonian has no time dependence, we define
\begin{equation}
 U_{eff}(t)=\exp{(-i_{eff}\hbar^{-1}tH_{eff})},
\end{equation}
so that,
\begin{eqnarray}
 \frac{d}{dt}U_{eff}(t)=-i_{eff}\hbar^{-1}H_{eff} U_{eff}(t),\hspace{0.2 in}\nonumber\\
 \frac{d}{dt}U^{\dagger}_{eff}(t)=i_{eff}\hbar^{-1} U^{\dagger}_{eff}(t)H_{eff}.\hspace{0.2 in}
\end{eqnarray}
From the Heisenberg picture time-independent state vector $\psi$ and time dependent operator $S_{eff}(t)$, in the Schr\"{o}dinger picture we perform the
construction
\begin{eqnarray}
 \psi_{Schr}(t)&=&U_{eff}(t)\psi \nonumber\\
 S_{eff \hspace{0.1 cm} Schr}&=&U_{eff}(t)S_{eff}(t)U^{\dagger}_{eff}(t),
\end{eqnarray}
which gives,
\begin{eqnarray}
i_{eff}\hbar \frac{d}{dt} \psi_{Schr}(t)&=&H_{eff}  \psi_{Schr}(t)\nonumber\\
\frac{d}{dt} S_{eff \hspace{0.1 cm} Schr}&=&0.
\end{eqnarray}
For obtaining the Schr\"{o}dinger equation, we make contact with space-time by taking the label $r$ as $\vec{x}$. In that case, the fermionic
anti-commutator becomes
\begin{equation}
 \{\Psi_{eff}(\vec{x}),\Psi^{\dagger}_{eff}(\vec{y})\}=\hbar\delta^3(\vec{x}-\vec{y}).
\end{equation}
We have assumed that $H_{eff}$ is bounded from below having the vacuum state $|{\cal V}\rangle$ as the lowest eigenvalue state and that $\Psi_{eff}$ should annihilate it,
$$\Psi_{eff}|{\cal V}\rangle=0.$$
Therefore,
\begin{eqnarray}
 \langle {\cal V}|\Psi_{eff}(\vec{x})\Psi^{\dagger}_{eff}(\vec{y})|{\cal V}\rangle=\hbar\delta^3(\vec{x}-\vec{y})\hspace{0.1 in}  \label{commutation}
\end{eqnarray}
Similarly, for the bosonic operator $A_{eff}$
\begin{equation}
 \langle {\cal V}|A_{eff}(\vec{x})A^{\dagger}_{eff}(\vec{y})|{\cal V}\rangle=\hbar\delta^3(\vec{x}-\vec{y}).\label{commutation1}
\end{equation}
Thus, an analysis of (\ref{commutation}) will analogously apply for (\ref{commutation1}) as well. 
Defining,
$$\hbar^{\frac{1}{2}}\Psi_{n}(\vec{x})=\langle {\cal V}|\Psi_{eff}(\vec{x})|n\rangle , $$
we obtain
\begin{eqnarray}
 \sum_n\Psi_{n}(\vec{x})\Psi^{*}_{n}(\vec{y})&=&\delta^3(\vec{x}-\vec{y}),\nonumber\\
 \text{and}\int d^3y \Psi^{*}_{n}(\vec{y})\Psi_{m}(\vec{y})&=&\delta_{nm},
\end{eqnarray}
from (\ref{commutation}).
Then, using the Heisenberg equation of motion,
\begin{eqnarray}
 \hbar^{\frac{1}{2}}\frac{d}{dt}\Psi_{n}(\vec{x}) \hspace{1.5 in}\nonumber\\
=\langle {\cal V}|i_{eff}\hbar^{-1}[H_{eff},\Psi_{eff}(\vec{x})]|n\rangle.    \label{Schr}
\end{eqnarray}
Again, defining
$$H_{eff}=\int d^3y\Psi^{\dagger}_{eff}(\vec{y}){\cal H}_{eff}(\vec{y})\Psi_{eff}(\vec{y}),$$
we get
$$[H_{eff},\Psi_{eff}(\vec{x})]=-{\cal H}_{eff}(\vec{x})\Psi_{eff}(\vec{x}),$$
thus modifying (\ref{Schr}) into a Schr\"{o}dinger equation
\begin{equation}
\boxed{i_{eff}\hbar \frac{d}{dt}\Psi_{n}(\vec{x})={\cal H}_{eff}(\vec{x})\Psi_{n}(\vec{x}). \label{Schrodinger}}
\end{equation}
%
However, in deriving (\ref{Schrodinger}) we have made certain approximations valid at equilibrium.
More explicitly, we replaced $\tilde{C}$ by its canonical average. If we also consider the fluctuations about the average quantities we have
possibilities of obtaining a stochastic equation of evolution by adding stochastic nonlinear terms to the Schr\"{o}dinger equation. Herein perhaps lies the greatest virtue of Trace Dynamics. By treating quantum theory as a thermodynamic approximation to a statistical mechanics, the theory opens the door for the
ever-present statistical fluctuations to play the desired role of the non-linear stochasticity which impacts on the measurement problem.\\
If we consider fluctuations in  $\tilde{C}$ to be described by
\begin{eqnarray}
 \Delta{\tilde{C}}\simeq \tilde{C} -i_{eff}\hbar \nonumber\\
\frac{1}{2} \{\tilde{C},i_{eff}\}=-\hbar+\frac{1}{2} \{\Delta\tilde{C},i_{eff}\} \nonumber\\
\frac{1}{2} \{\Delta\tilde{C},i_{eff}\} =-\hbar({\cal K}+{\cal N}),
\end{eqnarray}
with fluctuating $c$-number ${\cal K}$ and fluctuating matrix ${\cal N}$, we obtain the modified {\it linear} Schr\"{o}dinger equation (restricting to the $i_{eff}=i$ sector)
\begin{eqnarray}
|\dot{\Phi} \rangle &=&[i \hbar^{-1}\{-1+ ({\cal K}_0(t)+i{\cal K}_1(t))\}H_{eff}  \nonumber\\
&+&\frac{1}{2}i({\cal M}_0(t)+i{\cal M}_1(t))]|\Phi\rangle, 
\label{ModEq}
\end{eqnarray}
where $0,1$ label the real and imaginary parts of ${\cal K}$ and
$${\cal M}(t)=\sum_{r,l} m_r{\cal N}(t)_{r,l}.$$
In the above equations $m_r$ is the rest mass of the $r$-th species such that the Hamiltonian is
\begin{equation}
 H=\sum_r\sum_l \frac{1}{2} i m_r [\psi^{\dagger}_{rl},\psi_{rl}] + const
\end{equation}
with $l$ labeling a general complete basis set and we have used the correspondence proposed between trace-dynamics and quantum mechanics,
$$|\Phi\rangle =\prod_{r,l}\Psi^{\dagger}_{rleff}|{\cal V}\rangle.$$
As we will see next, with the added assumption of norm conservation through equation (\ref{ModEq}) Trace Dynamics connects with a CSL type non-linear stochastic equation. 
The weak link however, in the chain, at the present stage of our understanding, is the {\it assumption} that norm is conserved. Rather than being an assumption, this should 
follow from the underlying theory, and hopefully with improved understanding this will become possible in the future. On the other hand, the presence of anti-Hermitean modifications in the
Schr\"{o}dinger equation is inevitable from the Trace Dynamics viewpoint, since it is possible for $\tilde{C}$ to have a self-adjoint part as well.
\subsubsection{Stochastic modification of the Schr\"{o}dinger equation}
We further motivate that the fluctuations can be described by linear superposition of white noise terms owing to the hierarchy between the length scale
associated with the $\tilde{C}$ fluctuation and the length scale characterizing the emergent quantum degrees of freedom, much like in the case of
Brownian motion fluctuations. A Brownian motion is described by a stochastic process  $dW^n_t$ satisfying the following It\^{o} table 
\begin{eqnarray}
 (dW^n_t)^2={\cal \gamma}_n dt, \\
 dW^n_tdW^m_t=0, m\neq n, \\
 dW^n_t dt =dt^2 =0.
\end{eqnarray}
For our case we make the following identifications:\\
(i) For $c$- number fluctuations ${\cal K}_{0,1}$
\begin{eqnarray}
 i\hbar^{-1}{\cal K}_0dt=i\beta_IdW^I_t, \\
-\hbar^{-1}{\cal K}_1=\beta_RdW^R_t,
\end{eqnarray}
 with the following It\^{o} table
\begin{equation}
 (dW^R_t)^2=(dW^I_t)^2=dt,  dW^R_tdW^I_t=0. \nonumber
\end{equation}
(ii) For the fluctuating  matrix ${\cal M}_{0,1}$ having spatially correlated noise structure
\begin{eqnarray}
\frac{1}{2}i{\cal M}_0dt=i\int d^3x dW^I_t(\vec{x}){\cal M}^I_t(\vec{x}),\hspace{0.1 in} \\
-\frac{1}{2}{\cal M}_1=\int d^3x dW^R_t(\vec{x}){\cal M}^R_t(\vec{x}),\hspace{0.1 in}
\end{eqnarray}
with,
\begin{eqnarray}
 dW^I_t(\vec{x})dW^I_t(\vec{y})={\cal \gamma}dt \delta^3(\vec{x}-\vec{y})\\
 dW^R_t(\vec{x})dW^R_t(\vec{y})={\cal \gamma}dt \delta^3(\vec{x}-\vec{y})\\
 dW^I_t(\vec{x})dW^R_t(\vec{y})=0\\
 dW^I_tdW^I_t(\vec{x})= dW^I_tdW^R_t(\vec{x})=0\\
 dW^R_tdW^I_t(\vec{x})= dW^R_t dW^R_t(\vec{x})=0.
\end{eqnarray}
These identifications turn Eqn. (\ref{ModEq}) into a stochastic differential equation
\begin{eqnarray}
 |d\Phi \rangle =[-i \hbar^{-1} H_{eff}dt + i\beta_IdW^I_tH_{eff} \nonumber\\
+\beta_RdW^R_t H_{eff}+
i\int d^3x dW^I_t(\vec{x}){\cal M}^I_t(\vec{x})
 \nonumber\\
+\int d^3x dW^R_t(\vec{x}){\cal M}^R_t(\vec{x})]|\Phi\rangle. \nonumber 
\end{eqnarray}
The above evolution is not norm preserving.
The idea is to define a {\bf physical} state
$|\Psi\rangle\left(=\frac{|\Phi\rangle}{\langle \Phi|\Phi\rangle}\right)$
 with a conserved norm (as assumption)
which, along with the criterion of absence of superluminal signaling, after some calculation at length, gives
\begin{eqnarray}
 d|\Psi\rangle = \left[-i\hbar^{-1}H_{eff}dt+i\beta_I H_{eff}dW^I_t \hspace{0.5 in}\right.\nonumber\\
\left.-\frac{1}{2}[\beta_I^2H_{eff}^2+\beta_R^2(H_{eff}-\langle H_{eff}\rangle)^2]dt\hspace{0.5 in} \right.\nonumber\\
\left. +\beta_R(H_{eff}-\langle H_{eff}\rangle)^2 dW^R_t \hspace{0.5 in}\right.\nonumber\\
\left.+i\int d^3x {\cal M}^I(\vec{x})dW^I_t(\vec{x})-\frac{{\cal \gamma}}{2}dt\times\hspace{0.5 in}\right.\nonumber\\
\left. \int d^3x[{\cal M}^I(\vec{x})^2+({\cal M}^R(\vec{x})-
\langle{\cal M}^R(\vec{x})\rangle)^2] \hspace{0.5 in}\right.\nonumber\\
\left. +\int d^3x ({\cal M}^R(\vec{x})-\langle{\cal M}^R(\vec{x})\rangle)^2dW^R_t(\vec{x})\right]|\Psi\rangle.\hspace{0.25 in} \nonumber
\hspace{0.25 in} 
\label{CSL}
\end{eqnarray}
This equation  is a stochastic non-linear Schr\"{o}dinger equation which has the martingale structure\footnote{For stochastic $|\Psi\rangle,$ $$d\hat{\rho} = 
(d|\Psi\rangle)\langle \Psi| + |\Psi\rangle(d\langle \Psi|) + (d|\Psi\rangle)(d\langle \Psi|)$$}  of spontaneous collapse models,
 and is capable of explaining state vector reduction. In this sense, Trace Dynamics is an underlying theory for spontaneous collapse models.
 Of course, at the present stage of understanding, it cannot pick out one collapse model out of the many discussed, nor provide a theoretical 
origin for the values of the CSL parameters $\lambda$ and $r_C$. Nonetheless, one cannot escape the profound and natural hypothesis that on one hand 
thermodynamic equilibrium corresponds to quantum theory, and on the other hand fluctuations around equilibrium correspond to stochastic modifications 
of quantum theory. Why the effect of stochasticity must be larger for larger systems remains to be understood.  Nor is it understood why norm should be 
preserved during evolution in Trace Dynamics: one should not have to put this in as an assumption into the theory, but rather have it come out of the underlying theory as a consequence.

For explicit demonstration  of the collapse of the wave-function induced by stochasticity,
we study a simplified version 
\begin{eqnarray}
 d|\Psi\rangle &=& (-i\hbar^{-1}H_{eff}-\frac{1}{2}[\beta_R^2(A-\langle A \rangle)^2 \nonumber\\
 &+&\beta_I^2A^2])|\Psi\rangle dt + \beta_R(A-\langle A\rangle)|\Psi\rangle dW^R_t \nonumber\\
 &+& i\beta_I A|\Psi\rangle dW^I_t,  \label{CSL1}
 \end{eqnarray}
with
\begin{eqnarray}
 d\hat{\rho} =  i\hbar^{-1}[\hat{\rho}, H_{eff}]dt &-&\frac{1}{2}|\beta|^2[A, [A, \hat{\rho}]dt\nonumber\\
+\beta_R[\hat{\rho}, [\hat{\rho},A]]dW^R_t&+&i\beta_I[A, \hat{\rho}]dW^I_t. \hspace{0.1 in}
\end{eqnarray}
Defining $E[$ ] as expectation with respect to the stochastic process, ($E[dW^R_t]=0=E[dW^I_t]$) and variance of $A$,
$$V=\langle (A-\langle A\rangle)^2\rangle=Tr\hat{\rho}A^2-(Tr\hat{\rho}A)^2,$$
and using the It\^{o} product rules gives
\begin{equation}
 dE[V]=E[dV]=-4\beta_R^2E[V^2]dt.
\end{equation}
Therefore,
\begin{equation}
 E[V(t)]=E[V(0)]-4\beta_R^2\int_0^tdsE[V(s)^2]. \label{Rate}
\end{equation}
Using the inequality
$$0\leq E[(V-E[V])^2]=E[V^2]-E[V]^2,$$
this becomes
\begin{equation}
E[V(t)] \leq E[V(0)] -4\beta_R^2 \int_0^t ds E[V(s)]^2. \nonumber
\end{equation}
Non-negativity of variance suggests that $E[V(\infty)]=0$ and again as $V(t)$ is not supposed to be negative anywhere this will enforce
$$V[\infty]\rightarrow 0. $$
As the variance in expectation of $A$ goes to zero asymptotically, the system in this way results in one of the eigenstates\footnote{An open question is to make sure that the preferred basis for the collapse, as chosen by Trace Dynamics, corresponds to some sort of position basis, in order to guarantee that macroscopic objects are always localized in space.} of $A$. The demonstration of collapse using a system-apparatus interaction in the QMUPL model in Sec. II is a specific explicit application of this general analysis.

Also, we obtain from
(\ref{Rate})
$$E[V(t)]\leq \frac{V[0]}{1+4\beta_R^2V[0]t},$$
and hence a time scale of reduction as $\Gamma=4\beta_R^2V[0]$.

We can also see that in such a reduction scheme the Born probability
rule follows for the outcomes. To see that, let us take $\Pi_a$ as the projector into the $a$-th eigenstate of operator $A$,
$$\Pi_a=|a\rangle\langle a|.$$
Now for any operator $G$ commuting with $H_{eff}$ and $A$
\begin{eqnarray}
 E[d\langle G\rangle]= Tr(-i\hbar^{-1}[G,H_{eff}]E[\hat{\rho}]\nonumber\\
-\frac{1}{2}|\beta|^2[G,A][A,E[\hat{\rho}]])dt=0. \label{G}
\end{eqnarray}
If the initial state of the system is
$$|\Psi_i\rangle = \sum_ap_{ia}|a\rangle,$$
at $t=0$ when the stochastic evolution has not started,
\begin{equation}
 E[\langle \Pi_a \rangle]_{t=0}=\langle\Pi_a\rangle_{t=0}=|p_{ia}|^2.
\end{equation}
As we have argued, when the evolution is driven by $A$, the system results in a particular eigenstate $|f\rangle$ with some probability $P_f$.
Then, for the $a$-th eigenstate,
\begin{equation}
 E[\langle\Pi_{a}\rangle]_{t=\infty}=\sum_f \langle f|\Pi_{a}|f\rangle P_f=P_a.
\end{equation}
Now, since $A$ was taken to be commuting with $H_{eff}$, we can choose their simultaneous eigenstates, which we can call $|a\rangle$. Therefore,
operators $\Pi_a$ constructed from these eigenstates will commute with $H_{eff}$ and $A$, resulting in the time-independence of
$E[\langle\Pi_a\rangle]$ as evident from (\ref{G}).
Therefore,
\begin{equation}
E[\langle\Pi_a\rangle]_{t=0}= \langle\Pi_a\rangle_{t=0}=E[\langle\Pi_a\rangle]_{t=\infty}
\end{equation}
giving $P_a=|p_{ia}|^2$.
Thus, we have obtained the Born probability rule.

We have seen that when treated as a fluctuation around the thermodynamic limit of trace dynamics theory, the emergent non-linear equation captures the essential features of
CSL, and in a sense, can possibly be a theoretical motivation for the phenomenological CSL equation of evolution. Of course at the present stage of understanding, trace dynamics makes no definite prediction for the actual numerical values of the CSL parameters, and this remains a challenge for the theory.

\subsection{Gravity induced collapse}

The general theory of relativity dictates that gravity is the curvature of space-time. This curvature is produced by classical material bodies. However, even the motion of classical bodies possesses intrinsic quantum fluctuations, and these fluctuations imprint a small uncertainty on spacetime structure. When one considers the motion of quantum mechanical objects on such a fluctuating spacetime, the coherence of the quantum state can be lost, providing a possible mechanism for wave-function collapse in the macroscopic domain, while leaving microphysics untouched by gravity. Counterintuitive though it may seem, gravity possibly plays a profound role in bringing about wave-vector reduction, as the studies described below indicate.

\subsubsection{The model of Karolyhazy [K-model] }

The proposal of Karolyhazy: ~\cite{Karolyhazi:66, Karolyhazi:86} deals with a smearing of space-time which results from the  fundamental uncertainty in quantum theory being forced upon space-time structure.
It starts with a viewpoint that nature `somehow' tries to reconcile classical general relativity with quantum mechanics as much as possible. Space-time has in general, a fairly definite
metric structure mainly determined by classical massive objects with fairly definite positions. However, the metric should not be completely sharp, and
must have an in-built haziness to avoid contradiction with the fundamental quantum aspect of massive objects (spread in position and momenta). Even a
macroscopic massive body will have to satisfy
$$\delta x \; \delta v \geq \frac{\hbar}{2m}$$
where $m$ is the mass of the body. The resulting haziness in the metric produced by the body leads to a stochastic correction in the evolution of state-vectors in quantum theory.

The basic idea
of the approach is that when a wave packet of the center of mass of a body, sufficiently narrow in the beginning, spreads out in the Schr\"{o}dinger evolution,
into a space domain larger than a critical value (characteristic to the system), the coherence between distinct parts of the wave function gets
destroyed, owing to space-time haziness. This is interpreted as a signal for stochastic reduction of the extended wave function to one of its smaller,
coherent parts.

\smallskip

\noindent{{ \it Quantum imprecision of space-time} }

Let us consider a world-line segment $s=cT$, in a flat space-time. We wish to estimate the precision with which we can realize this segment. Thus, the
segment of $t$-axis is to be realized by the narrowest possible tube formed by a standing wave packet. Let, at the start (i.e. at the bottom of the
world-line segment) the width of the wave packet be $\triangle{x_0}$. For mass $M$ of the wave packet the velocity spread is
$$\triangle V= \frac{\hbar}{2M\triangle x_0}.$$
The corresponding spread at the end (i.e. at the top of the line segment) will be
\begin{equation}
\triangle x=\triangle V T=\frac{\hbar}{2M\triangle x_0c}cT. \label{unC1}
\end{equation}
The uncertainties $\triangle x$ and $\triangle x_0$ are the uncertainties in the top and bottom of the segments as well as in the length of the
segments. Minimum amount of uncertainty in the length of segment will be introduced if we choose
\begin{equation}
 \triangle x=\triangle x_0. \label{unC2}
\end{equation}
Clearly the uncertainty in the length of segment decreases with increasing $M$ and point-like description becomes progressively more valid. Now, the
gravitational radius of the mass $M$ is bounded by the fact that it should not be greater than the spread $\triangle x$,
\begin{equation}
\triangle x\approx GM/c^2. \label{unC3}
\end{equation}

From Eqns. (\ref{unC1}), (\ref{unC2}) and (\ref{unC3}), the uncertainty in the length of the segment is given by
\begin{eqnarray}
  (\triangle s)^2= (\triangle x)^2=\frac{\hbar}{2Mc}cT  \nonumber\\
  =\frac{\hbar}{2Mc}s=\frac{G\hbar}{2\triangle sc^3}s. \nonumber\\
(\triangle s)^2 =\left(\frac{G\hbar}{2c^3}\right)^{2/3}s^{2/3}. \label{smear}
\end{eqnarray}
This relation, giving the minimum amount of uncertainty in space-time structure is often known as the Karolyhazy uncertainty relation.
Therefore, we should be careful in using classical space-time considerations once the length of the segment starts approaching its uncertainty value; thus
providing a critical length scale for the system.

Next we consider a physical space-time domain of nearly Minkowski metrics with a corresponding
smear structure as argued in ~\cite{Karolyhazi:66, Karolyhazi:86}.
We introduce a family $\{g_{\mu\nu}^{\beta}\}$ of matter-free metrics very close to the Minkowski metric,
where different $\beta$ mark different members (hence different metrics) of the family. The proper length $s=cT$ between two world points $x_1$ and $x_2$  will be
defined as the mean
value of the lengths $s_{\beta}$ corresponding to different $g_{\mu\nu}^{\beta},$
\begin{equation}
 s=\bar{s}_{\beta}, \label{AvgS}
\end{equation}
with the bar describing average over $\beta$. The uncertainty in the line segment is defined as
\begin{equation}
 \triangle s=[\overline{(s-s_{\beta})^2}]^{1/2}, \label{AvgUnc}
\end{equation}
In the family of metrics $\beta=0$ gives the Minkowski metric. In the present analysis attention will be confined to the case in which we
do not have macroscopic bodies moving relatively to each other with a velocity
near that of light. The co-ordinate system will therefore be assumed to be
one relative to which all macroscopic bodies move slowly. This will enable
us to confine our use of the set $\{s_{\beta}\}$ to non-relativistic many-particle wave
equations in spite of the fact that by invoking curved manifolds we are employing
the language of general relativity.

Since we are considering only slowly moving particles, $v\ll c$, only $(g_{00})_{\beta}$ part
 of the metric will be required for the analysis.
 The general form of the metric in the family is of the form
$$(g_{00})_{\beta}(x)=-1+\gamma_{\beta}(x); \quad (\beta \neq0).$$

Since the space-time is matter-free apart from the test particle, we have
$$\Box \gamma_{\beta}=0.$$
Now, the idea is to  fix the set $\gamma_{\beta}$ in such a way that  the length of the world-line
\begin{equation}
 s_{\beta}=\int dt\left[g_{\mu\nu}^{\beta}\frac{dx^{\mu}}{dt}\frac{dx^{\nu}}{dt}\right]^{1/2}
\end{equation}
is averaged to (\ref{AvgS}),
and the uncertainty obtained from (\ref{AvgUnc})
is the same as obtained in (\ref{smear}). We thus do not regard
the functions $\gamma_{\beta}$ as dynamical variables, rather we represent physical
space-time by the whole set $\{g_{\mu\nu}^{\beta}\}$ at once.
In this spirit we construct $\gamma$s through their Fourier series,
\begin{eqnarray}
\gamma_{\beta}(x) =
\frac{1}{L^{3/2}}\sum_{\vec{k}} \nonumber\\
\left(c_{\beta}(\vec{k})\exp{[i(\vec{k}\cdot \vec{x} -\omega t)}]+c.c\right),
\end{eqnarray}
where $L$ is the length of an arbitrarily chosen large box (for normalization),
$$\vec{k}=\frac{2\pi}{L}\vec{n} \hspace{0.1 in}\text{ and }\omega=c|\vec{k}|.$$
We now choose an integer
$N_{\vec{k}}>2$
for each $\vec{k}$ and introduce a random variable $\alpha(\vec{k})$, such that
$$\alpha(\vec{k})\in\frac{2\pi}{N_{\vec{k}}}[0,1,2,....,N_{\vec{k}}-1].$$
For a particular $\alpha(\vec{k})$ a particular Fourier coefficient $c_{\beta}(\vec{k})$ is given as
\begin{equation}
 c_{\beta}(\vec{k})=f(k)\exp{[i\alpha(\vec{k})]}.
\end{equation}
The unknown function $f(k)$ is obtained from the scheme proposed above and is found out to be
\begin{equation}
 f(k)=\left(\frac{G\hbar}{2c^3}\right)^{1/3}k^{-5/6},
\end{equation}
using (\ref{smear}).
The contribution to $f_k$ for large values of $k$ comes from the requirement that  Eqn.
(\ref{smear}) should be valid even if $s$ is very small. Clearly, (\ref{smear}) is not meaningful in the limit
$s\rightarrow 0$ and a cut-off is assumed: $f(k)=0 \ {\rm for}\ k>10^{13}\ {\rm cm}^{-1}, \ s < 10^{-13} {\rm} cm$. It is asserted that details of the cut-off are not important, and only long-wave components are relevant.  This has been contested in ~\cite{Diosi:93} where it has been claimed that this cut-off is at a very high physically unacceptable value of $k$ and leads to absurd situations such as neutron star scale densities all over space. However, it seems that this objection can possibly be avoided by working entirely in real space, without going to Fourier space ~\cite{Frenkel:2002}. The analysis of Karolyhazy has been repeated by Frenkel, according to whom some of the Fourier sums diverge in some intermediate expressions, but ``in the formulae for physical quantities these sums are convergent''. In this work, the impact of the Karolyhazy uncertainty relation is realized, not by introducing a family of metrics, but by introducing a local time operator, and a corresponding phase operator in the wave-function describing the quantum state. The final results on wave-vector reduction are the same as those described below.

For considering wave propagation (Schr\"{o}dinger type evolution) in this `hazy' space-time,
we introduce a family $\{\psi_{\beta}\}$ of wave-functions corresponding to the metric family $\{g_{\mu\nu}^{\beta}\}$. For a single scalar elementary particle, via the
relativistic Klein-Gordon equation
\begin{equation}
 \frac{1}{\sqrt{-g_{\beta}}}\frac{\partial}{\partial x^{\mu}}(\sqrt{-g_{\beta}}g^{\mu\nu}_{\beta}\frac{\partial \phi}{\partial x^{\nu}})-
 \left(\frac{mc}{\hbar}\right)^2\phi=0,\nonumber
\end{equation}
we obtain  the non-relativistic generalization
\begin{equation}
 i\hbar\frac{\partial}{\partial t}\psi_{\beta}=\left(-\frac{\hbar^2}{2m}\triangledown^2+V_{\beta}\right)\psi_{\beta}.
\end{equation}
The small perturbation $V_{\beta}$ is given as
$$V_{\beta}(\vec{x},t)=\frac{mc^2\gamma_{\beta}(\vec{x},t)}{2}.$$
More interesting is the case for many particles: the multi-particle equation, where $V_{\beta}$ is replaced by
\begin{eqnarray}
 U_{\beta}(\{\vec{X}\},t)=\sum_i\frac{m_ic^2\gamma_{\beta}(\vec{x}_i,t)}{2},\nonumber\\
\{\vec{X}\}=\{\vec{x}_i\}.
\label{multi-potnal}
\end{eqnarray}
To see the qualitative effect of such a smearing we start with an initial `composite wave-function' $\Psi_{0}(\{\vec{X}\},0)$ for all the metrics
 $\{g^{\mu\nu}_{\beta}\}$. After the evolution different $\Psi_{\beta}(\{\vec{X}\},t)$ will become different. We can write to a good approximation,
\begin{equation}
 \Psi_{\beta}(\{\vec{X}\},t)\approx\Psi_{0}(\{\vec{X}\},t)e^{i\phi_{\beta}(\{\vec{X}\},t)},
\end{equation}
with
\begin{equation}
 \phi_{\beta}(\{\vec{X}\},t)=-\frac{1}{\hbar}\int_{0}^{t}dt' U_{\beta}(\{\vec{X}\},t).
\end{equation}
Let us choose and fix an $\vec{X_1}$ and an $\vec{X_2}$, and calculate the difference in phase between these two points in configuration space 
for different $\beta$. The answer will depend on $\beta$ and on time. The root mean square spread in the phase (average is over $\beta$)
can be estimated as a function of $\{\vec{X_1},\vec{X_2}\}$ and time $t$. The uncertainty in the relative phase depends only on the separation 
between the two points in configuration space,and  for a sufficiently large separation can reach the value $\pi$. 

\noindent{\it Microscopic and Macroscopic behavior:}

For a single quantum particle of mass $M$ for small values of $a\equiv|\vec{x}_1-\vec{x}_2|$ the spread in the phase
$$\triangle(a)\ll \pi,$$
and only for a large critical value $a_c$
$$\triangle(a_c)\approx\pi$$
will be achieved. The spread in the phase and the separation for which the critical value is reached can be calculated, as described above.
 We next discretize the space in terms `{\it coherence cells}' of dimension $a_c$. If initially the particle is confined to a single cell, 
then Schr\"{o}dinger
evolution will try to spread the wave packet, resulting in the wave function extending over to different cells and the set $\{\psi_{\beta}\}$ will
no longer behave as a single coherent wave function. When the original coherent set develops incoherent parts of comparable weights, it is taken as
signal for stochastic reduction of $\{\psi_{\beta}\}$ to a single cell.
Therefore, this stochastic reduction scheme is governed by Schr\"{o}dinger evolution and stochastic part comes through smearing of space-time metric.
This process provides us with a description of a physical phenomenon taking place regardless of the presence of any observer. Still this formalism indicates towards  but
does not provide any
formal embedding of idea of stochastic jumps into evolution in a consistent mathematical framework. It can be heuristically argued that microscopic quantum particles will
take astronomically large time before their wave-functions ``spill over'' a single coherence shell, making the possibility of stochastic reduction very
remote.

For an elementary particle of mass $m$, it can be shown that ;~\cite{Frenkel:2002};
\begin{equation}
a_c\approx \frac{\hbar^2}{G}\frac{1}{m^3} \approx \left( \frac{L}{L_p} \right)^2 L  ;\hspace{0.2 in}
L\approx  \frac{\hbar}{mc},
\label{micro:trans}
\end{equation}
and the critical time of reduction can be shown to be
\begin{equation}
\tau_c \approx\frac{ma_c^2}{\hbar}.
\label{treductn}
\end{equation}
For a proton one finds
$$a_c\approx 10^{25}cm,\hspace{0.5 cm} \tau_c\approx 10^{53}s $$
thus showing that one can never observe wave-packet reduction for a proton. The origin of the expression for the reduction time lies in the fact that according to the Schr\"{o}dinger equation, a wave-packet initially spread over $a_c$ will spread to a size $2a_c$ over time $\tau_c$. When this happens, we could take that as an indicator of loss of coherence, and hence stochastic reduction. The dynamics thus consists of cycles of
deterministic Schr\"{o}dinger evolution followed by stochastic jumps - something fully reminiscent of the GRW model; a comparison to which we will return shortly. In fact, $\tau_c$ is analogous to $\lambda^{-1}$ in GRW, and $a_c$ is analogous to $r_c$.

For more complex systems, such as a macroscopic body,  one works with the center of mass co-ordinate. However, care is needed because the gravitational perturbation described by the multi-particle potential
(\ref{multi-potnal}) depends on extended region of space. Still it can be shown that only the phase of the wave-function of the center of mass is affected, and the already introduced concepts of coherence cells and coherence length $a_c$ can be applied to the center of mass coordinate.
  In such cases not only mass $M$ of the system but the size $R$ also enters
in the expression for $a_c$ (as arbitrariness in metric will be experienced throughout the size). It can be shown that
\begin{equation}
 a_c\approx\left(\frac{\hbar^2}{G}\right)^{1/3}\frac{R^{2/3}}{M} = \left(\frac{R}{L_p}\right)^{2/3} L\ ; \quad
L = \frac{\hbar}{mc}
\label{macro:trans}\nonumber
\end{equation}
The reduction time is again given by (\ref{treductn}).
For a ball of $R=1$ cm and for terrestrial densities this gives $a_c \approx 10^{-16}$ cm and $\tau_c \approx 10^{-4}$ cm. The wave-function undergoes $10^{4}$ expansion-reduction cycles per second, and at the end of each cycle the momentum performs a jump $\Delta p_c$ of the order $\hbar/a_c$ which corresponds to a velocity shift of the order $a_c/\tau_c\sim 10^{-12}$ cm/sec.  These repeated kicks amount to an anomalous Brownian motion and a tiny associated energy non-conservation of the order
$\hbar^2/Ma_c^2$, another feature in common with spontaneous collapse models.

One can try to understand the transition region from micro- to macro- behavior. We have seen that for $R\approx 1 cm$ we have $a_c\ll R$. Furthermore, the expressions (\ref{macro:trans}) and
(\ref{micro:trans}) become the same when $a_c=R$. So one can now classify \\
(i) $a_c\gg R \quad \left(i.e. \ \hbar^2/G \gg M^3 R \right)$\quad  micro-behavior regime \\
(ii) $a_c\approx R \quad \left(i.e. \ \hbar^2/G \approx M^3 R \right)$\quad transition region\\
(iii) $a_c\ll R \quad \left(i.e. \ \hbar^2/G \ll M^3 R \right)$\quad macro-behavior regime.

If $a_c\gg R$ it can be shown that ~\cite{Frenkel:2002}   the expression
(\ref{micro:trans}) continues to hold, for a micro-object having an extended linear size $R$.

Setting $a_c=R$ in (\ref{macro:trans})   and assuming density to be about 1 gram/cc for terrestrial bodies gives for the transition region
\begin{equation}
a^{tr} \approx 10^{-5} {\rm cm}, \ \tau^{tr} \approx 10^{3} {\rm s}, \ M^{tr}
\approx 10^{-14} {\rm g}
\label{transvalues}
\end{equation}
It is significant that $a^{tr}$ coincides with the favored value for $r_C$ in the GRW and CSL model. The transition mass corresponds 
to about $10^{10}$ amu. Note that because Planck length and the size of the body also enter the picture, the transition occurs at a mass
 much lower than the simplistic but much higher Planck mass ($10^{-5}$ grams).

Interestingly, the jump velocity $\Delta v_c = a_c/\tau_c$ in a reduction cycle takes its maximal value in the transition region $a_c \approx R$
 and decreases on either side away from this transition region ~\cite{Karolyhazi:86}.

Thus we 
obtain a transition point which, in principle can be tested upon. The measurement
process can be argued as interactions resulting in significant change in mass distribution of the whole set up (system+surrounding) making the definite 
and
different outcome states incoherent thereby reducing the state of set-up to a particular outcome. Still the formal mathematical framework of this
idea is missing making it a challenge to precisely calculate the characteristics of quantum state reduction.

The K-model has also been discussed in ~\cite{Karolyhazy:74, Frenkel:2002, Frenkel:77, Karolyhazy:1982,
Frenkel:90, Frenkel:97, Karolyhazy:90, Karolyhazy:95, Frenkel:95}.

\noindent{\it Comparison with GRW model}

There is a fascinating similarity between the K-model and GRW model, despite significant differences in detail. 
The overall picture of Schr\"{o}dinger evolution interrupted by stochastic reduction is the same. In the K-model, the origin of
 stochasticity lies in the intrinsic uncertainty of space-time structure, whereas in GRW the origin is left unspecified. However, both models have 
 length/time scales ($a_c/\tau_c$ in K-model, $r_C/\lambda^{-1}$ in GRW). There are no free
 parameters in the K-model, whereas GRW introduce new parameters $\lambda$ and $r_C$. Thus it is entirely possible that gravity might provide the 
fundamental underpinning for models of spontaneous collapse. Of course a mathematically rigorous treatment of gravity in the K-model remains 
to be developed, but the physical principles and semi-rigorous results already obtained are highly suggestive by themselves.

An important early study comparing the K-model and GRW was made in ~\cite{Frenkel:90}. It should be noted that while in both cases the reduction
 time decreases with increasing mass, the quantitative dependence is different. In the K-model, for $a_c\geq R$ the reduction time falls as $1/m^5$,
 and if $a_c \ll R$ it falls as
$1/m^{5/9}$ assuming a fixed density. In GRW, the reduction time simply falls as $1/m$, whereas we have seen that in the CSL model the dependence 
is more complex. Similarly, $a_c$ falls with increasing mass. While $r_c$ in GRW does not depend on mass, the linear size to which the stochastic
 reduction confines an expanding wave-packet does depend on mass - this linear size is the analog of the coherence cell $a_c$ ~\cite{Frenkel:90}. 
In the light of modern experiments there is perhaps need for a more careful comparison between the quantitative predictions of the K-model and GRW/CSL.
 Also, a careful quantitative description of the quantum measurement process, showing the emergence of the Born rule, seems to not yet have been 
developed in the K-model. A time-evolution equation for the density operator in the K-model, analogous to the corresponding equation in GRW, has been
 discussed in ~\cite{Frenkel:90}. See also related discussions in
~\cite{Unturbe:92}.

\subsubsection{The model of Di\'osi}

Di\'osi's approach ~\cite{Diosi:87}, while being similar to Karolyhazy's, is inspired by the famous work of Bohr and Rosenfeld ~\cite{Wheeler-Zurek:1983} which investigated the principles of measuring the electromagnetic field by apparatuses obeying quantum mechanics. It was argued in ~\cite{Diosi:87a} that if a Newtonian gravitational field ${\bf g}=-\nabla \phi$ is measured by a quantum probe over a time $T$, then its average
$\tilde{\bf g}({\bf r}, t)$ over a volume $V$ exhibits an uncertainty which is universally bounded by
\begin{equation}
(\delta \tilde{g})^{2} \geq \hbar G / VT
\label{grav-uncertain}
\end{equation}
This is Di\'osi's analog of the Karolyhazy uncertainty relation, and the idea now is to see how how this intrinsic quantum  imprecision in the space-time metric affects the Schr\"{o}dinger evolution of a quantum state in quantum mechanics.

To this effect, Di\'osi introduces the concept of a universal gravitational white noise, by proposing that the gravitational field possesses universal fluctuations [in other words the potential $\phi({\bf r}, t)$ is a stochastic variable] whose stochastic average equals, up to numerical factors of order unity, the intrinsic uncertainty given by (\ref{grav-uncertain})
\begin{equation}
\langle [ \nabla \tilde{\phi}({\bf r}, t)]^2\rangle - [ \langle\nabla \tilde{\phi}({\bf r}, t) \rangle]^2
= {\rm const} \times \hbar G / VT
\end{equation}
From here, it can be shown that, assuming $\langle \phi ({\bf r}, t)\rangle \equiv 0$, the correlation function of $\phi({\bf r}, t)$ is given by
\begin{equation}
\langle\phi({\bf r}, t) \phi({\bf r}', t')\rangle = \hbar G |{\bf r} - {\bf r}'|^{-1} \ \delta(t-t')
\end{equation}
The probability distribution of the stochastic variable $\phi({\bf r}, t)$ is completely specified by this correlation function if the distribution is assumed to be gaussian [gaussian white noise].

Next, one asks for the effect of the stochastic fluctuations in $\phi$ on the propagation of the quantum state $\psi$ of a system whose evolution is assumed to be described by the Schr\"{o}dinger equation
\begin{equation}
i\hbar\dot{\psi} = \left( \hat{H}_{0} + \int \phi \hat{f}({\bf r})\ d^3 r \right) \psi(t)
\end{equation}
where $\hat{f}({\bf r})$ stands for the operator of the local mass density of the system.

$\psi$ is now a stochastic variable, and the corresponding density operator
$\hat{\rho}=\langle \psi(t) \psi^{\dagger}(t)\rangle$ obeys the following deterministic master equation for the assumed gaussian white noise
\begin{eqnarray}
& & \dot{\hat{\rho}} = -\frac{i}{\hbar} \left[ \hat{H}_0, \rho(t)\right] \nonumber \\
& - & \frac{G}{2\hbar}\int\int \frac{d^3 r \ d^3 r'}{|{\bf r} - {\bf r}'|}
\left[ \hat{f}({\bf r}), \left[\hat{f}{(\bf r')}, \hat{\rho}(t)\right]\right]\nonumber\\
\end{eqnarray}
The second term on the right hand side is the damping term which represents the universal violation of quantum mechanics.

To compute the nature of violation, denote the configuration coordinates of a dynamical system by $X$, and denote the corresponding mass density at a point ${\bf r}$ by $f({\bf r}|X)$. Given a pair of configurations, a characteristic damping time $\tau_d(X,X')$ is defined by
\begin{eqnarray}
& & [\tau_d(X,X')]^{-1} 
=  \frac{G}{2\hbar} \int\int d^{3}r\ d^{3}r'   \ \times\nonumber\\
& &\frac {[f({\bf r}|X) - f({\bf r}|X')][f({\bf r}'|X) - f({\bf r}'|X')]  }{|{\bf r} - {\bf r}'|}\nonumber\\
\end{eqnarray}
Introducing the coordinate eigenstates $|X\rangle$ the master equation can be written as
\begin{eqnarray}
& & \langle X|\dot{\hat{\rho}}| X'\rangle = -\frac{i}{\hbar}
\langle X |\left[ \hat{H}_0, \rho(t)\right] |X'\rangle\nonumber \\
&-& [\tau_d(X,X')]^{-1} \langle X|\hat{\rho}(t)|X'\rangle
\end{eqnarray}
Just like in decoherence and in models of spontaneous collapse, the second term on the right hand side destroys interference between the states $|X\rangle$ and $|X'\rangle$ over the characteristic time $\tau_d$, and this effect can become significant if the difference between the mass distributions $f({\bf r}|X)$ and
$f({\bf r}|X')$ is significant.

To estimate the scale of the gravitationally induced violation Di\'osi considers a dynamical system consisting of a rigid spherical ball of homogeneously distributed mass $m$ and radius $R$, so that the configuration $X$ is represented by the center of mass coordinate ${\bf x}$. The characteristic damping time $\tau_d$ is shown to be
\begin{equation}
\tau_d ({\bf x}, {\bf x}') = \hbar [ U(|{\bf x} - {\bf x}'|) - U(0)]^{-1}
\end{equation}
where $U$ is the gravitational potential between two spheres, each of mass $m$ and radius $R$. The master equation can now be written as
\begin{eqnarray}
& & \frac{d\ }{dt} \langle {\bf x} | \rho | {\bf x}' \rangle = \frac{i\hbar}{2m} (\nabla^2 - \nabla'^2)
\langle {\bf x} | \rho | {\bf x}' \rangle \nonumber\\
& - &
\frac{1}{\hbar} [ U(|{\bf x} - {\bf x}'|) - U(0)] \langle {\bf x} | \rho | {\bf x}' \rangle
\end{eqnarray}

We define the coherent width $l$ of a given state as the characteristic distance $l=|{\bf x} - {\bf x'}|$ above which the off-diagonal terms $\langle {\bf x}| \rho | {\bf x}'\rangle$ become negligibly small. The time-scale $t_{kin}$ over which kinetic changes are introduced due to ordinary quantum evolution given by the first term on the right hand side is of the order $ml^{2}/\hbar$. A crititical length $l_{crit}$ is defined by equating $t_{kin}(l_{crit})$ and the damping time $\tau_{d}(l_{crit})$
\begin{equation}
ml_{crit}^{2}/\hbar = \hbar [ U(l_{crit}) - U(0)]^{-1}
\end{equation}
If the coherent width $l$ of the quantum state
is much smaller than the critical value $l_{crit}$  then the standard quantum kinetics dominates and damping is not effective. On the other hand if $l\gg l_{crit}$ then the coherence of the state will be destroyed by the gravitational damping term in the master equation. $l_{crit}$ is the analog of the phase coherence length $a_c$ of the K-model, and the length parameter $r_C$ of GRW. Also, there clearly are analogs of
$\tau_d (l_{crit})$ in the other two models.

One can now show that in two limiting cases $l_{crit}$ takes the following form:
\begin{eqnarray}
l_{crit}&\sim& (\hbar^2/Gm^3)^{1/4}R^{3/4}, \quad {\rm if} \quad Rm^3 \gg \hbar^2/G,\nonumber\\
&\sim& (\hbar^2/Gm^3)^{1/2}R^{1/2}, \quad {\rm if} \quad Rm^3 \ll \hbar^2/G\nonumber\\
\end{eqnarray}
These expressions are similar to, though not identical with, those in the K-model. The fact that they are  similar but not identical suggests that the involvement of gravity in wave-vector reduction is strongly indicated, but  the exact mathematical treatment remains to be found. Importantly, the transition $l_{crit}=R$ happens at the same value $l_{crit}=\hbar^{2}/Gm^{3}$ in both the models. Notice though that for small masses $l_{crit}$ is not independent of $R$, unlike in the K-model. For a proton, taking $R$ to be the classical radius $10^{-13}$ cm, Di\'osi estimates $l_{crit}$ to be $10^{6}$ cm, which is curiously much smaller than the prediction  $10^{25}$ cm for the K-model. Also, the reduction time is $10^{15}$ sec, much smaller than in the K-model. However, the models are in better agreement in the macro- region, and in Di\'osi's model too, the transition parameters are the same as that given by Eqn. (\ref{transvalues}).

Subsequently, Di\'osi took the inevitable step of casting the master equation in the equivalent language of a stochastic Schr\"{o}dinger equation ~\cite{Diosi:89}. He called this model QMUDL (Quantum mechanics with universal density localization). It is similar to his QMUPL model, which we reviewed earlier in this article, except that the localization is not in the position operator $q$, but in the mass density operator
$\hat{f}({\bf r})$ introduced above.  The universal free parameter $\lambda$ of QMUPL is now replaced by the gravitational constant, so that the theory becomes parameter free.

As was discussed in ~\cite{Ghirardi3:90} the QMUDL model has certain limitations - it cannot deal with point particles (for which case it leads to divergent densities) and restricts itself to extended objects. The model parameters are such that it leads to an unacceptably high rate of energy increase during reduction.
Furthermore, for microscopic dynamics the reduction and localization process can lead to unacceptable processes such as excitation or dissociation of nuclei. To avoid these problems. Ghirardi et al. proposed a CSL type modification of QMUDL, and the introduction of a new universal length parameter. It is suggested that it does not seem possible to have a parameter free theory for reduction, such as gravity induced collapse. An alternate way out of the difficulties of the otherwise very attractive model of Di\'osi has been suggested by Penrose - we will recall this proposal next, but find that here one is faced with a possibly new set of difficulties. Thus it would seem that at present CSL might be the best model at hand, even though its fundamental origin remains to be understood, and it yet may have a strong connection with gravity whose proper implementation remains to be achieved.

\subsubsection{The model of Penrose}

Penrose ~\cite{Penrose:96,  Penrose:98} addressed the question of the stationarity of a quantum system  which consists of a linear superposition
 $|\psi\rangle = a|\alpha\rangle + b|\beta\rangle$ of two well-defined stationay states $|\alpha\rangle$ and $|\beta\rangle$, having same energy $E$
If gravitation is ignored, as is done in standard quantum theory, the superposition $|\psi\rangle = a|\alpha\rangle + b|\beta\rangle$ is also
 stationary, with the same energy $E$,
\begin{equation}
i\hbar \frac{\partial|\psi\rangle}{\partial t} = E. |\psi\rangle
\end{equation}

However, the inclusion of gravitation raises a new question: what is the meaning of the Schr\"{o}dinger time-evolution operator $\partial/\partial t$? There will be a nearly classical spacetime associated with the state $|\alpha\rangle$, and a Killing vector associated with it which  represents the time displacement of stationarity.  And there will be a different nearly classical spacetime associated with the state $|\beta\rangle$, and a {\it different} Killing vector associated with it which  represents the associated time displacement of stationarity.  The two Killing vectors can be identified with each other only if the two space-times can be identified with each other point by point. However, the principle of general covariance in general relativity forbids that, since the matter distributions associated with the two states are different, in the presence of a  background gravitational field. On the other hand,
unitary evolution in quantum theory requires and assumes the existence of a Schr\"{o}dinger operator which applies to the superposition in the same way that it appies to the individual states, and its action on the superposition is the superposition of its action on individual states. There is thus a conflict between the demands of quantum theory and of general relativity.

A tentative resolution is to make an approximate point-wise identification between the two spacetimes, which in turn corresponds to a slight error in the identification of the Schr\"{o}dinger operator for one spacetime with that for the other. This corresponds to a slight uncertainty in the energy of the superposition, for which it is possible to make an estimate in the case when the superposition amplitudes are nearly equal in magnitude. In the Newtonian approximation, this energy uncertainty $E_G$ is of the order of the gravitational self-energy of the mass distribution in the two superposed states. In accordance with the Heisenberg uncertainty principle, the superposition lifetime can be taken to be $\hbar / E_G$, beyond which time the superposition will decay. In concept and in detail, this is quite like the damping time $\tau_d$ in Diosi's model. It is not clear here though, as to how the Born rule will be recovered dynamically.

Penrose notes the commonality with Di\'osi's ideas, the difficulties encountered by Di\'osi, and the resolution proposed by Ghirardi et al. by way of introducing a fundamental length scale. Penrose observes that essentially the same difficulty would arise in his own approach too, because if one were dealing with point particles, the gravitational self-energy difference can become infinitely high, implying instantaneous
reduction, which is clearly unreasonable. While Ghirardi et al. avoid this problem by introducing a new length scale, Penrose proposes  a different way out. The way out is in particular based on noting that one needs to specify which states are the basic [stable] states, to which superpositions of basic states decay.

It is proposed that the basic stationary states to which a general superposition will decay by state reduction are  stationary solutions of the so-called Schr\"{o}dinger-Newton equation (SN-equation). This equation is actually a pair of coupled differential equations which are set up as follows, for a quantum mechanical particle of mass $m$ moving in its own gravitational field ~\cite{Diosi:84}
\begin{eqnarray}
& & i\hbar \frac{\partial\psi}{\partial t} = - \frac{\hbar^2}{2m} \nabla^{2}\Psi + m\Phi \Psi\nonumber\\
& & \nabla^{2}\Phi = 4\pi G m |\Psi|^2
\end{eqnarray}
This system of equations has been analyzed in ~\cite{Moroz:98, Moroz:99, Harrison:2003, giulini2011gravitationally, Bernstein:98, Ruffini:69}. These equations are closely related to the Schr\"{o}dinger-Poisson equations which have been studied for much longer ~\cite{Lange:95}.

At this stage an important difference with the models of Karolyhazy and Di\'osi seems to be that that unlike in the latter two models, where an intrinsic uncertainty in spacetime structure is assumed, here the impact on the evolution of the quantum state is due to the particle's own gravitational field. Also, the system seems to be set up deterministically and the presence of a stochastic element is not evident, at least a priori. Thus one could ask as to the origin of the stochastic feature which actually drives the system to one of the stationary states, and the accompanying Born rule. Also, if the evolution is deterministic and non-linear, there appears to be present the possibility of superluminal propagation.

These issues apart, the SN system of equations yields some very interesting results. Spherically symmetric stationary solutions have been found and their stability has been investigated. A comprehensive recent analysis is given in ~\cite{giulini2011gravitationally}. Their study was motivated in response to
~\cite{carlip2008quantum, salzman2006possible} - the SN equation induces a gravitational suppression of expanding
Gaussian wave-packets, and it was suggested by Carlip and Salzman that the suppression [and hence wave-vector reduction] becomes significant already at $m\sim 1600$ amu. This surprisingly low value is at variance with the much higher estimates coming from simple analytical estimates [and also from the work of Karolyhazy and Di\'osi] and prompted ~\cite{giulini2011gravitationally} to look at the problem closely.

Various numerical studies, as well as heuristic estimates, show that the ground state energy is of the order
\begin{equation}
E\sim - \frac{1}{8} \frac{G^2 m^5}{\hbar^2}
\end{equation}
The width $a$ of the mass distribution in the ground state is
\begin{equation}
a_0 \approx \frac{2\hbar^2}{Gm^3}
\end{equation}
which we immediately notice coincides with the phase coherence cell length in microscopic limit of the K-model.

By introducing a length scale $l$ the SN equation can be written in terms of a dimensionless coupling constant
\begin{equation}
K = 2 \frac{Gm^3 l}{\hbar^2} = 2 \left(\frac{l}{L_p}\right) \left(\frac{m}{m_P}\right)^3
\label{SNcoupling}
\end{equation}
One considers the time-dependent SN equation for initial values given by a spherically symmetric gaussian wave-packet of width $a$
\begin{equation}
\psi(r, t=0) = \left(\pi a^2\right)^{-3/4} \exp \left( - \frac{r^2}{2a^2}\right)
\end{equation}
There are thus two free parameters $a$ and $m$, and one asks for the regions in this parameter space where significant inhibitions of the usual free quantum dispersion occur. In an important analysis
~\cite{giulini2011gravitationally} give four different analytical arguments to show that inhibition of the dispersion becomes significant when the dimensionless coupling constant $K$ of (\ref{SNcoupling}) becomes of order unity. This conclusion coincides with that of Karolyhazy and Diosi, and we believe it leads us to an important inference:

{\it The models of Karolyhazy, Di\'osi and Penrose all agree that if the width of the quantum state
associated with an object of mass $m$ becomes greater than of the order  $\hbar^2 / Gm^{3}$, the quantum-to-classical transition sets in.}

For the experimentally interesting $a=0.5\ \mu$m this gives  $m$ to be about $10^{9}$ amu.

These results are further supported by numerical investigations of the SN equation in
~\cite{giulini2011gravitationally}. The authors also note that the coherence time, the time beyond which collapse takes place, can be brought down by reducing the grating period in a molecule interferometry experiment. For instance, for a mass of $10^{11}$ amu and grating period of  0.5 $\mu$m they report a coherence loss time of 300 ms.

It is significant that while the Penrose approach does not directly address the emergence of the Born rule, it correctly predicts the regime where the quantum to classical transition takes place, in agreement with the other gravity models.

This essentially completes our brief review of the three well-known models of gravity-induced collapse. Other considerations of gravity induced collapse have been made in ~\cite{Ellis:84} and ~\cite{Percival:95}. A brief but elegant summary of gravity models and trace dynamics is given by
~\cite{Diosi:05}.

In our view, gravity induced collapse is a promising physical mechanism for a physical realization of spontaneous collapse. Furthermore, Trace Dynamics and its extension to space-time structure [treating space and time as operators] provide a plausible mathematical avenue for rigorously developing the stochastic theory of gravity induced spontaneous collapse.

\section{Experimental Tests of the Theoretical Predictions}

\subsection{Introduction}

We have considered two classes of underlying theories for dynamical collapse: Trace Dynamics, and gravity induced collapse. The phenomenology of Trace Dynamics manifests itself through models of spontaneous collapse. If spontaneous collapse or  gravity induced collapse is a possible explanation for the measurement problem, then the experimental predictions of these models differ from those of standard quantum theory. Bounds can be set on the parameters of these models by requiring that their predictions should not disagree with those observations which are well-explained by the standard quantum theory. On the other hand one can perform new experiments, such as diffraction experiments with large molecules, for which the predictions of these experiments differ appreciably from those of quantum theory. The results of such experiments could vindicate the modified quantum dynamics [and specific values of the associated parameters] or rule it out. This section reviews the bounds on model parameters which come from known physical and astrophysical processes, and from diffraction experiments that have been carried out in the laboratory or are planned for the near future.   Experiments are discussed through Sections IV.B to IV.G. Bounds on spontaneous collapse models will be discussed in
Sec. IV.H and IV.I and those on gravity based models in Sec. IV.J.

As we saw in Sec.~II, a large variety of collapse models has been proposed: QMUPL, GRW, CSL, dissipative and non-dissipative, white, colored,  Markovian and non-Markovian. The overall task of constraining these models is very extensive, given their large variety, and considering that a large variety of observations [laboratory and astrophysical] as well as table-top experiments, has to be considered.  The subject is in a state of rapid flux, and still in a developmental stage. Here we try to do as complete a job as possible, relying on the analysis in  a host of important papers that have appeared in the last few years ~\cite{Adler2:07, Adler3:07, Adler:09, nimmrichter2011testing, Romero2012decoherence, feldmann:11}. It should be noted though that the CSL model has received the maximum amount of attention, and we will focus mainly on CSL.

The CSL model introduces two new parameters, the rate constant $\lambda$ and the correlation length $r_C$. If spontaneous collapse is a correct theory of nature, the values of these parameters must follow from some underlying fundamental principles and/or be determined by experiments.
As mentioned earlier, GRW chose $\lambda \simeq 10^{-16}$ sec$^{-1}$ and $r_{C} \simeq 10^{-5}$ cm, in order to be consistent with observations, while Adler chose $\lambda \simeq 10^{-8\pm2}$ sec$^{-1}$ and $r_{C} \simeq 10^{-5}$ cm. However, there is room for more general considerations and for establishing the allowed part of the parameter space in the $\lambda - r_{C}$ plane. In Sec. IV.G we will consider bounds coming from cosmology and in Sec. IV.H we will summarize other physical processes which constrain $r_C$ and $\lambda$.

In Sec. IV.C  we will summarize experiments which directly test quantum superposition by interferometers. These experiments also test collapse models and gravity models. The rationale is that if we can observe superposition at mesoscopic and possibly even macroscopic scale, the quantum dynamics does not need alteration. If instead we experimentally observe a quantum to classical transition such as the collapse of the wave function while convincingly reducing all potential sources of noise, this would strongly hint that  an alteration of the fundamental equations of quantum mechanics is needed. The CSL model with Adler's value for $\lambda$ predicts a quantum to classical transition at only two or three orders of magnitude away from present molecule matter-wave experiments. This and new proposals for optomechanics experiments with trapped bead particles [see Sec. IV.D and IV.E] bring experimental tests of CSL (amongst other proposals) within reach. The main focus of the following sections will be on those table-top matter-wave and optomechanics experiments.

While we do not aim to give a complete review of all possible experimental tests of collapse models, we present the strongest current bounds in Table 1.

\subsection{Possible experimental tests of CSL basing on quantum superposition}
Matter-wave interference experiments such as molecule interference are approaching the mass limit for the quantum to classical transition in a 'bottom-up' fashion, starting with particles, where quantum superposition is existing and pushing the limit upwards step by step. Nano- and micro-mechanical devices cooled to the quantum-mechanical ground state within optomechanics approach the problem from the top, starting at very massive objects namely mechanical cantilevers of hundreds of nanometer and even some micrometer size, which are sometimes even visible with the naked eye. 
The range in which both types of experiments will probably meet [to combine techniques, knowhow and ideas to overcome experimental hurdles and
to switch-off known decoherence mechanisms] is from
10 nm to 100 nm in size [mass - $10^6$ - $10^9$ amu].
The experimental aim is to show quantum superposition by negativity in the Wigner function of the motional states or by proving the wave-nature of such particles by single-particle interference. Interestingly, this size range is vital to test non-standard quantum theories such as CSL and gravity induced collapse.

\subsubsection{Collapse theory for diffraction experiments}
In order to understand how collapse models differ from standard quantum mechanics, when applied to interferometric experiments, let us consider once again the QMUPL model of Sec. II.E, due to its simplicity. The multi-particle dynamics is given by Eqn.~(\ref{nlemp}). By using It\^o calculus, it is easy to show that the master equation for the statistical operator $\rho_t = \mathbb{E}[|\psi_t\rangle\langle\psi_t|]$ is:
\begin{equation}
\frac{d}{dt}\rho_t = - \frac{i}{\hbar} [H, \rho_t ] - \frac{1}{2} \sum_{i=1}^{n} \lambda_i [q_i,[q_i, \rho_t]].
\nonumber
\end{equation}
Suppose - for simplicity sake - that all particles are identical, and that during the time of measurement, the free evolution (given by $H$) can be neglected. Then, according to the above equation, the density matrix in the position representation evolves as follows:
\begin{equation}
\rho_t(x,y) = \rho_0(x,y) e^{- \lambda N (x-y)^2 t/2}.
\end{equation}
This equation contains everything there is to know about the effect of collapse models on interferometric experiments, at least from the conceptual point of view. Different models differ only in the technical details. The above equation tells that, in order to measure a collapse effect, corresponding to a significant damping factor, the following criteria must be met: the system should be as big as possible (large $N$); it should be created in a ``large'' superposition (large $|x-y|$), which is monitored after a time as large as possible (large $t$). This is the goal that all interferometric experiments aim at reaching, in order to test the validity of the superposition principle, and thus also of collapse models.

\begin{figure*}[!htb]
  \centerline{
  \includegraphics[width=9.0 cm]{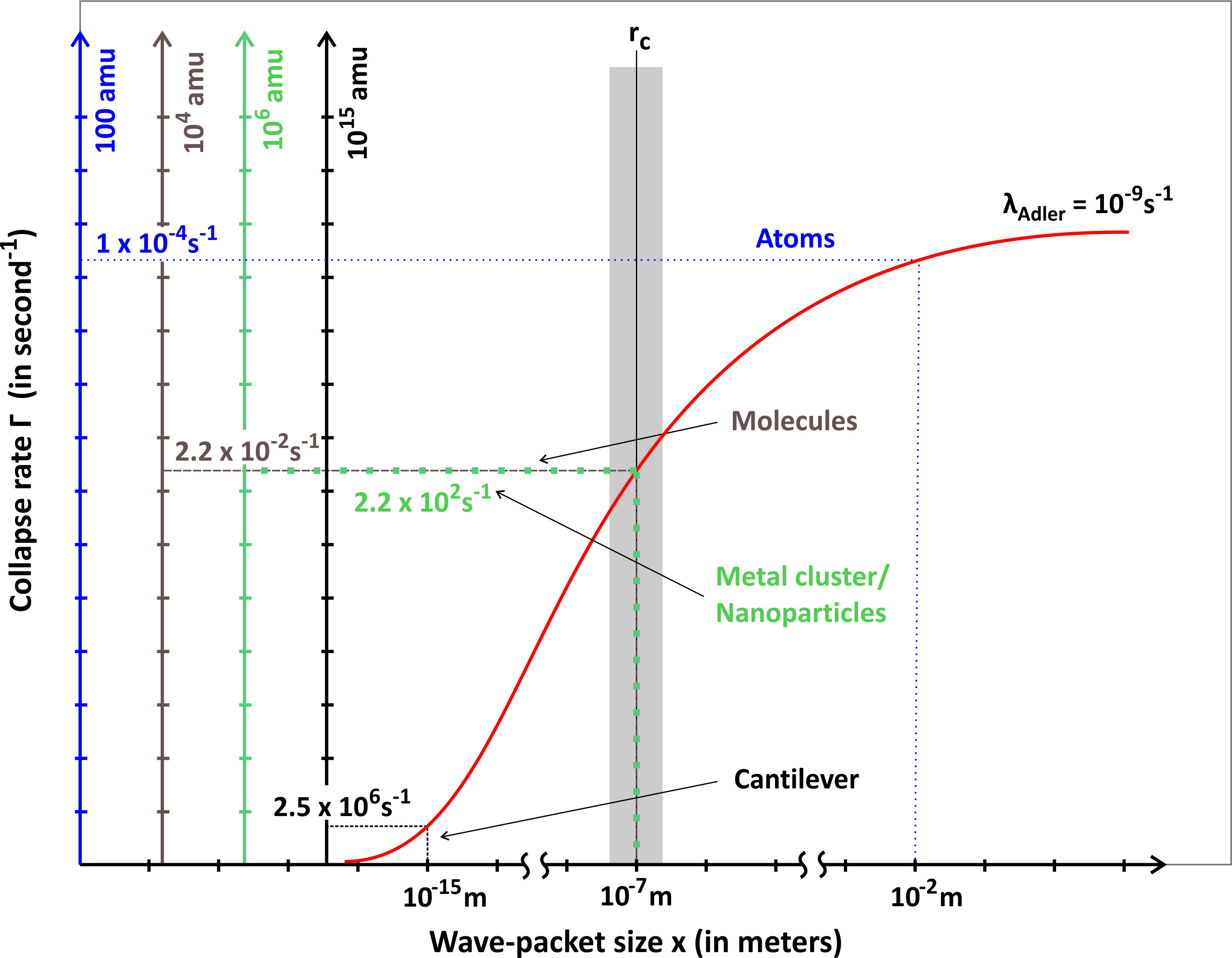}}
  \caption{This plot aims to compare experiment with model to map the significance of different experiments to test collapse models. The CSL decay function $\Gamma$ is shown against spatial dimension/separation
in case of four different mass systems (represented here as four different $y-$ axes), using 
$\lambda=\lambda_{\text{\tiny Adler}}$. The x-axis is not continuous, so as to permit combination of all experiments of very different mass and size scales in one plot. We aim for experimental parameters to fit to the part of the $\Gamma$ curve where we observe a significant chance to test collapse models. Essentially we need a good mixture of mass, spatial separation and duration of superposition.  For example, for molecule ($10^4$ amu, experiment done) and OTIMA metal cluster ($10^6$ amu, experiment proposed; bead experiments such as MERID are at the same range in the plot) interferometer which both give spatial separation on the same scale as $x=r_C$, decay rates are $2.2 
\times 10^{-2} s^{-1},$ and $2.2 \times 10^{2} s^{-1}$ respectively. In the case of the cantilever (experiment proposed) the experimental bound to $\Gamma$ would be quite strong, but the spatial separation is small and quite far away from $x=r_C$. Atom interferometers (experiment done) have an exceptionally large spatial separation, but the mass is small and therefore the bound to $\Gamma$ is very weak.} \label{fig.gammavsx}
\end{figure*}

We now come back to the CSL model, which we are primarily interested in. In this case, the damping behavior is less trivial than that of the QMUPL model. As we have seen in Sec. II.I, for small distances there is a quadratic dependence of the decay function on the superposition distance $|x-y|$, while for large distances such a dependence disappears. The intermediate behavior is not easy to unfold, but it can be conveniently modeled by the following ansatz. Recall first that for a single constituent the master equation
\begin{equation}\label{eq:decoherence}
\frac{d}{dt}\rho_t(x,y) = - \frac{i}{\hbar} [H, \rho_t(x,y) ] - \Gamma_{\makebox{\tiny CSL}}(x,y) \rho_t(x,y)
\nonumber
\end{equation}
implies that the decay function is
\begin{equation}\label{eq:decayrate}
\Gamma_{\makebox{\tiny CSL}}(x) = \lambda [ 1 - e^{-x^2/4 r_c^2} ]
\end{equation}
for one single constituent. See Fig. \ref{fig.gammavsx} for a plot of $\Gamma_{\makebox{\tiny CSL}}$ vs $x$. Here we see how the two fundamental parameters of the CSL model enter into play. For a many-particle system, one makes an ansatz and assumes that the above expression for the decay function holds, except that one has to multiply $\lambda$ by the appropriate numeric factorial, as described in Sec. II.I. The numerical factor is $n^2 N$ where $n$ is the number of nuclei [called cluster] within a volume of linear size $r_C$, and $N$ is the number of clusters in the many-particle system.

In all interferometric experiments so far realized, the period of the grating is comparable to
$r_C\sim 100$ nm. Therefore, to extract the significant order of magnitude, it is sufficient to work in the regime $x \gg r_C$. Taking into account Eqn. (\ref{eq:ridert}) ~\cite{Adler3:07} and that in the case of macromolecules $N = 1$ (typical molecule size being about 1 nm), we have:
\begin{equation}
\Gamma_{\makebox{\tiny CSL}} \simeq \lambda n^2,
\end{equation}
Since no interferometry-based experiments have so far detected any spontaneous collapse effect, this implies that the damping factor $\exp[- \Gamma_{\makebox{\tiny CSL}}  t]$ must be insignificant. We then have:
\begin{equation}
\lambda \leq 1/n^2 t,
\end{equation}
where $n$ measures the number of nucleons in the system, and $t$ the duration of the experiment. This is the type of bound that interferometric experiments place on the collapse rate $\lambda$. The experiments {\it do not} provide a bound on the second parameter of the CSL model $r_C$, for the reasons explained above. More general situations could be considered, but they have not been analyzed so far. It will be desirable to carry out a careful analysis of the allowed part of the $\lambda-r_C$ plane, based on the data available from experiments, and to understand what role the grating period and size of the macromolecule will eventually play in bringing experiment and theory closer.

The latest situation on the results from diffraction experiments has been discussed in ~\cite{nimmrichter2011testing, Romero2012decoherence, feldmann:11}. The strongest current bound on $\lambda$ seems to be from the experiment of ~\cite{Gerlich2011} which sets $\lambda < 10^{-5}$ sec$^{-1}$ for $n=7,000$. Adler estimates that an experiment with $n=500,000$ will confront the enhanced CSL value proposed by him, based on reduction in latent image formation ~\cite{Adler3:07}. Interestingly, in the same paper he also proposes to test whether `latent image formation' constitutes a measurement, by using a photographic emulsion as a `which path' detector in one arm of a quantum interferometer.

Spatial or centre of mass motion superposition is needed to be demonstrated in experiments to test the quantum to classical transition. As described previously atoms are so light that even the very large areas in today's atom interferometers do not increase the chance to test CSL (see Fig. 2). On the other hand the very massive cantilevers do not possess a large enough spatial separation (spatial size of superposition) to become good test embodiments for the quantum to classical transition. It seems that the size range of particles of 10~nm to 100~nm, which corresponds to a mass range of $10^6$amu to $10^9$amu are ideal for such tests in matter-wave interference experiments.

In this section we focus on possible experimental tests of the CSL model, while similar or quite different experiments are possible to test different collapse models.

\subsection{Matter-wave interferometry: Molecule interferometry}
Experiments with matter waves exist since 1927 when Davissson and Germer diffracted a beam of electrons. It was the first proof of de Broglie's hypothesis on particle-wave duality. Since then matter-wave interferometry of electrons~\cite{hasselbach2010progress}, neutrons~\cite{rauch2000neutron}, atoms~\cite{cronin2007atom} and molecules~\cite{Hornberger2011review} has a long and successful history to investigate fundamental physics, and has been applied for metrology and sensing~\cite{Arndt2011focusissue}. Interestingly, a recent interpretation of atom interferometry experiments resulted in a debate on the possible detection of gravitational red shift by such tabletop experiments~\cite{Muller2010}.

We are here interested in centre-of-mass-motion interferometry (or de Broglie interference) of very massive particles, as these experiments are promising to test modifications of Schr\"{o}dinger dynamics such as collapse models predicting a quantum to classical transition at mesoscopic length and mass scales. The appearance of a single-particle interference pattern demonstrates wave-like behaviour of the particles and can be seen as an indication for superposition. The full beauty of this particle position superposition can be seen from reconstruction of the Wigner function of the motional quantum state by tomography~\cite{kurtsiefer1997measurement}.

Technically, to perform de Broglie interference experiments, one has to overcome challenges of preparation of intense gas-phase beams, of preparation of spatial and temporal coherence of the matter wave, and of the efficient detection of the particles. Central to all experimental demonstrations of matter wave interference are optical elements which serve to coherently manipulate wave phases, and in particular to divide the wave fronts, thus creating different possible interference paths. While bulk and surface crystals are well-adapted to diffract electrons and neutrons with de Broglie wavelengths in the range of 1~pm to 10~pm, it is impossible to use the same structures for atoms or molecules as those would stick to the surfaces. Typically beam splitters for molecules are realized by gratings. Gratings are nanofabricated highly ordered periodic structures of freestanding nanowires made from metal or semiconductor materials or realized by standing light fields using the Kapitza-Dirac effect~\cite{kapitza1933reflection}. Today the tightest bound for the quantum to classical transition comes from molecule interferometry. We shall give a brief history of molecule interferometry before we describe more details of the work horse of molecule interferometry - the Talbot-Lau interferometer.

Beams of small molecules were first scattered at surfaces in the experiments by Estermann and Stern in 1930~\cite{Estermann1930} followed by interferometry experiments with di-atomic molecules in the 1990s. In 1999, matter-wave interferometry with large neutral molecules was first demonstrated with the $C_{60}$ fullerene in Vienna~\cite{Arndt:99}. Fraunhofer far-field interference was shown by using molecular diffraction at a single nano-fabricated silicon nitride grating with a grating constant of 100~nm. The beam was collimated by a series of 5~$\mu$m slits to a beam divergence smaller than the expected beam diffraction angle of about 10~$\mu$rad. Only very few molecules originally in the beam reached the diffraction grating and the detector and typical count rates were of only very few molecules per second with a detection efficiency of around 10~$\%$. The resulting long integration time to resolve the interference pattern makes such experiments susceptible to noise. Prospects for large particle far-field interferometry and the related Poisson spot experiments can be found elsewhere~\cite{Juffmann2010} as well as new developments of promising techniques for far-field experiments~\cite{Juffmann2012}.

\textbf{Talbot-Lau interferometer:} Later molecule interferometry experiments were done with a so called Talbot-Lau interferometer (TLI) to increase the beam intensity of the diffracted beam. A TLI is operating in the near-field diffraction regime described by Fresnel integrals, where the spatial period of the diffraction grating and the interference pattern are on the same size scale. The scheme has been introduced by Clauser to cope with beams of low intensity and low collimation in interferometry experiments~\cite{clauser1992new}. An advantage of a TLI with respect to a Fraunhofer single-grating far-field interferometer is that the scaling of the distance between the gratings (Talbot length: $L_T$) is inversely proportional to the de Broglie wavelength $\lambda_{dB}$ but quadratic with the grating period $d$, $L_{T}=d^2/\lambda_{dB}$. This helps to compensate for small de Broglie wavelength by increasing the distance between the gratings.

\begin{figure}[!htb]
  \centerline{\includegraphics[totalheight=.7\textheight, width=.5\textwidth]{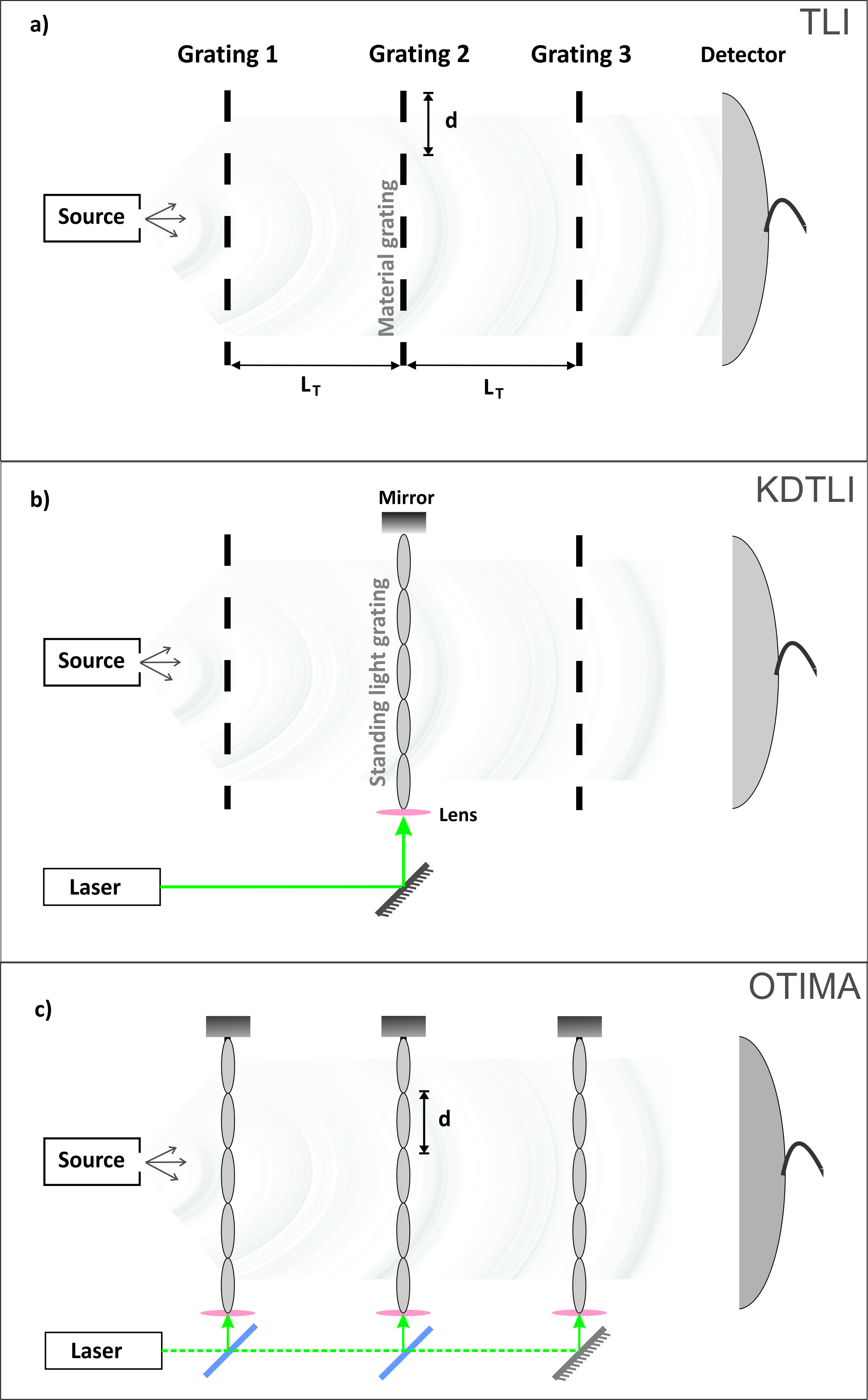}}
  \caption{Different configurations of the Talbot-Lau interferometer are shown. (A) Three material grating as experimentally realized in ~\cite{Brezger2002matter}. (B) Kapitza-Dirac-Talbot-Lau interferometer realized in ~\cite{Gerlich2007kapitza}. (C) Optical Time-Domain Ionizing Matter Interferometer
[OTIMA] as proposed in ~\cite{Nimmrichter2011concept}}\label{fig.TLI}
\end{figure}

In more detail, the three grating TLI operates with weakly collimated molecular beams with divergence of about 1~mrad and accepts a large number of molecules in the initial beam contributing to the final interference pattern. The first grating prepares the beam coherence, while imprinting a spatial structure on the molecular beam (see illustration of the Talbot-Lau interferometer in Fig \ref{fig.TLI}) acting as an absorptive mask. The second grating - the diffraction grating - is then simultaneously illuminated by $10^4$ individual coherent molecular beams. The second grating generates a self-image at the Talbot distance $L_T$.  Therefore each of the $10^4$ initial sources will be coherently mapped to the Talbot distance after the second grating. This Talbot effect results in a self-image of the second grating at half-integer multiples of $L_{T}$. The Lau effect makes an incoherent summation of the individual coherent beam sources located at the first grating to an integrated signal. Basically, the number of molecules contributing to the final interference pattern is multiplied by the number of illuminated slits of the first grating. The third grating is then placed close to this Talbot position after grating two and scans over the diffraction pattern perpendicular to the molecular beam to enable integrated signal detection. This is enabled as the period of the scan grating exactly matches the period of the Talbot self-images of the diffraction grating. Talbot, Lau and Talbot-Lau effects have been nicely illustrated by recent optical experiments~\cite{case2009realization}. The successful implementation of  different Talbot-Lau interferometers for molecules has been summarized in a recent review article~\cite{Hornberger2011review}, where more and detailed information about techniques and requirements can be found.

A recent version of the TLI is the so-called Kapitza-Dirac-Talbot-Lau interferometer (KDTLI), which has been used to demonstrate interference of a 3~nm long di-azobenzene molecule~\cite{Gerlich2007kapitza}. Here the second grating was realized by an optical phase grating, where molecules are diffracted at periodic optical potentials due to the Kapitza-Dirac effect~\cite{kapitza1933reflection}. The use of light gratings avoids the dispersive van der Waals (vdW) or Casimir-Polder (CP) attraction between molecules and gratings~\cite{hornberger2009theory}, which is known to phase shift the interference pattern but also to reduce the visibility~\cite{hackermuller2003wave} due to dispersive effects for molecular beams with finite velocity spread. The interaction effect scales with the particle velocity and particle polarisability as well as the dielectric properties of the grating material. The details of the interaction potentials and related vdW-CP effects and how those can be investigated by molecule interferometry experiments are still under intense investigation~\cite{Buhmann2012, Canaguir2012, Canaguir2012b}. Estimates show that even with improved velocity selection schemes, where the width of the selected velocity is below one percent of the mean velocity at full width half maximum, it is expected to disable interference with particles with masses beyond $10^5$~amu. Presently, the largest de Broglie interfered particle is an about 7000~amu massive perfluoro-alkylated $C_{60}$ molecule~\cite{Gerlich2011}, which gives currently the strongest bound on collapse models. For comparison all present experimental bounds are listed in Table 1.

The specifications of a TLI can be estimated easily. For instance the specifications for interference of $10^6$~amu massive particle in a Talbot-Lau type interference with gratings of period $d$=100~nm:  at a Talbot distance of $L_T$=2.5~cm a particle velocity of v=1~m/s would be needed to be constant over this distance $L_T$. [The size of the grating opening is limited to about 50nm by the size of the nanoparticles which have to transmit as well as by available technologies for grating realization by light and from material nanostructures].
For a higher mass the particle would need to be slower at the same Talbot distance or alternatively the Talbot distance would need to be extended for the same particle speed. Simple estimates show that for a particle of $10^8$~amu we find already $L_T$=2.5~m at the same speed and grating constant. However any particle traveling over that distance, even if it starts at zero velocity, will be accelerated to higher speed than 1~m/s (namely ~7~m/s over 2.5m) by Earth's gravity $g$ acting over that distance on the particle. Slowing or compensation of acceleration by additional carrier fields or in space experiments would be needed to overcome this limitation while presenting a significant experimental challenge. Therefore, TLI experiments (where the speed of the particle or equivalently it's wavelength has to have a certain value between the gratings) without compensation of Earth's gravity are limited to a particle mass of around 10$^{7}$~amu. That limit exists for all possible orientations of the interferometer to $g$. While that is true the alternative single grating far-field interferometry is not limited by $g$. Diffraction of the matter-wave at the location of the grating, the separation of maxima and minima of the interference pattern does not depend on the speed of the particles, but only on the distance between the grating and the particle detector.

\textbf{Technical challenges for mass scaling:} The quest is for new technologies which can efficiently control and manipulate the centre of mass motion of heavy particles. The mature techniques of ion manipulation and of optical tweezing are of particular  interest to scale the mass up to particles of 10~nm to 1~um (mass of $10^6$amu to $10^{10}$amu) in diameter. All experiments have to be performed under ultra-high vacuum conditions to avoid decoherence by collision. We will come back to this in Sec. \ref{sec.decoherence}. In particular the challenges are:

 1) \emph{Generation of intense particle beams:} Particles need to be slow if massive to keep the de Broglie wavelength within the range for experimental possibilities (not much smaller than pm). The ideal particle beam has a high phase space density, which means that many (ideally all) particles propagate at the same speed. The beam needs to be highly collimated, which means that the transverse velocity needs to be as small as possible, ideally zero. [Very high beam collimation
($<10 \mu$rad) would enable the conceptually simpler far-field single grating interferometry.] In the wave picture this means that the transverse or spatial coherence needs to be high. All this could be achieved by cooling techniques, that effect the centre of mass motion of particles, which yet have to be developed for complex particles. Very interesting and promising approaches have been followed in the last few years. This especially includes the collisional buffer gas cooling \cite{Maxwell2005} as well as optical cooling techniques ~\cite{shuman2010laser}. Both techniques have so far been demonstrated for diatomic molecules. Interestingly, a feedback cooling technique has been realized for optically trapped beads of 1$\mu$m in diameter in the field of optomechanics \cite{li2011millikelvin}. We will come back to this in
Sec. \ref{sec.optomech}. Furthermore particles need to be structurally stable to survive launch and detection procedures. This includes techniques to generate gas-phase particles such as by thermal or laser induced sublimation, laser desorption or ablation, but also sprays of particles from solutions, etc. and the subsequent manipulation of such particles to meet the coherence requirements of matter-wave experiments.

 2) \emph{Beam path separation:} This is the need for coherent beam splitters and other matter-wave optical elements. While different realisations of beam splitters are known for cold atoms \cite{cronin2007atom}, material and optical gratings are only existing option for large particles. The challenging part is the realization of gratings with a high enough precision in periodicity. The demand on the periodicity is very high for the TLI scheme, where the average grating pitch has to be accurate within sub-nanometer scales between all gratings. This can only be realized so far by sophisticated optical interference lithography techniques. For far-field gratings the demand is lower and electron-beam lithography with alignment pattern to avoid stitching errors is possible for fabrication. The ability to form laser light gratings from retro-reflection or other superposition of laser beams depends on the intensity and frequency stability of the laser. The power of the laser needs to be sufficient to form an optical potential strong enough to act as a phase grating. This is on the order of some 1W continuous power for fullerenes. The limited availability of stable and medium power UV and XUV (wavelength $< 200$nm) lasers limits the fabrication of grating periods by optical lithography as well as the optical grating periodicity to about 100nm (opening of about 50nm). About an order of magnitude smaller grating periods can be possibly fabricated by electron-beam lithography or direct focus ion beam (including the novel He-ion direct write) milling. Another limitation is that the grating area has to fit the size of the particle beam diameter which is on the order of 1mm. Not many fabrication techniques are capable of manufacturing precise gratings on that size scale. However, in combination with an efficient detector this dimension can be decreased.

3)  \emph{Efficient detection of large particles:}  Ideally, we want single particle detection resolution. For example in the most recent molecule interferometry experiments detection is realized by ion counting after electron impact ionization, which is known to have a very low ionisation efficiency ($10^{-4}$). This has to be at least kept at the same level for particles of increased size and mass. To resolve the interference pattern a spatial resolution on the order of the grating period is needed for near-field interferometry experiments, which is elegantly realized in the case of the TLI by the third grating. Also a high spatial resolution of the detection is needed if the particle beam is not velocity selected before entering the interferometer gratings. This is important to select the temporal coherence which is given by the distribution of de Broglie wavelengths of matter waves emitted by the source. Fluorescent molecules can be detected with single particle resolution and sufficient spatial resolution~\cite{Juffmann2012}.

In the following, we will discuss different alternative approaches on how to possibly implement experiments to probe the quantum superposition of particles. We will summarize different proposals for such experiments.

\subsubsection{Neutral particles vs charged particles}
Quantum superposition experiments per se need to avoid any decoherence effect which is able to read out which-way information and to localize the particle. A neutral particle is a natural choice for superposition experiments as the number of possible interactions, which would enable a readout of which-way information, is reduced in comparison to charged particles. That is especially true for superposition of slow particles. Therefore all interference experiments with molecules have been performed with neutrals. On the other hand the centre of mass motion of charged particles can be manipulated and controlled to a higher degree by external electric and magnetic fields. This would be handy to prepare coherent particle beams. Here we will discuss the benefits and possibilities for charged particle interferometry as neutrals have been covered in the previous section. From the matter wave point of view we have to achieve the same parameter values: de Broglie wavelength, periodicity of the diffractive element, etc. to observe an interference pattern. This especially means that for a given (high) mass the speed needs to be rather low: for m=$10^6$~amu requires around v=1~m/s.

Trapping charged particles such as electrons or ions in Paul and Penning traps has a long and successful history \cite{Paul1990}. It has been used for studies of fundamental physics such as the precise evaluation of physical constants~\cite{brown1986geonium}, for quantum information processing with one or many ions \cite{Liebfried2003, Singer2010, Duan2010}, in chemical physics to investigate the kinetics and dynamics of chemical reactions on the few molecule level under controlled conditions \cite{Mikosch2010, Willitsch2008, kreckel2005high} such as with buffer gas cooled polyatomic ions in multi-pole traps\cite{Gerlich2009, gerlich1995ion}. The obvious benefit of using charged particles for matter-wave experiments is the higher control over the motion of the particles. Guiding, trapping, and cooling is possible even for massive ions. For instance 200 bio-molecules of 410 amu have been co-trapped with laser cooled atomic ions (Ba$^+$) and cooled to 150mK \cite{ostendorf2006sympathetic}. This sympathetic cooling via Coulomb interaction of laser-cooled atomic ions with molecular ions has been demonstrated to be efficient, however a difficulty which remains is to realize an ion trap which is stable for both species. The mass over charge ratio $m/q$ must not be too different for both particles, which demands also a high control on the ionisation technique for atom and molecule.

Most of these techniques aim to spatially fix the ion in the trap to increase interaction times for spectroscopic and collision studies or to cool the ions, while we are interested in well controlled centre of mass motion for interference. It might be difficult to achieve a coherent centre of mass motion manipulation, but seems not impossible also with respect to exciting new guiding techniques such as the recently demonstrated microwave-manipulation~\cite{hoffrogge2011microwave}, the manipulation of ions by light~\cite{schneider2010optical}, or multi-pole trap techniques~\cite{gerlich1995ion}.

To scale up the mass of ions for experiments in order to test collapse models the very mature techniques of gas-phase cluster sources are available, such as sputter magnetron sources \cite{Haberland1994} or other noble gas aggregation sources with pick up for large molecules \cite{Goyal1992, Toennies1998}. Beams of such sources are intense since  they are cooled by the supersonic expansion and the mass of a single cluster can be $10^9$ amu and beyond \cite{Issendorf1999}. In combination with quadrupole mass filters, which work very similar to ion Paul traps, metal clusters of very narrow mass distributions can be realized with m/$\Delta$m=25 \cite{Pratontep2005}. Additional techniques will need to be realized for deceleration of such big clusters, but as long as the particle is charged a high degree of control is guaranteed.

While this is true massive ion interference has yet to be shown to work. A recent review article on ion interferometry lists that so far only electrons and the He$^+$ ion showed quantum interference~\cite{hasselbach2010progress}. Typically electrons are diffracted at bi-prisms or solid surfaces as if applied for holography \cite{Tonomura1987}, but also light gratings are possible utilizing the Kapitz-Dirac effect \cite{Batelaan2007}. Electron interferometry has been for instance used to investigate the Aharonov-Bohm effect \cite{Tonomura2009}. The general understanding, which is supported by experiments on for instance image charge decoherence effects, is that ions have to be very fast to not decohere via one of the multiple interaction channels with the environment~\cite{sonnentag2007measurement}. The challenge will be to avoid and shield all possible interactions of ions with surrounding matter and fields, such as for instance the coupling of the ion to its own image charge in a metal surface.

As for the neutral particles in the case of a TLI the acceleration by Earth's gravitational field has to be compensated. Guiding potentials have to be extremely flat to not influence the superposition state, to not localize the particle. External electro-magnetic fields have to be shielded by a Faraday cage of the right dimensions and materials, where recent technological progress has been made for the stabilization of magnetic fields in atom experiments~\cite{gross:08919}. Very stable electrical power supplies will be needed for the cold Paul trap for ion beam generation and an ion guide field. Electric stray fields from patch effects of adsorbed atoms and molecules at the shielding and elsewhere may be avoided as well as time varying electronic inhomogeneities in the shielding material. Edge fields of the guiding electrodes and other parts inside the shielding have to be carefully considered.

However, a simple estimate shows that for instance all applied voltages would have to be stabilized to the level of below $10^{-10}$~V for the time of interference which seems to be impossible to be achieved at the moment. At present only the neutrals show success for large particle centre of mass motion interference. On the other hand interference attempts with larger particles suffer from the non-existence of guiding, slowing and cooling techniques for neutrals. Therefore a clever solution for now is to try to take the best of both the worlds: manipulation of charged particles and interfering after neutralization, which we describe in the following section.

\subsubsection{The compromise - combination of techniques for charged and neutral particles: OTIMA}
A novel three light grating Talbot-Lau scheme in the time domain aims towards the interference of particles of up to $10^9$ amu as proposed by the Vienna molecule interferometry group and described in~\cite{Nimmrichter2011concept}. This interferometer is called optical time-domain matter-wave (OTIMA) interferometer. The charged particles will be provided by a mass filtered metal cluster aggregation source as mentioned before. A further cooling/deceleration device will reduce the velocity of the big clusters which is an existing technology for charged particles. A chopper modulated particle beam can be used for mass as well as velocity selection of the clusters in combination with a time of flight mass spectrometer (TOF-MS) detector.

The main invention is a neutralisation/ionization scheme implemented as the interferometer. The neutralization of the clusters to enable a coherent propagation of the superposition state is planned to be achieved by light-matter effects directly at the light gratings. The scheme makes use of a sequence of three vacuum ultra-violet (VUV, $\lambda$=157nm) ns long light pulses to realize the interferometer gratings. The energy of a single photon of about 8 eV is sufficient to ionize or neutralize metal clusters by photo-detachment  \cite{Haberland1994}. These processes are also applicable to large bio-molecular complexes~\cite{Marksteiner2009ionisation}. The light intensity pattern realized by three retro-reflected laser pulses hitting the propagating particles transversely at precisely timed locations with respect to each other realize the TLI gratings with grating period of $d=\lambda/2$. The standing wave normal mode pattern forms on one hand the gratings but is also a spatial resolved ionisation/neutralization device: the intensity in the antinodes is sufficient to ionize or neutralize the particles while it is not in the nodes. Therefore clusters which pass through the antinode will get ionized while others in nodes will not. This in combination with electrodes to divide the beams of neutrals and ions is the realization of an absorptive grating. The first and third gratings need to be absorptive gratings, which means they need to spatially mask out parts of the cluster beam and are realized by intensity dependent ionisation of the molecules. The second grating needs to be a phase grating and will be realized by the optical dipole force acting on the particle by making use of the Kapitza-Dirac effect \cite{Batelaan2007}.

Charged clusters will be neutralized by the first grating, diffracted at the second phase grating and ionized again for detection at the third grating. This is a promising attempt to realize matter wave experiments with very massive particles to test collapse models. More details about the OTIMA approach to test CSL can be found in \cite{nimmrichter2011testing}. The spatial size of the superposition is estimated by the grating constant and is on the order of the CSL parameter, namely $r_{C}$=100~nm. Fig. 2  illustrates the bound on $\Gamma_{CSL}$ while choosing Adlers’s value for $\lambda$ and a cluster mass of $10^{6}$~amu.

\subsection{Optomechanics: Cantilever}\label{sec.optomech}
Here we describe an experimental approach which is alternative to matter wave interferometry. While the aim to understand the limitations of quantum mechanics is as old as quantum mechanics itself, the first proposals for a table-top experimental test by using the superposition or other non-classical states of massive mesoscopic or even macroscopic mirrors have been published in the late 1990's \cite{Bose1997, Bose1999, Marshall:03}.

The mechanical motion - the vibration - of the mirror, which was later realized by a nano-mechanical or micro-mechanical cantilever, has to be prepared in the quantum mechanical ground state which is modeled by a simple harmonic oscillator ($k_{B}T<\hbar\omega$). The readout of the vibrating mirror is done by coupling it to a sensitive optical interferometer to compare the light phase with a stabilized cavity. The conceptual idea is to first prepare the mechanical oscillator in the vibrational or phononic ground state $|0\rangle$ by cooling and then generate a coherent superposition of or with the first  excited vibrational state $|1\rangle$ by single photon excitations.

A high mechanical as well as a high optical quality (Q) factor is needed to reach the regime of low dissipation to strongly couple optics to mechanics and to cool the device ultimately to the ground state. While optical control of cantilevers was under investigation for quite some time it was only in 2006 that two groups reported the successful optical cooling of mechanical cantilevers \cite{gigan2006self, schliesser2006}. Interestingly, the cooling mechanism is very similar to the optical cooling of atoms: The optical resonance - in most cases an optical cavity resonance - is slightly detuned to the cooling sideband of the mechanical resonance, which can be in the range of 1MHz to 10GHz. This opens a cooling channel for the mechanics of the cantilever though the optical leakage of the cavity.  Achieved temperatures corresponded still to a high phononic occupation - to many vibrational states occupied, but it boosted the rapid development of an exciting new field of research, namely opto-mechanics. This is summarized elsewhere \cite{revKippenberg2008, revMarquardt2009, revAspelmeyer2010}. Also the first schemes on how to generate and probe the superposition state of a cantilever appeared \cite{Kleckner1999}.

\begin{figure}[!htb]
  \centerline{\includegraphics[width=8.0cm]{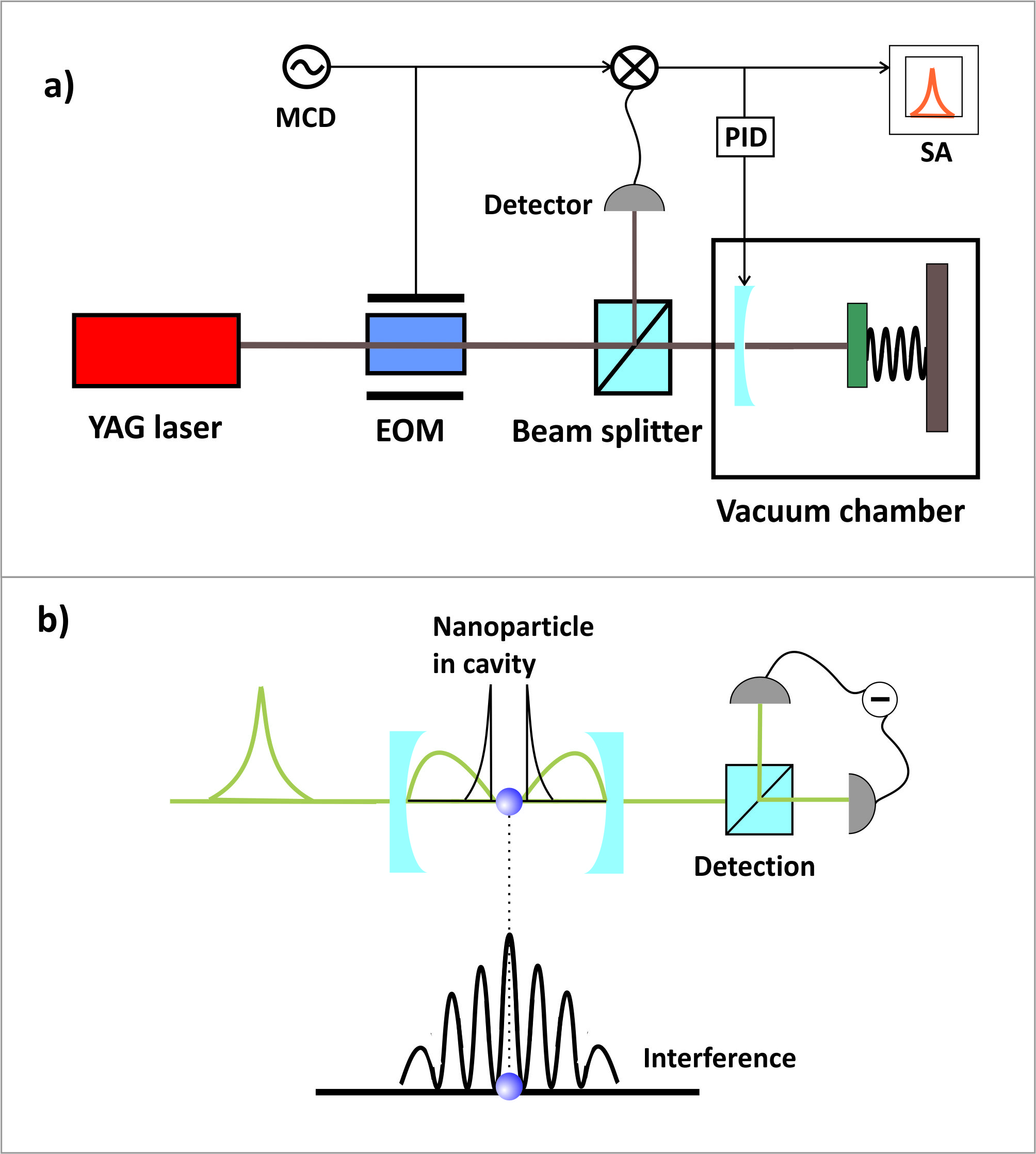}}
  \caption{Optomechanics. Upper Panel:  Prototype of optomechanically cooled cantilever as realized in ~\cite{gigan2006self}. Quantum optical detection techniques enable the sensitive read out of vibrations as they couple to light fields.
  Lower Panel:  Mechanical Resonator Interference in a Double Slit (MERID) as proposed in ~\cite{Romero2012decoherence}. The centre of mass motion of a single optically trapped nanoparticle is first cooled and then superimposed by an optical double potential. The interference pattern evolves in free fall after switching off the trapping field.}\label{fig.optomech}
\end{figure}

Only a few years later ground state cooling of micrometer sized structures has been achieved by Cleland's group~\cite{o2010quantum} and by the groups of Aspelmeyer and Painter~\cite{chan2011laser}. Advanced nanofabrication technology enabled  realization of structures with both high mechanical and high optical Q factors ($10^5$) in addition to clever optical or electronic readout techniques. This opens the door to many exciting quantum information processing and sensing experiments in the near future, but let us go back to our initial question if those structures can test our beloved collapse models.

These structures are very massive, $10^{6}$ amu to $10^{15}$ amu depending on their size, but the vibration amplitudes when compared to the parameter $r_C$ of the CSL model are very small. This limits their ability to test collapse models and the parameter range to test CSL by such systems is indicated in Fig. \ref{fig.gammavsx}. To investigate this a little further we estimate the spatial size $x_0$ of this position superposition state by using the size of the zero point motion of a simple harmonic oscillator $x_{0}=\sqrt{\hbar/2m\omega}$, where $m$ is the mass of the cantilever and $\omega$ its frequency in a harmonic potential. This spatial size of the ground state at 25~$\mu$K is $1\cdot 10^{-15} m$ for a typical micro-mechanical oscillator of a mass of 50~ng resonating at about 1~MHz~\cite{gigan2006self}.

However, spatial superpositions to tests CSL have to be on the order of 10 nm or larger ($r_C=100$~ nm), which is roughly seven orders of magnitude away from what micromechanical oscillators can achieve at the moment. Mass or frequency or combinations of both have to be improved by that amount, which is very difficult as for most materials mass and resonance frequency are coupled and depend on the spatial dimensions of the cantilever. To see vibrational state superposition larger than the quantum mechanical ground state the optomechanical device has to be driven in an extreme regime: eight orders of magnitude in mass or frequency at optical finesse of 10$^6$. But there are very interesting systems providing a larger zero point motion such as carbon materials with exceptional mechanical properties \cite{iijima1991, Novoselov2004}: e.g. individual single-wall carbon nanotube oscillators generate $x_0=$1 pm, with  m=$8\times10^{-18}$ g and  at $\omega$=100 MHz at a ground state temperature of T=2.5 mK \cite{Sazonova2004}. Such systems have been used for mass sensing with hydrogen mass resolution~\cite{chaste2012}. One big challenge remains for such carbon materials, which is their very small absorption and reflection cross sections. This means it is not clear how to realize the needed high optical quality factor for opto-mechanics. But hope is to cool via other interaction channels possibly in the electronic regime \cite{Eichler2011, Chen2009,  Brown2007}.

Another difficulty for the test of collapse models by cooling mechanical cantilevers to the ground state 
$|0\rangle$, is that the light field has to be switched on all the time. Otherwise the substrate to which the cantilever is coupled  will rapidly heat back which is probably much faster than the collapse time. There is not much time for 'free propagation' of the superposition. New ideas on pulsed optomechanics may help to prepare and reconstruct quantum states of the mechanical motion \cite{Vanner2011} faster. So ideally we would prefer to use a massive harmonic oscillator which is realized without a link to any substrate. This is what we discuss in the next section.

Optomechanical superposition using single photon postselection and their detection with nested interferometers has been discussed in ~\cite{Pepper:2011,Pepper:2012}.

Interestingly, mirror stabilization ideas are linked to the much larger interferometers for the detection of gravitational waves \cite{Braginsky2002}, while in a different parameter range due to the much higher mass of the mirrors in use.

\subsection{Micro-spheres and nanoparticles in optical potentials}
Here we describe a new and very promising route to test collapse models by generating spatial superposition states of the centre of mass motion of very massive nanoparticles and possibly even microspheres. This is a combination of optomechanics with centre of mass motion superposition states as in matter-wave interferometry. Optically trapped particles represent an almost ideal realization of a harmonic oscillator as already mentioned by Ashkin \cite{ashkin1970acceleration} and more recently re-discovered for cavity optomechanics~\cite{chang2010cavity, romero2010toward}. In comparison to mechanical cantilevers as discussed in the previous section there is no mechanical link acting as a dissipation channel for the mechanical oscillation in such systems if implemented in a vacuum chamber to avoid collisions with background gas particles. Therefore the mechanical quality factor is very large. Such trapped particles can be seen as an optomechanical system and techniques such as for cooling the oscillation - which is now the centre of mass motion of the particle in the optical trap - need to be implemented. 
Two criteria to test CSL and other collapse models, are fulfilled: a high mass of the particle and a large size of the superposition which can be comparable to $r_{C}$. This makes such experiments a strong competitor to the OTIMA cluster interferometer, see Fig.2.

In addition these systems enable free centre of mass motion of the initially trapped particles after switching off the trapping field and after generation of the spatial superposition. This allows for combination with matter-wave interferometric techniques and schemes. Recently, some ideas have been put forth how to perform tests of quantum superposition with so-called beads (balls of diameter 10~nm to 10~$\mu m$ made of glass or polystyrene)~\cite{romero2011large, romero2011optically}. The basic sequence for such experiments is to first optically dipole trap a single particle, use optical techniques to cool the centre of mass motion of the bead in the optical trapping potential. The next step is to generate a superposition state of the particle position by a double-well optical potential by single photon addressing of the first excited vibrational state, as theoretically described within cavity quantum electrodynamics (QED). Cooling must be sufficient to increase the size of the particle wave-packet to overlap with both wells, so that there is an equal probability to find the particle left or right - the coherent superposition by a measurement of a squared position observable. After switching off the trapping potential in free fall the spatial density distribution of the particle in multiple subsequent experiments can be mapped and evaluated for a quantum signature by for instance state tomography through Wigner function reconstruction \cite{romero2011optically} or much simpler by interference pattern detection at a fixed detector position. This is basically a double-slit experiment applied to very massive objects (polystyrene bead of about 30 nm diameter has a mass of $10^6$ amu). To get a significant detection statistics the single particle experiment has to be repeated many times. The connection to test collapse models is worked out in detail in \cite{Romero2012decoherence} and the experiment is called MERID which is the shortcut for mechanical resonator interference in a double slit.

The manipulation of microscopic particles as silica and polystyrene sphere but also biological cells and even living organisms as viruses by optical fields has been pioneered by Ashkin and others since the 1970s and is now very broadly applied in many fields of science~\cite{ashkin1970acceleration, ashkin1987}. Techniques which are typically summarized by the term optical tweezing include the broad fields investigating optical angular momentum \cite{allen2003}, optimizing the trapping, levitating and guiding of single dielectric particles by optical gradient and scattering forces in various geometries \cite{Ashkin2006book, Chu1998} including the guiding through hollow core photonic crystal fibres \cite{Benabid2002} and optical binding \cite{Dholakia2010}. Ashkin and co-workers demonstrated already the trapping of polystyrene and glass microspheres, of viruses and bacteria and even of complete cells in solutions and high-vacuum. They developed a vacuum loading system and they  demonstrated the stable levitation of particles at a vacuum of $10^{-6}$~mbar for half an hour by a feedback stabilization technique. A very detailed summary of this field can be found in Ashkin's book \cite{Ashkin2006book}. The particle size is typically limited to be not smaller than 1 $\mu$m to form a stable trap, while optical near-field techniques have been very recently used to trap single nanoparticles with the help of plasmonic \cite{juan2009self} or photonic crystal structures \cite{Rahmani2006} in solution.  An application of such advanced trapping techniques in vacuum has to engineer the challenge of particle-surface van der Waals (vdW) and Casimir-Polder (CP) interactions or in turn could be used to investigate those interactions. However it has been demonstrated very recently that even 30 nm particles can be optically trapped in tightly focused free beams under vacuum conditions when gradient forces dominate scattering forces and with parametric stabilization \cite{Gieseler2012}.

Experimental challenges: Cooling is again the key for this experiment. Here one needs to cool the centre of mass motion of a bead in an optical field ideally to the ground state: Ashkin pioneered the feedback stabilization \cite{Ashkin2006book}. The Doppler cooling using whispering gallery modes of the particle has been proposed~\cite{barker2010doppler}. Recently, the cooling of the centre of mass motion of a single 1 $\mu m$ glass bead to 1 mK has been achieved by a fast feedback stabilization technique~\cite{li2011millikelvin} as well as the optical parametric stabilization of a single silica nanoparticle (30nm) under vacuum conditions ($10^{-4}$mbar) at 400mK \cite{Gieseler2012}. These are the first promising steps to realize the proposed experiments to test superposition of such large and heavy particles. Importantly, feedback stabilization techniques will enable to trap beads under vacuum conditions to dissipate the kinetic energy of the trapped particle. All experiments - such as the competing cluster and molecule interferometry experiments - have to be performed at ultra-high vacuum (UHV) conditions (p $<$ 10$^{-10}$mbar) to avoid collisional decoherence of the superposition state~\cite{hornberger2003collisional}. A further challenge is that the interferometer has to be stable over the duration of many single particle experiments. One idea is for experiments in space~\cite{Kaltenbaek2012}. Centre of mass motion trapping would also be possible with ions \cite{Liebfried2003}. The electric trapping fields would replace the optical trap, but optical fields would still be needed for the cooling. While this is true free propagation of the charged particles, such as after switching off the trap in the protocols for bead superposition experiments as explained above - would not be possible. The Coulomb interaction will certainly dominate the motion of the particle - it will not be a free motion. On the other hand a recent proposal with magnetic levitated superconducting particles claims feasibility for large superpositions \cite{Romero2012magneto}.

To avoid the difficulty of ground state cooling, one possibility is to use the Talbot-Lau interferometer scheme. Here, as we know from molecule interferometry, the requirements on cooling are lowered as a quantum interference effect can be observed at low spatial coherence of the matter wave. Centre of mass motion temperatures of 1mK (for a given TLI geometry) would be sufficient to observe interference. This will work with a single particle source, but also many particles in parallel traps would be possible which could significantly reduce the operation time of the interferometer and therefore lower the stabilization criteria on the interferometer. We are looking forward to see more exciting developments in this rapidly progressing field of research.

One significant advantage of cavity optomechanics with trapped particles is the in principle very large separation of left and right for the superposition state which can be tuned by the optical field. Furthermore the optical field can be switched off and the particle can propagate in free space - showing the signature of superposition: an interference pattern in the spatial distribution. The size of the beads in the proposed experiments is on the order of 10nm to 100nm, exactly the same size and mass range where cluster matter-wave experiments such as OTIMA are heading towards. 

\subsection{Environmental Decoherence}\label{sec.decoherence}
While the aforementioned  and discussed collapse models can be seen as an exotic decoherence mechanism we here discuss decoherence effects of the environment interacting with the particle in superposition. Collisions with background particles and thermal radiation of the superimposed particle itself are counted as the major processes to localize the superimposed particle. Both decohering effects affect all the different experimental schemes to perform mesoscopic quantum superposition experiments and set limits on particle (and experimental setup) temperature as well as background pressure inside the vacuum chamber depending on the size of the particle. According to decoherence theory the superposition state is destroyed and the particle is entangled with the environment whenever any interaction of the superimposed particle with the environment has the sufficient resolution to localize - to measure the position of - the particle, which-way information is read out.

We note, that an intrinsic difficulty with the test of collapse models is that it is clear how to falsify a proposed model with respect to predicted parameters. If on the other hand no interference pattern is shown by the experiments all systematic effects related to environmental decoherence have to be excluded as reason for the quantum to classical transition. Here tuning of one of the test parameters such as mass of the particle or size of the spatial superposition will help to study the environmental decoherence effects.

Mathematically, decoherence is described (as for the case of collapse models) by the effect on the off diagonal elements of density matrix of the system including the particle and the environment which are reduced by the decoherence effect as given by the master equation very similar to Eqn. \ref{eq:decoherence}. The effect is evaluated by the decoherence rate function $\Gamma$ as given in Eqn. \ref{eq:decayrate}. More details on the concept and formalism of standard decoherence theory can be found in the references given in Section I. More details on estimations of decoherence effects and associated decay rates for superposition experiments can be found in \cite{Romero2012decoherence} and \cite{nimmrichter2011testing}. Both processes, collision and black body photon decoherence, have been experimentally investigated and compared to theory with fullerene interference \cite{Hackermuller2004decoh, hornberger2003collisional}. We will here summarize the most recent estimate from this literature to give boundaries to the experiments.

\subsubsection{Thermal decoherence} The emission, absorption, scattering of thermal - black body - radiation by the particle in superposition can localize the particle if the wavelength of that light is comparable or smaller than the size of the superposition. The emission of thermal photons is seen as the most important effect as the internal temperature of the particle is typically higher than the temperature of the environment.  As an example for $C_{70}$ fullerenes there is still full quantum contrast for emission of thermal photons by the fullerene at about 1500K, as experimentally observed~\cite{Hackermuller2004decoh, Hornberger2005}. The interference visibility is rapidly reduced for temperatures at above 2000K where the wavelength of the emitted photons is comparable to the size of the superposition which was about $1\mu$m in this experiment. For a more detailed discussion of this long and short wavelength regimes see \cite{chang2010cavity, Romero2012decoherence}.

Romero-Isart estimates in \cite{Romero2012decoherence} an emission localization time, which is inverse to the superposition decay rate, of 100ms at a temperature of 100K for a 50nm particle, but claims that this time is independent of the particle size. If this claim is correct it contradicts the observation of fullerene interferometry at 1500K. An extrapolation of this relation to mesoscopic particles ($10^6$amu - $10^8$amu) gives temperatures between 800K and 200K \cite{nimmrichter2011testing}.  In any case the predicted temperatures will have to be reached for the particle and the environment. This may require the cooling of the internal degrees of freedom of the particle which is an experimental challenge, but buffer gas techniques are in principle applicable to any particle and cool all degrees of freedom \cite{Maxwell2005, Gerlich2009}.

\subsubsection{Collision decoherence} Here the collision of the superimposed particle with any other particle present will read out which-way information. Such collision decoherence processes have been studied in depth for fullerene experiments and an elaborate theory has been developed \cite{hornberger2003collisional, Hornberger2004, Hornberger2006}. Applied to the mesoscopic range of $10^6$amu to $10^8$amu particles in OTIMA this gives minimum required pressures between $10^{-8}$mbar and $10^{-11}$mbar \cite{nimmrichter2011testing}.

Romero-Isart estimates for a 100nm sized particle and collisions with (N$_{2}$)-molecules at $10^{-11}$mbar, a decoherence time of about 100ms in MERID. A more detailed parameter set is given in \cite{Romero2012decoherence} and is in agreement with the above values for OTIMA. This pressure is possible to achieve in ultra-high vacuum experiments. The parameter set also means that single experimental sequence from preparation of the coherence through superposition and detection has to be done in 100ms, which seems feasible in OTIMA as well as MERID.  This estimate strongly depends on the mean free path of the particle under the given vacuum conditions and therefore the size of the particle.

Certainly, more work on the theoretical side is needed to investigate those decoherence effects further. This will be an important guidance for experiments. For now it seems that collision decoherence can be controlled for mesoscopic particles while maintaining extreme UHV conditions in the experiments. On the other hand thermal radiation can become a more serious issue for larger particles of $10^8$amu.

\subsection{Concluding Remarks on Laboratory Experiments}
Interference of beads levitated in optical fields and interference of large metal clusters are both promising experimental routes to test collapse models. Clearly there is a certain possibility that other experimental routes or variations and combinations of the two main proposals: OTIMA \cite{Nimmrichter2011concept} and MERID \cite{Romero2012decoherence} are successful in observing a mesoscopic single particle superposition state. OTIMA and MERID are the most advanced experimental attempts reported in the literature at this time.

In our opinion an experimental test of collapse models such as CSL with the Adler value for $\lambda$ and a mass bound of $10^6$amu is within reach in the next 5 to 10 years.
This will only be possible with intense research and development of new technologies for the handling of mesoscopic - 10nm to 100nm sized - neutral and charged particles. Conditions to control environmental decoherence seem feasible to be reached in the experiments. We hope to see a scientific competition to probe this quantum to classical transition in the coming years. It will be interesting to see if quantum mechanics again survives.

\subsection{Cosmological Bounds}
As we saw in Sec. II.F, stochastic collapse leads to a secular increase in the energy of a system. For a group of particles of mass $M$ the rate of energy increase is given by ~\cite{Adler3:07}
\begin{equation}
\frac{dE}{dt} = \frac{3}{4}\frac{\hbar^2}{r_C^2} \frac{M}{m_N^2}.
\end{equation}
See also ~\cite{Pearle:94, Bassi:03}. If there is no dissipation in the stochastic collapse model, such an energy deposit will heat the system and the absence of the observed heating can be used to put upper bounds on $\lambda$.

An important case is the ionized intergalactic medium (IGM), which has a temperature of about $2\times 10^{4}$ between the redshifts of $z=2$ and $z=4$. The IGM is kept in thermal equilibrium because the cooling due to the adiabatic expansion of the Universe and the recombination of the plasma is balanced by the energy input into the IGM that comes from astrophysical processes such as supernova explosions and quasars. An upper bound on the stochastic parameter $\lambda$ can be obtained by assuming that all the heating of the IGM is from the stochastic heating of protons and this gives that $\lambda$ should be smaller than about $10^{-8}$. More detailed discussions of cosmological and astrophysical bounds can be found in ~\cite{Adler3:07, feldmann:11}.

A subject that is recently beginning to draw attention ~\cite{Perez:2006, Sudarsky:2007, Unanue:08, Sudarsky:11, Landau:2011, Leon:2011} is the possible role of wave-function collapse in the very early Universe. A possible mechanism for the generation of primordial density fluctuations which eventually grow to form large scale structures is provided by the hypothesized inflationary epoch in the very early history of the Universe, just after the Big Bang. Inflation may have been driven by a scalar field and the zero point fluctuations of the quantized scalar field serve as a possible source for generating the requisite density inhomeogeneities ~\cite{Lyth:2009}. But how do these quantum fluctuations become classical, as the Universe evolves?  Decoherence accompanied by the many worlds interpretation has been proposed as one possible solution ~\cite{Kiefer:2009}.  Another possibility is that classicality is introduced by the models of stochastic collapse reviewed here, and it will be important and interesting to understand what sort of bounds are placed on the CSL parameters by the quantum-to-classical transition of density fluctuations in the very early Universe.

\subsection{Bounds from other physical processes}

The standard GRW and CSL values for the model parameters were reviewed in Sec. II.I [including the enhanced value for $\lambda$ proposed by ~\cite{Adler3:07} based on latent image formation in a photograph, and ~\cite{Bassi2:10} based on image formation in the eye]. Earlier in this section we discussed bounds coming from diffraction experiments and from cosmology. A few other upper bounds have been placed too, taking into account how some other processes would be affected ~\cite{Adler3:07}. In so far as $r_C$ is concerned, tentative but plausible arguments have been given that it should be in the range $10^{-5}$ to $10^{-4}$ cm.

Amongst the processes studied thus far are : (i) decay of supercurrents induced by stochastic collapse, giving $\lambda < 10^{-3}$ sec$^{-1}$, (ii) excitation of bound atomic and nuclear systems [cosmic hydrogen should not decay during the life-time of the Universe] : $\lambda < 1$ sec$^{-1}$; (iii) proton does not
decay : $\lambda < 10$; (iv) rate of spontaneous 11 keV photon emission from Germanium :   $\lambda < 10^{-11}$ sec$^{-1}$, (v) effect on the rate of radiation from free electrons : $\lambda < 10^{-5}$ sec$^{-1}$.

Another interesting result is that of ~\cite{PEARLE4}, which uses the Sudbury Neutrino Observatory data to place a limits on the ratio of collapse rates of neutron and proton. The result this analysis is that the ratio of neutron to proton collapse rates is equal to (neutron mass/proton mass)$\pm.008$. That is, mass proportionality to 1\% accuracy. We also mention  a proposal of an experiment to test the anomalous random walk due to collapse ~\cite{PEARLE5}, which however has not been performed so far.

{\it Spontaneous photon emission}: According to standard quantum mechanics, a free charged particle travels along a straight line and does not emit radiation. According to collapse models, the same particle---though being ``free''---always interacts with the noise field. It undergoes a random motion and, being charged, it emits radiation. In a similar way, also a stable atom emits radiation. Not only because it undergoes a Brownian motion in space, but also because its electrons have a non-negligible probability of being excited and subsequently de-excited with the emission of photons. Therefore, collapse models predict the spontaneous emission of radiation from matter.

The emission rate has been computed to first order perturbation theory using the mass-proportional CSL model, both for a free particle ~\cite{Fu:97} (see ~\cite{PandS,PEARLE3} for a previous analysis) and for an hydrogen atom 
~\cite{Adler2:07}. In the first case, the photon emission rate per unit photon's momentum is:
\begin{equation}
\left.\frac{d\Gamma_k}{dk}\right|_{\text{\tiny free}} = \frac{e^2 \lambda \hbar}{2 \pi^2 \epsilon_0 m_0^2 c^3 k},
\end{equation}
where $e$ is the electric charge, $\epsilon_0$ the vacuum permittivity, $m_0$ the nucleon's mass and $k$ the emitted photon's momentum. In the second case, the formula changes as follows:
\begin{equation}
\left.\frac{d\Gamma_k}{dk}\right|_{\text{\tiny H}} = 2 \left[ 1 - \left( 1 + \left(\frac{k a_0}{2}\right)^2 \right)^{-2} \right] \cdot \left.\frac{d\Gamma_k}{dk}\right|_{\text{\tiny free}},
\end{equation}
where $a_0$ is Bohr's radius. For small $k$ this expression is suppressed with respect to the rate of a free particle (the electron and the proton radiation rates add incoherently), while for large $k$ is approaches twice the free-particle's rate.

Comparison with experimental data ~\cite{Fu:97} (see ~\cite{PandS,PEARLE3} places a very strong upper bound on the collapse parameter $\lambda$ of the CSL model, only 6 orders of magnitude away from the GRW value. Therefore, it excludes an enhancement of this value of 8 orders of magnitude proposed by Adler. However, as proven in~\cite{Adler2:07} the emission rate strongly depends on the type of noise. In particular, for a colored noise the emission rate is equal to that of the white noise, times the Fourier transform  $\gamma(\omega_k)$ of the correlation function of the noise:
\begin{equation}
\left.\frac{d\Gamma_k}{dk}\right|_{\text{\tiny Colored Noise}} = \gamma(\omega_k) \cdot \left.\frac{d\Gamma_k}{dk}\right|_{\text{\tiny White Noise}},
\end{equation}
where $\omega_k$ is the frequency of the emitted photon. For example, a cut off in the frequency spectrum of the noise field at $10^{18}$Hz would highly suppress the emission rate and restore compatibility between Adler's value. This is a rather high cut off, much higher than typical cosmological ones ($\sim 10^{11}$Hz). Therefore, it is reasonable to assume that the `physical' emission rate---assuming that collapse models provide a correct description of physical phenomena---is lower than predicted by the standard mass proportional CSL model.

Table I summarizes the bounds on the CSL parameter $\lambda$ coming from various laboratory experiments and cosmological data.

{\setstretch{0.9}
\begin{center}
\begin{table*}[ht]
{
\begin{tabular}{|c|c|c|c|} 
\multicolumn{4}{c}{\bf } \\
\multicolumn{4}{c}{\bf Upper bounds on the collapse parameter $\lambda$ of the CSL model} \\
\multicolumn{4}{c}{\bf (with noise correlation length $r_C \sim 10^{-5}$ cm)} \\
\multicolumn{4}{c}{\bf } \\
\hline \hline
& & {\footnotesize\bf Distance (orders of} & {\footnotesize\bf Distance (orders of} \\
$\qquad${\footnotesize\bf Laboratory}$\qquad$ & {\footnotesize\bf Upper Bound}& {\footnotesize\bf
magnitude) from the} & {\footnotesize\bf
magnitude) from}  \\
{\footnotesize\bf Experiments} &{\footnotesize\bf on $\lambda$} & {\footnotesize\bf CSL
value} & {\footnotesize\bf Adler's
value} \\
& (unit = s) &{\footnotesize $\lambda_{\text{\tiny CSL}} \sim  10^{-17} \rm{s}^{-1}$} &
{\footnotesize $\lambda_{\text{\tiny Adler}} \sim  10^{-9} \rm{s}^{-1}$}  \\
\hline
{\footnotesize Matter-wave} & & &\\
{\footnotesize  interferometry experiments} & $10^{-5}$ & 12 
& 4 \\
\hline
{\footnotesize Decay of} & & &\\
{\footnotesize supercurrents} & $10^{-3}$ &14 & 6 \\
{\footnotesize (SQUIDS)} & & &\\ \hline
{\footnotesize Spontaneous} & & &\\
{\footnotesize X-ray emission} & $10^{-11}$ & 6 & {\footnotesize Excluded} \\
{\footnotesize from Ge} & & &\\ \hline
{\footnotesize Proton} & & & \\ 
{\footnotesize decay} & 10 & 18 & 10 \\ 
& & &\\ \hline
& & {\footnotesize\bf Distance (orders of} & {\footnotesize\bf Distance (orders of} \\
$\qquad${\footnotesize\bf Cosmological}$\qquad$ &{\footnotesize\bf Upper Bound} &{\footnotesize\bf
magnitude) from the} & {\footnotesize\bf
magnitude) from} \\
{\footnotesize\bf Data} &{\footnotesize\bf on $\lambda$} &{\footnotesize\bf CSL
value} & {\footnotesize\bf Adler's
value} \\
& (unit = s) & {\footnotesize $\lambda_{\text{\tiny CSL}} \sim  10^{-17} \rm{s}^{-1}$} &
{\footnotesize $\lambda_{\text{\tiny Adler}} \sim  10^{-9} \rm{s}^{-1}$} \\
\hline
{\footnotesize Dissociation} & & &\\
{\footnotesize  of cosmic} & 1 & 17
& 9 \\
{\footnotesize Hydrogen} & & &\\
\hline
{\footnotesize Heating of} & & &\\
{\footnotesize intergalactic medium} & $10^{-9}$ & 8 & 0 \\
{\footnotesize (IGM)} & & &\\ \hline
{\footnotesize Heating of} & & &\\
{\footnotesize Interstellar dust} & $10^{-2}$ & 15 & 7 \\
{\footnotesize grains} & & &\\ \hline
\end{tabular}}
\vskip 1.0cm \caption{The table gives upper bounds on $\lambda$ from
laboratory experiments and cosmological data, compared both with the
CSL value $\lambda_{\text{\tiny CSL}} \sim  10^{-17} {\rm s}^{-1}$ and Adler's value $\lambda_{\text{\tiny Adler}} \sim  10^{-9} {\rm s}^{-1}$ (see Section II.I). The
X-ray emission bound excludes the value $\lambda_{\text{\tiny Adler}}$ for white noise,
but this constraint is relaxed if the noise spectrum is cut off
below $10^{18} {\rm s}^{-1}$. Therefore, the bound coming from X-ray emission is very sensitive to the type of noise, i.e. to the type of collapse model. 
Large molecule diffraction would
confront $\lambda_{\text{\tiny CSL}}$ for molecules heavier  than
$\sim 10^9$ Daltons, and would confront $\lambda_{\text{\tiny Adler}}$ for
molecular weights greater than $\sim 10^5$ Daltons. (The molecular
diffraction bound on $\lambda$ decreases as the inverse square of
the molecular weight, provided the molecular radius is less than
$r_C$; see Section II.H.)}
\end{table*}
\end{center}}

As a final note, although spontaneous photon emission currently provides the strongest upper bound on the collapse parameter $\lambda$, macromolecule diffraction experiments seem to represent the most significant type of tests of collapse models, not only because they directly test the superposition principle of quantum mechanics, but also because they are less sensitive to the type of collapse model (dissipative or non-dissipative, with a white or colored noise field).

\subsection{Tests of gravity induced collapse}

Experiments on molecule interferometry and optomechanics are per se also test of gravity based collapse models: if a violation of quantum superposition were to be observed, the next task would of course be to analyze  which of the collapse models is indicated - CSL, gravity, or perhaps something entirely different. Another test, which has received considerable attention in the literature on K-model cited above, is to look for the anomalous Brownian motion induced by the stochastic reductions. Such motion, which of course could also be induced by spontaneous collapse, seems too tiny to be detectable by present technology, but further careful investigation into current technological limitations is perhaps called for ~\cite{PEARLES}.

The optomechanical cantilever experiment proposed in ~\cite{Marshall:03} and discussed above in Sec. IV.D has received particular attention with regard to gravity induced collapse. Related discussions on this experiment can be found in ~\cite{Adler:05, Bernad:06, Bassi:05}. 

An experiment to establish violation of Bell inequalities has been carried out ~\cite{Salart:2008}, assuming that the time of collapse is as determined by gravity induced collapse ~\cite{Diosi:87}.

It is not clear at this stage that there is a unique experimental signature of gravity models which will distinguish it from gravity-independent models of spontaneous collapse.

\section{Summary and Outlook}

In the early years following the development of quantum theory in the 1920s the Copenhagen interpretation took shape. Dynamics is described by deterministic Schr\"{o}dinger evolution, followed by a probabilistic evolution when the quantum system interacts with a classical measuring apparatus and quantum superposition is broken. An artificial divide was introduced between a {\it quantum} system and a
{\it classical} measuring apparatus, in order for one to be able to interpret results of experiments on atomic systems. While widely accepted, even in its early years the Copenhagen interpretation had worthy detractors including Einstein and Schr\"{o}dinger, to whom it was immediately apparent that quantum theory by itself never says that it does not apply to large macroscopic objects, and a direct consequence is paradoxes such as Schr\"{o}dinger's cat.

Two broad classes of attitudes developed towards the theory. One, given the extraordinary success of the theory, was to not question it at all: since no experiment to date contradicts the theory, one should accept the Copenhagen interpretation and the associated probability interpretation as a recipe for making predictions from theory and comparing them with experiment.

The other was to take serious note of the following difficulties : (i) classical macroscopic systems are {\it also}
quantum systems, and the quantum-classical divide introduced by the Copenhagen interpretation is vague;
(i) the observed absence of macroscopic position superpositions is in conflict with a straightforward
interpretation of the quantum superposition principle; (iii) Schr\"{o}dinger evolution being deterministic, it is `paradoxical' that probabilities should show up when one tries to describe the outcome of a measurement.

As appreciation of these difficulties grew, the Copenhagen interpretation took a back-seat, and today it is perhaps fair to say that the interpretation is no longer considered viable, and should be permanently put to rest, having well served its purpose in the early phase of quantum theory.

What has emerged on the scene instead, is three classes of explanations which address the difficulties mentioned in the previous paragraph:

\smallskip

[i] {\bf Do not modify quantum theory, but change its interpretation}: This is the many-worlds interpretation. Quantum linear superposition is never broken, despite appearances. The different outcomes of a measurement are realized in `different' universes, which do not interfere with each other because of decoherence. It seems to us that in this interpretation it is not easy to understand  the origin of probabilities and the Born probability rule. Moreover, it is not clear when the multifurcation occurs.

\smallskip

[ii] {\bf Do not modify quantum theory, but change its mathematical formulation}: This is Bohmian
mechanics. There are additional degrees of freedom---particles' positions in space---whose introduction implies that outcomes of measurements can in principle be predicted beforehand, and probabilities can be avoided.

\smallskip

[iii] {\bf Modify quantum theory}: Replace quantum theory by a different theory, which agrees with quantum theory in the microscopic limit, agrees with classical mechanics in the macroscopic limit, quantitatively and dynamically  explains the absence of macroscopic superpositions and the emergence of probabilities, and whose experimental predictions differ from those of quantum theory as one approaches the mesoscopic and macroscopic regime.

In so far as the empirical situation is concerned, all three explanations are acceptable today. The
many-worlds interpretation is in fact perhaps the favoured establishment viewpoint, because it involves  minimal change in standard quantum theory: everything can continue to be as such, and that which is not observed is attributed to parallel branches of the Universe which cannot be observed.

We have here proposed that the third avenue mentioned above be pursued: modify quantum theory. What happens during a quantum measurement is a stochastic process. Even though the initial conditions and evolution for a microscopic system, successfully described by Schr\"{o}dinger evolution, are completely deterministic, the outcome of a measurement is completely random! A straightforward resolution would be to face the evidence head-on and declare that in the dynamics, deterministic Schr\"{o}dinger evolution competes with stochastic evolution/reduction. For micro-systems Schr\"{o}dinger evolution completely dominates over stochastic reduction. For macro-systems stochastic reduction dominates Schr\"{o}dinger evolution, giving evolution the effective appearance of Newtonian mechanics. Somewhere between the micro- and the
macro- the Schr\"{o}dinger evolution becomes comparable in strength to stochastic reduction. In this regime, which experiments are now beginning to probe, new physical phenomena are predicted, which can be explained neither by quantum theory, nor by classical mechanics. These predictions, which are vulnerable to falsification, are also the strengths of a modified quantum theory. They are benchmarks against which the domain of validity and accuracy of the standard theory can be verified in the laboratory.

To this effect, the quantitative phenomenological models of Spontaneous Collapse, such as GRW, CSL, QMUPL and others,  have been rigorously defined within the well-defined mathematical framework of stochastic dynamics. The models successfully incorporate a Schr\"{o}dinger type evolution, and a stochastic evolution - the demanding requirements of non-linearity, causality, non-unitarity and norm preservation are successfully fulfilled. Two new universal parameters are introduced. One is a strength parameter which scales with mass, and ensures that stochastic reduction is negligible for microsystems, but significant for macrosystems. The other is a localization length scale which defines the linear extent of the region to which stochastic reduction localizes an expanding wave-function. While known physical and astrophysical processes put upper and lower bounds on these parameters, there is still a large permitted part of the parameter space and it will now be up to future laboratory experiments to confirm or rule out these parameter values.

Keeping in view the phenomenological nature of these models, which have been devised especially to resolve the quantum measurement problem, it is highly desirable to search for underlying physical principles and theories for these models. Theories which emerge for reasons of their own, and which are not designed for the explicit purpose of explaining measurement. Trace Dynamics does well in this regard: its goal is to derive quantum theory from a deeper level, instead of arriving at quantum theory by `quantizing' its own limiting case [classical dynamics]. It is an elegant structure in which Schr\"{o}dinger evolution is the equilibrium thermodynamics of a `gas' of classical matrices, and the ever-present Brownian motion fluctuations of the gas provide the stochastic process which competes with the equilibrium Schr\"{o}dinger evolution. Under appropriate circumstances, the Brownian motion becomes important enough to be noticeable, and is responsible for the breakdown of quantum superposition. There perhaps could not be a more compelling representation of `determinism + randomness' than `statistical equilibrium + statistical fluctuations'.
What is missing still are two important pieces of the puzzle: why do the Brownian motion fluctuations become more important for larger systems, and what is the origin of norm-preservation?

Keeping Trace Dynamics aside for a moment, one turns to investigate if gravity could couple with quantum effects and lead to an intrinsic uncertainty in space-time structure in such a way as to enable stochastic reduction in macro-systems. At first glance, this seems not possible at all: quantum gravitational effects can only be important at the Planck scale, and Planck length is too small to be of interest in laboratory physics, whereas Planck mass is too large to play a role in the quantum-classical transition. However, as more than one analysis shows, a subtle combination of linear extent of the object [measured in Planck units] and its mass [again measured in Planck units] allows gravity to bring about stochastic reduction. Gravity predicts the quantum-classical transition very much in the domain in which it is expected on other grounds.

Gravity provides a much needed physical mechanism which could underlie spontaneous collapse models. However, a proper mathematical treatment for building a gravity based theory of reduction is not yet available. It is quite possible that a generalization of Trace Dynamics that includes gravity could unify spontaneous collapse and gravity models. Doing so could also explain why the Brownian fluctuations in Trace Dynamics and spontaneous collapse become larger for larger systems. For, we have indeed explicitly seen in gravity models that the stochastic effect increases with mass.

The need for inclusion of gravity in Trace Dynamics also stems from reasons having to do with space-time structure. With hindsight, it is apparent that only when position localization is complete for nearly all objects in the Universe, it becomes meaningful to talk of a background classical spacetime geometry. If position localization is not achieved, and quantum coherence is significant, indeed that would prevent a meaningful definition of classical spacetime. Under such circumstances, and if one does not want to use classical physics as a starting point for quantization, one would have to include in trace dynamics, a matrix structure not only for the matter degrees of freedom, but also for space-time and gravity. Doing so holds the promise that one will be naturally led to a concrete mathematical formalism for describing gravity induced collapse. Investigation and development of these ideas is currently in progress.

A big stumbling block is the construction of relativistic models of spontaneous collapse. It is difficult to say at this stage whether this block will eventually be overcome, or is an indicator of some incompatibility between dynamical models of wave-function collapse and special relativity. The collapse of the wave-function is an instantaneous process, and is said to violate the `spirit' of relativity [like in an EPR experiment]. Radical though it may seem, we should eventually not be averse to a possible modification of special relativity to make it consistent with spontaneous collapse theories. Nevertheless, at the moment there is no matter-of-principle reason why collapse models should be incompatible with a fully relativistic scenario. Perhaps a generalized trace dynamics in which space and time are non-classical might have something useful to contribute here.

The development of modified quantum theory has received great impetus from the arrival of pioneering experiments on molecule interferometry and optomechanics which can test these modifications. Prime amongst these is perhaps the 1999 discovery of interference and the verification of superposition in the fullerene diffraction experiment. This paved the way for the developments that took place in the next two decades. Interference has now been observed in molecules with 7,000 amu and tremendous effort is afoot to push this frontier to a million amu and beyond. Great ingenuity is being invested in devising new experimental techniques and technology which help advance this frontier. These experiments undoubtedly hold a place beside experiments which ushered in quantum theory a century ago: the spectrum of black-body radiation, atomic spectra, photoelectric effect, and matter interferometry with electrons. A broad class of theories predict that new physics will be seen in the range $10^{6}$ amu to $10^{9}$ amu. Perhaps in two decades from now, this range will have been tested. If quantum theory is found to hold good through this regime, then chances are good that linear quantum theory is universally valid on all mass scales: we must then be content with many-worlds / Bohmian mechanics, lest a more convincing interpretation of the standard theory should emerge by then. If confirmation of the predicted modifications is found, this will be nothing short of a revolution; a new theory of dynamics will have been born, to which quantum theory and classical mechanics will be approximations.

\section*{Acknowledgements}
\noindent This publication was made possible through the support of a grant
from the John Templeton Foundation. The opinions expressed in this publication are
those of the authors and do not necessarily reflect the views of the John Templeton
Foundation.
AB is deeply indebted to S. L. Adler, D. D\"urr and G. C. Ghirardi for the long-standing collaborations on collapse models and quantum foundations. 
He wishes also to thank: M. Bahrami, C. Curceanu, D.-A. Deckert, S. Donadi, L. Ferialdi, E. Ippoliti, P. Pearle, B. Vacchini and the other co-authors of this review,
 for useful discussions and their help. He has been financially supported by the University of Trieste, MIUR (PRIN 2008), INFN, and acknowledges also 
support from  COST (MP1006). Moreover, he acknowledges kind hospitality and partial financial support from the Tata Institute for Fundamental Research 
(Mumbai), where part of this work was done. 
HU would like to thank for discussions: Markus Arndt, Klaus Hornberger, Stefan Nimmrichter, Markus Aspelmeyer, Nikolai Kiesel, Anton Zeilinger, Domenico
 Giulini, and Carsten Gundlach. HU has been financially supported by the South-English Physics network (SEPNet) and the Foundational Questions Institute 
(FQXi) and acknowledges hospitality of University of Trieste and TIFR. For useful discussions, TPS would like to thank Suratna Das, Gerald Goldin, Claus Kiefer, 
Kalle-Antti Suominen, Cenalo Vaz;  the participants of the Project Review Meeting on "Quantum Physics and the Nature of Reality" held from July 3-7 at 
the International Academy Traunkirchen (IAT), Austria, including Markus Aspelmeyer, Caslav Brukner, Thomas Durt, Simon Kochen and Anton Zeilinger; and
 the participants of the conference Quantum Malta 2012, including Markus Arndt,  Lajos Di\'osi, Detlef D\"urr,  Thomas Filk, Adrian Kent, 
 Daniel Sudarsky, Ward Struyve and Andre Xuereb. He also acknowledges support from FQXi, the kind hospitality of the Department of Theoretical Physics
 at the University of Trieste where part of this work was done, the Turku Center for Quantum Physics, Finland, and the visit to the laboratories of
 Markus Arndt and Markus Aspelmeyer at IQOQI, Vienna. KL wishes to thank Cenalo Vaz, S. Sahu and S. Barve for insightful discussions. The authors are grateful to
 S. L. Adler and G. C. Ghirardi for critically reviewing the manuscript. We would also like to thank Lajos Di\'osi, Detlef D\"urr and Ward Struyve and Adil Gangat for
 giving comments and suggesting improvements. 
\bibliography{biblioqmts3}

\end{document}